\documentclass[reprint,
 amsmath,amssymb,
 aps,
floatfix,
]{revtex4-2}

\usepackage{graphicx}% Include figure files
\usepackage{dcolumn}% Align table columns on decimal point
\usepackage{bm}% bold math
\usepackage{float}
\usepackage{mathtools}
\usepackage{makecell,tabularx}
\usepackage{array}
\usepackage{tabularray}
\usepackage{hyperref}
\usepackage{placeins}
\usepackage{titlesec}
\newcommand{\Xdot}{\dot{X}}
\newcommand{\vecXdot}{\mathbf{\dot{X}}}
\newcommand{\vecXddot}{\mathbf{\ddot{X}}}
\newcommand{\vecXdotpm}{\mathbf{\dot{X}_{\pm}}}
\newcommand{\Xddot}{\ddot{X}}
\newcommand{\Xdddot}{\dddot{X}}

\renewcommand{\Im}{\operatorname{Im}}
\renewcommand{\Re}{\operatorname{Re}}
\newcommand{\sigstar}{\sigma^{*}}

\graphicspath{{Plots/}}
\begin{document}

\preprint{APS/123-QED}

\title{Modeling the Beam of Gravitational Radiation from a Cosmic String Loop}% Force line breaks with \\

\author{Namitha Suresh}
 \email{ns873@cornell.edu}%Lines break automatically or can be forced with \\
\affiliation{
 Department of Physics, Cornell University, Ithaca, New York 14850, USA
}
\author{David F. Chernoff}
 \email{chernoff@astro.cornell.edu}
\affiliation{
 Department of Astronomy, Cornell University, Ithaca, New York 14850, USA
}

\date{\today}% It is always \today, today,

\begin{abstract}

We investigate exact and approximate techniques to calculate the emission of gravitational radiation from cosmic string loops in order to generate beam models covering the entire celestial sphere for a wide range of modes $m$. One approach entails summing over contributions of stationary and nearly-stationary points of individual, factorized, left- and right-moving modes. This ``multipoint method" generalizes traditional methods that rely on expansions around exact stationary points of mode products. A second complementary approach extends the method of steepest descent to generate an asymptotic description of the beam as $m \to \infty$. We present example calculations of the emission of power from cusp-containing loops and compare the results with those obtained by numerically exact techniques as well as by previous approaches. The multipoint method achieves its best results at an intermediate range of modes, improving over previous methods in terms of accuracy. It handles the emission from all regions of the loop not just those near cusps. We demonstrate this capability by making a detailed study of the ``pseudocusp" phenomenon.

\end{abstract}

\maketitle

%\tableofcontents

\section{\label{sec:intro}Introduction}

In certain well-studied cosmological models in string theory
\cite{Jones_2002,Sarangi_2002,Kachru_2003,Copeland_2004}, a period of
inflation may stretch microscopic theory constituents such as
F-strings (fundamental strings) or D-branes (branes wrapped on small
dimensions) to macroscopic size. Some of these entities span the
horizon of the visible universe once inflation ceases, hereafter
denoted as ``superstrings''\footnote{Similar effectively
one-dimensional objects arise from symmetry breaking phase transitions
in the early universe in Grand Unified Theories (GUTs). The tension of
such strings is set by the GUT energy scale.  We will denote these
strings as ``cosmic strings.''  Many comments apply equally well to
superstrings and cosmic strings. GUT cosmic strings are disfavored by
a variety of upper limits on the string tension and by detailed
observations of the cosmic microwave background (consistent with
adiabatic fluctuations, \cite{Bouchet_2001,Bevis_2007}). Superstrings
in braneworld scenarios may possess low tensions that satisfy the
known constraints. The stability of different types of superstrings is
model dependent.}. The horizon-crossing superstrings are the
progenitors to a network composed of long strings and smaller closed
loops. Minimally coupled strings behave in a relatively simple
fashion: a statistical description of the network converges to an
attractor solution that is realized when the universe experiences
power law expansion as it does during the Big Bang's conventional
radiation and matter eras
\cite{Vachaspati_Vilenkin_1984,Albrecht_1985,Bennett_1988,Sakellariadou_2005,Urrestilla_2008,Blanco_Pillado_2011}.
The attractor is a scaling solution: long strings collide and form
sub-horizon closed loops, the loops oscillate and emit gravitational
radiation and decay. All properties measured relative to the horizon
are fixed, e.g. the density parameter
$\Omega$ \footnote{$\Omega=\frac{8\pi G}{3H_{0}^{2}}\rho$ where $\rho$
is the energy density of the component.} for each individual component
(long strings, loops and emitted gravitational radiation) remains
constant. The statistical properties of the scaling solution are fixed
by a few parameters such as the string tension $\mu$, the number of
species or types of different string and the propensity for string
intersections to break and rejoin (intercommutation) string
elements. The indirect detection of network elements relies on string
generated gravitational perturbations that affect Standard model
components. One possible scenario is measuring lensing of
light-emitting background sources. Direct detection, on the other
hand, might involve observing the characteristic gravitational
radiation emitted by evaporating strings. The unresolved emission
produced by all the network elements in the visible Universe
contributes to the stochastic gravitational wave background (SGWB).

Each string loop emits at its fundamental frequency and all higher
harmonics. Given a loop's motion it is straightforward to compute the
gravitational radiation emitted mode by mode in different directions
on the sky \cite{Weinberg:1972kfs}. Low frequency emission from
parsec-sized loops might match the bandpass of pulsar detectors
whereas high frequency emission from the same loop might fall in the
band of frequency sensitivity of a space-based gravitational wave
instrument. Low frequency emission is generally directed over a
wide angle on the sky, high frequency emission tends to be more
narrowly collimated. The goal of this paper is to develop useful
approximate tools for determining the gravitational wave emission
beamed to the sky by string loops over the entire plausible frequency
range of interest. We unite efficient existing methods of calculation at low frequencies with new, more accurate techniques at higher frequencies.

This paper focuses on calculations for loops with cusps and without
kinks. It is known that smooth string loops
that do not self-intersect generically possess cusps \cite{Thompson}, small bits of
the string loop that move at ultra-relativistic speeds and beam high
frequency gravitational wave emission \cite{TUROK1984520,Vachaspati:1987sq,Damour_Vilenkin_2000}. If cusps are present on
string loops then their emission is expected to dominate the SGWB at
high frequencies. The SGWB is an especially important and promising
target for space-based gravitational wave detectors like LISA \cite{Auclair_2020}. Nonetheless, the methodology introduced in this paper is
potentially applicable to a wider variety of loops such as those that possess
kinks \cite{Garfinkle_1988,Allen_Ottewill_2001}. In fact, cosmic string simulations produce mostly
kink-filled loops \cite{Bennett_1988,Blanco_Pillado_2011} and it will be important to return to apply the new
methods to loops with kinks.

The exact calculation of gravitational wave emission must generally
be done numerically. However, approximate descriptions \cite{Damour_2001,Blanco_Pillado_2017} are well-developed, especially for the
power emitted by high modes. These formalisms consider the
emission from a small region on the worldsheet near cusps which is
expected to dominate at asymptotically
high frequencies. More extensive regions of the worldsheet are important at low frequencies. The methods outlined here build upon and extend
existing formalisms to systematically include
multiple regions of the loop worldsheet in addition to those near the
cusps.

Section \ref{sec:formalism} reviews the analytic description of string
loop motion, the conditions for cusp formation and the rate at which
energy, momentum and angular momentum are carried off by the emission of
gravitational radiation. The rates
depend upon one-dimensional integrals obtained by
factorization of the loop dynamics into the left- and right-moving
modes \cite{Damour_2001, Garfinkle_1987, Allen_Shellard_1992}. Finding effective approaches, both analytic and numerical, to
evaluate these integrals is the key purpose of this paper. In Section
\ref{sec:Techniques}, we describe two exact methods (real direct,
complex direct) and two approximate ones (real multipoint, complex
asymptotic) for studying the integrals of interest. In this paper we
concentrate on
the real multipoint method, providing a detailed analytic
exposition. The formalism reproduces existing
expressions for emission in the vicinity of a cusp when one is present
but applies to a much broader set of physical conditions.  We
examine necessary criteria for validity. We show that the multipoint
method works best in conjunction with other approaches, especially
the direct numerical and complex asymptotic methods. In Section \ref{sec:Turokloop} we examine the
application of the multipoint method to a representative set of string
loops \cite{KIBBLE1982141} for a range of emission directions to
highlight the loop features that impact accuracy. We show which
parts of the mode spectrum and which parts of the celestial
sphere are best covered by the various techniques we study. Section
\ref{sec:Pseudocusp} applies the multipoint method to investigate
pseudocusp emission by both analytic and numerical means. Section
\ref{sec:Results} compares three approaches (exact numerical
integration; cusp-centered calculation -- hereafter, ``single-point'' method;
and the new multipoint method) by quantifying
accuracy and performance.
Section \ref{sec:Discussion} gives a general discussion and the applications of the methods and
Section \ref{sec:Conclusion} summarizes our conclusions.

Conventions used: Throughout the paper, we have assumed
$\hbar=c=1$. The Greek letters $\mu,\nu,...$ denote
spacetime indices of 4-vectors (running from 0 to 3), raised
and lowered using the metric with signature (-,+,+,+).  In this paper we work to lowest order in the string tension and the required form of the metric is just the Minkowski metric
$\eta^{\mu\nu}=\text{diag}\left(-1,+1,+1,+1\right)$. The 3-vectors are
written in bold (eg: $\mathbf{X_{\pm}}$ is the 3-vector
$\vec{X}_{\pm}$). The Roman letters $i,j,k$ denote spatial indices running from 1 to 3 while $p,q$ run from 2 to 3. The letters $\tau$ and $\sigma$ are conformal coordinates which parameterize the cosmic string worldsheet.

\section{\label{sec:formalism} Gravitational Wave Emission from a Cosmic String Loop}
Cosmic string loops are effectively one-dimensional objects when the curvature scale of the strings is much larger than their thickness. The unperturbed dynamics is described by the Nambu-Goto equation of motion in flat spacetime. The string worldsheet is parameterized by two conformal coordinates $\tau$ and $\sigma$, and represented by $X^{\mu}(\tau,\sigma)$. In the conformal gauge, the Nambu-Goto equation of motion is the two-dimensional wave equation,
\begin{equation}
    \left(\frac{\partial^{2}}{\partial \sigma^{2}}-\frac{\partial^{2}}{\partial \tau^{2}}\right)X^{\mu}(\tau,\sigma)=0,
\end{equation}
with constraints set by the Virasoro conditions,
\begin{subequations}
\label{eq:Virasoro}
\begin{align}
    \eta_{\mu\nu}\left(\frac{\partial X^{\mu}}{\partial\tau}\right)\left(\frac{\partial X^{\nu}}{\partial\sigma}\right)&=0,
    \label{Virasoro1}
    \\
    \eta_{\mu\nu}\left(\frac{\partial X^{\mu}}{\partial\tau}\right)\left(\frac{\partial X^{\nu}}{\partial\tau}\right)+\eta_{\mu\nu}\left(\frac{\partial X^{\mu}}{\partial\sigma}\right)\left(\frac{\partial X^{\nu}}{\partial\sigma}\right)&=0.
    \label{Virasoro2}
\end{align}
\end{subequations}
For a loop of ``invariant" length $l$ (defined as $l=E/\mu$, where $E$ is the energy of the loop in the center-of-mass frame and $\mu$ is the tension of the loop) the motion is periodic in $\tau$ with a period of $l/2$ and periodic in $\sigma$ with a period of $l$. The fundamental domain may be taken to be $0 \leq \tau < l/2$ and $0 \leq \sigma < l$.

The general solution to the Nambu-Goto equation of motion is a superposition of left-moving and right-moving modes where
\begin{equation}
  \label{eqn:spacetimeX}
    X^{\mu}(\tau,\sigma)=\frac{1}{2}\left[X^{\mu}_{-}(\sigma_{-})+X^{\mu}_{+}(\sigma_{+})\right]
\end{equation}
where $\sigma_{\pm}=\tau \pm \sigma$. Here, $X^{\mu}_{\pm}(\sigma_{\pm})$ are periodic with a period $l$, i.e.
\begin{equation}
    X^{\mu}_{\pm}(\sigma_{\pm}+ l)=X^{\mu}_{\pm}(\sigma_{\pm}).
\end{equation}

In the center-of-mass frame, we choose the worldsheet coordinate $\tau$ to coincide with the coordinate time i.e. $X^{0}(\tau,\sigma)=\tau$, yields $X^{0}_{\pm}=\sigma_{\pm}$. In this gauge, the Virasoro conditions Eq. \eqref{eq:Virasoro} become
\begin{equation}
    \vecXdotpm^{2}=1,
    \label{VirasoroCOM}
\end{equation}
where the overdot denotes derivative with respect to the corresponding $\sigma_{\pm}$ variable. The spacetime vectors obey the relation
\begin{equation}
    \Xdot_{\pm}^{2}=0.
    \label{eq:Virasoro4vec}
\end{equation}
Differentiating the above equation leads to the conditions
\begin{subequations}
\label{eq:diff Virasoro}
\begin{align}
    \Xdot_{\pm}.\Xddot_{\pm}&=0,\label{diff Virasoro1}\\
    \Xdot_{\pm}.X^{(3)}_{\pm}+\Xddot_{\pm}^{2}&=0, \label{diff Virasoro2}
\end{align}
\end{subequations}
which will be used in Section \ref{sec:MultipointMethod}.
\subsection{Cusps}
The space of the tangent vectors $\vecXdotpm$ is the surface of a unit sphere and $\vecXdotpm$ trace out two separate curves on this unit sphere. If two tangent curves intersect they give rise to a cusp. When that happens a part of the string loop momentarily doubles back on itself in spacetime. Suppose the two tangent curves intersect at
\begin{equation}
    \mathbf{\Xdot_{-}}(\sigma^{c}_{-})=\mathbf{\Xdot_{+}}(\sigma^{c}_{+}).
\end{equation}
Without loss of generality, define $\tau^{c}=\left(\sigma_{+}^{c}+\sigma_{-}^{c}\right)/2$ and $\sigma^{c}=\left(\sigma^{c}_{+}-\sigma^{c}_{-}\right)/2$. It follows
from Eq. \eqref{eqn:spacetimeX} that
\begin{equation}
    \mathbf{\Xdot}^{2}(\tau^{c},\sigma^{c})=1,
\end{equation}
where $\mathbf{\Xdot}=d\mathbf{X}/d\tau$. The above result implies that the string moves at the speed of light at the cusps in the tangent vector direction. The high Lorentz boost in the regions near the cusp leads to beamed gravitational radiation and quite possibly emission of other particles for non-minimally coupled strings \cite{Vilenkin_1987,Blanco_Pillado_2001,Peloso_2003,Long_2014}.

\subsection{Energy, Momentum and Angular Momentum Emitted}

A cosmic string loop of length $l$ has fundamental frequency $f_{1}= 1/{T_{1}}=2/l$ where $T_{1}$ is the oscillation period of the loop. The undamped loop motion is periodic in the center of mass frame and all relevant functions may be expanded in terms of sinusoidal modes with frequencies $f_{m} = m f_{1}$ where $m\in\mathbb{Z}^{+}$. Since we work to lowest non-vanishing order in the string tension the waveform is a linear sum over the sinusoidal modes.

\begin{figure}[h]
    \centering
    \includegraphics[width=0.9\linewidth, height=8 cm]{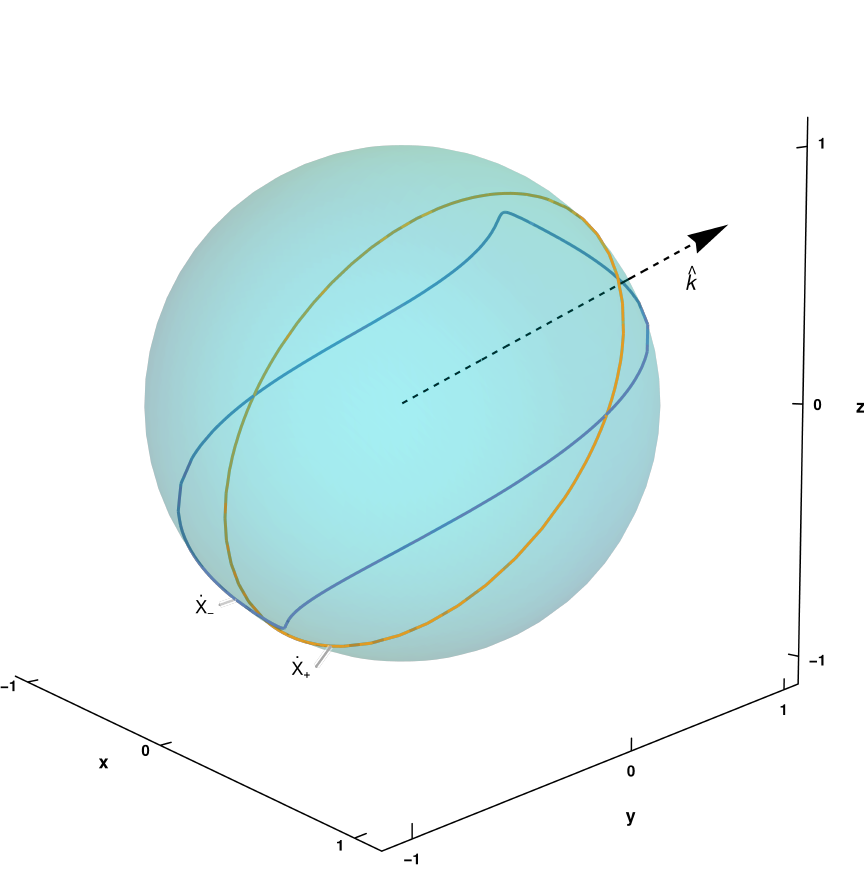}
    \caption{The unit sphere and the curves traced by the tangent vectors $\vecXdotpm$. The arrow points in the direction $\hat{k}$. (This set of tangent curves corresponds to a specific case of the ``Turok loop" which we shall examine in more detail in Section \ref{sec:Turokloop}).}
    \label{fig:KTsphere}
\end{figure}

One way to characterize the gravitational wave emission is in terms of the power emitted. The time-averaged power emitted is quadratic in the wave contributions mode-by-mode since all cross terms between different modes average to zero. We write the time-averaged differential power emitted at mode $m$ per solid angle as $dP_{m}/d\Omega$ which is a function of $\hat{k}$, the direction of emission. Figure \ref{fig:KTsphere} illustrates the celestial sphere for emission direction $\hat{k}$ overlaying
the tangent sphere for tangent curves $\mathbf{\Xdot_{\pm}}$.
Usually we will use spherical polar coordinates and write $\hat{k}= \left(\sin\theta\cos\phi,\sin\theta\sin\phi,\cos\theta\right)$. The power emitted in a single mode, $P_{m}$, is the integral of this differential power over the celestial sphere
\begin{equation}
    P_{m}=\int d\Omega \frac{dP_{m}}{d\Omega}.
\end{equation}

The details of the calculation of power emitted have been laid out in \cite{BURDEN1985277,DURRER1989238,Vilenkin:2000jqa,Allen_2001}. 
For a cosmic string loop of length $l$, define the one-dimensional integrals,
\begin{equation}
  I^{\mu}_{\pm} \equiv \frac{1}{l}\int_{-l/2}^{l/2}  d\sigma_{\pm} \Xdot^{\mu}_{\pm}e^{-\frac{i}{2}\omega_{m}k.X_{\pm}},
  \label{Imu}
\end{equation}
where $k^{\mu}=\left(1,\hat{k}\right)$ and $\omega_{m}=4\pi m/l$ for a given mode number $m$. The vectors $\hat{u}$ and $\hat{v}$
\begin{align}
    \hat{u}&=\left(-\sin\phi,\cos\phi,0\right)\\
    \hat{v}&=\left(-\cos\theta\cos\phi,-\cos\theta\sin\phi,\sin\theta\right)
\end{align}
form a right-handed orthonormal basis with $\hat{k}=\hat{u} \times \hat{v}$. We will label the 3 elements by $\left(n_{1},n_{2},n_{3}\right)=\left(\hat{k},\hat{u},\hat{v}\right)$ below.

The Fourier Transform of the energy-momentum tensor \footnote{For the energy-momentum tensor of the string $T_{\mu\nu}$, the Fourier Transform is obtained as $\tau_{\mu\nu}(k^{\lambda})=\frac{1}{T_{1}}\int_{0}^{T_{1}}dt\;\int d^{3}\mathbf{x}e^{-ik.x}T_{\mu\nu}(x^{\lambda})$ where $x^{\lambda}$ denotes the spacetime point.} of the string loop is
\begin{equation}
    \tau_{ij}=\frac{\mu l}{2}\left[I_{-}^{(n_{i})}I_{+}^{(n_{j})}+I_{-}^{(n_{j})}I_{+}^{(n_{i})}\right]
    \label{eq:EMtensor}
\end{equation}
where 
\begin{equation}
    I_{\pm}^{(n_{i})}=\mathbf{I}_{\pm}.n_{i}.
\end{equation}
The power emitted per unit solid angle in mode $m$ along direction $\hat{k}$ is \cite{DURRER1989238}
\begin{align}
    \frac{dP_{m}}{d\Omega}=\frac{G\omega_{m}^{2}}{\pi}\left[\tau^{*}_{pq}\tau_{pq}-\frac{1}{2}\tau^{*}_{qq}\tau_{pp}\right]
\end{align}
where each of $p$ and $q$ take the values 2 and 3 (repeated indices are summed over). In terms of $I_{\pm}$, the differential power is
\begin{align}
    \frac{dP_{m}}{d\Omega}&=8\pi G\mu^{2}m^{2}\left[\left(|I_{-}^{(u)}|^{2}+|I_{-}^{(v)}|^{2}\right)\left(|I_{+}^{(u)}|^{2}+|I_{+}^{(v)}|^{2}\right)\right.\nonumber\\
    &\left.+4\Im(I_{-}^{(u)}I_{-}^{(v)*})\Im(I_{+}^{(u)}I_{+}^{(v)*})\right].
    \label{dPdOmegaIpm}
\end{align}

Generic oscillating string loops radiate not only energy but also momentum
and angular momentum. 
The radiation of net momentum results in a gravitational rocket effect on the string loops which may play an important role in their dynamical motions \cite{Hogan:1984is,Vachaspati_Vilenkin_1985}. The differential rate of radiation of momentum from a cosmic string loop is 
\begin{align}
    \frac{d\mathbf{\dot{p}}_{m}}{d\Omega}&=\frac{dP_{m}}{d\Omega}\hat{k} .
\end{align}
Note that the momentum density follows directly from the flux of energy.

The calculation of the rate of radiation of angular momentum from a cosmic string loop is more complicated and has been discussed in detail in \cite{DURRER1989238}. Introducing the integrals
\begin{equation}
    M_{\pm}^{\mu\nu} \equiv \frac{1}{l}\int_{-l/2}^{l/2}d\sigma_{\pm} \Xdot^{\mu}_{\pm}X^{\nu}_{\pm}e^{-\frac{i}{2}\omega_{m}k.X_{\pm}}
    \label{Mmunu}
\end{equation}
and defining 
\begin{equation}
    M_{\pm}^{(n_{i},n_{j})}\equiv \frac{1}{l}\int_{-l/2}^{l/2}d\sigma_{\pm} \left(\mathbf{\Xdot}_{\pm}.n_{i}\right)\Bigl(\mathbf{X}_{\pm}.n_{j}\Bigr)e^{-\frac{i}{2}\omega_{m}k.X_{\pm}},
\end{equation}
the Fourier Transform of the first moment of the string energy-momentum tensor is
\begin{align}
    \tau_{ijk}&=\frac{\mu l}{4}\left[I_{-}^{(n_{i})}M_{+}^{(n_{j},n_{k})}+I_{-}^{(n_{j})}M_{+}^{(n_{i},n_{k})}\right.\nonumber\\
    &\left.+I_{+}^{(n_{i})}M_{-}^{(n_{j},n_{k})}+I_{+}^{(n_{j})}M_{-}^{(n_{i},n_{k})}\right].
\end{align}
The radiated angular momentum is orthogonal to $\hat{k}$.
The differential rate of radiated angular momentum per unit solid angle of mode number $m$ is
\begin{equation}
    \frac{d\mathbf{\dot{L}}_{m}}{d\Omega}=\frac{d\dot{L}_{m,u}}{d\Omega}\hat{u}+\frac{d\dot{L}_{m,v}}{d\Omega}\hat{v},
\end{equation}
where
\begin{align}
    \frac{d\dot{L}_{m,u}}{d\Omega}&=\frac{G}{2\pi}\left[-i\omega_{m}\left(3\tau^{*}_{13}\tau_{pp}+6\tau^{*}_{3p}\tau_{p1}\right)\right.\nonumber\\
    &\left.-\omega_{m}^{2}\left(2\tau^{*}_{3pq}\tau_{pq}-2\tau^{*}_{3p}\tau_{pqq}-\tau^{*}_{pq3}\tau_{pq}+\frac{1}{2}\tau^{*}_{qq3}\tau_{pp}\right)\right.\nonumber\\
    &\left.\hspace{5.5 cm} + c.c.\right],\\
    \frac{d\dot{L}_{m,v}}{d\Omega}&=\frac{G}{2\pi}\left[i\omega_{m}\left(3\tau^{*}_{12}\tau_{pp}+6\tau^{*}_{2p}\tau_{p1}\right)\right.\nonumber\\
    &\left.+\omega_{m}^{2}\left(2\tau^{*}_{2pq}\tau_{pq}-2\tau^{*}_{2p}\tau_{pqq}-\tau^{*}_{pq2}\tau_{pq}+\frac{1}{2}\tau^{*}_{qq2}\tau_{pp}\right)\right.\nonumber\\
    &\left.\hspace{5.5 cm}+c.c.\right].
\end{align}

Finally, the total radiated energy, momentum and angular momentum are
\begin{eqnarray}
  P & = & \sum_m P_m = \sum_m \int d\Omega \frac{dP_m}{d\Omega} \label{eq:integratedEconservedquantities},\\
  \mathbf{\dot{p}} & = & \sum_m \mathbf{\dot{p}}_m = \sum_m \int d\Omega \frac{d{\mathbf{\dot{p}}_m}}{d\Omega} \label{eq:integratedPconservedquantities},\\ 
    \mathbf{\dot{L}} & = & \sum_m \mathbf{\dot{L}}_m = \sum_m \int d\Omega \frac{d{\mathbf{\dot{L}}_m}}{d\Omega} . \label{eq:integratedLconservedquantities}
\end{eqnarray}
All three conserved quantities depends upon one dimensional integrals Eq. \eqref{Imu} and Eq. \eqref{Mmunu}. Our focus is on the calculation of these basic quantities. It is important to recognize that the one-dimensional integrals are gauge-dependent but that the combinations giving the physical quantities above are gauge-independent.

\subsection{Example Calculations for the Vachaspati-Vilenkin Loop}
To sketch the larger context, we examine the power emitted $P_{m}$, the rate of emission of momentum $\mathbf{\dot{p}}_{m}$ and the rate of emission of angular momentum $\mathbf{\dot{L}}_{m}$ as a function of the mode number $m$ for an example cosmic string loop. 

For the calculations we select
a particular family of loops (which we will refer to as ``Vachaspati-Vilenkin loops" to distinguish from other, more symmetric classes of loops introduced in the later sections) with three frequencies and characterized by two parameters $\alpha$ and $\Phi$, introduced in \cite{Vachaspati_Vilenkin_1985},
\begin{subequations}
\label{eq:exampleloop}
\begin{align}
    X_{-}(\sigma_{-})&=\frac{l}{2\pi}\left\{\frac{2\pi}{l}\sigma_{-},\right.\nonumber\\
    &\left.-\left(1-\alpha\right)\sin \left(\frac{2\pi\sigma_{-}}{l}\right)+\frac{\alpha}{3}\sin\left(\frac{6\pi\sigma_{-}}{l}\right),\right.\nonumber\\
    &\left.-\left(1-\alpha\right)\cos\left(\frac{2\pi\sigma_{-}}{l}\right)-\frac{\alpha}{3}\cos\left(\frac{6\pi\sigma_{-}}{l}\right),\right.\nonumber\\&
    \left.-\sqrt{\alpha \left(1-\alpha\right)}\sin\left(\frac{4\pi\sigma_{-}}{l}\right)\right\},
    \label{Xminus_exampleloop}
    \end{align}
    \begin{align}
    X_{+}(\sigma_{+})&=\frac{l}{2\pi}\left\{\frac{2\pi}{l}\sigma_{+},\sin\left(\frac{2\pi\sigma_{+}}{l}\right),\right.\nonumber\\&\left.-\cos\Phi\cos\left(\frac{2\pi\sigma_{+}}{l}\right),-\sin\Phi\cos\left(\frac{2\pi\sigma_{+}}{l}\right)\right\}.
    \label{Xplus_exampleloop}
    \end{align}
\end{subequations}

\begin{figure*}
\centering
\includegraphics[width=0.475\linewidth, height=6cm]{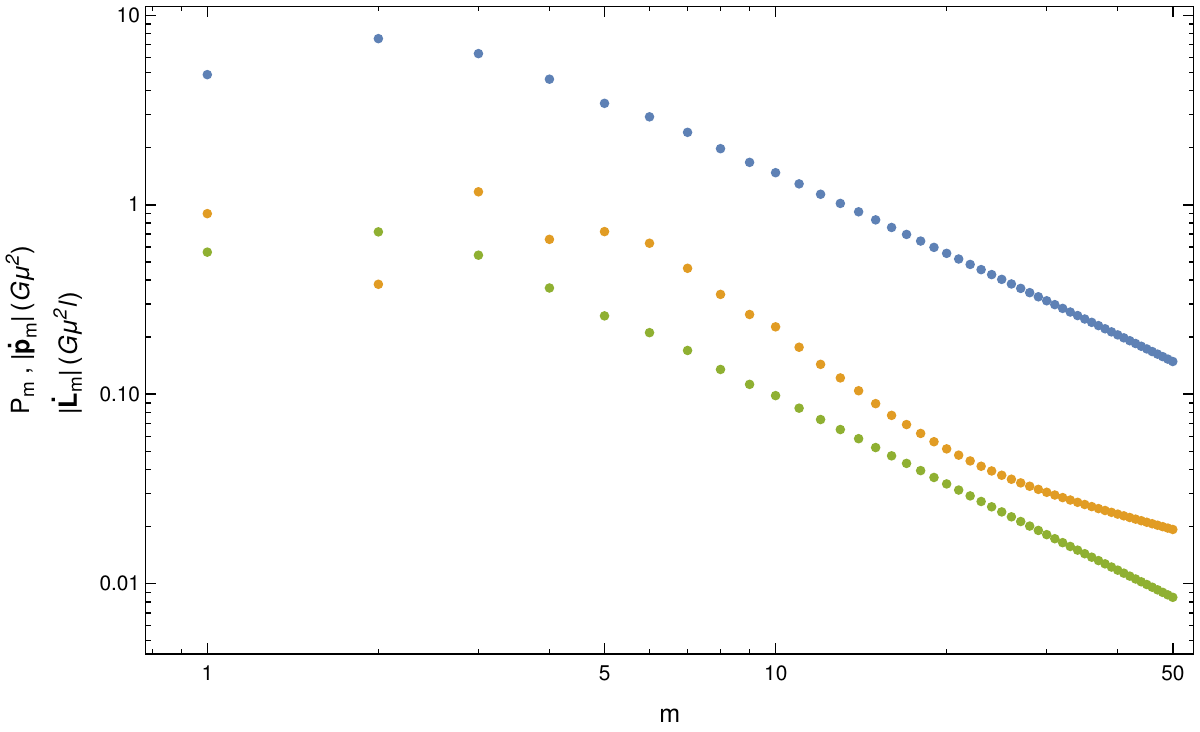} 
\quad\quad
\includegraphics[width=0.475\linewidth, height=6cm]{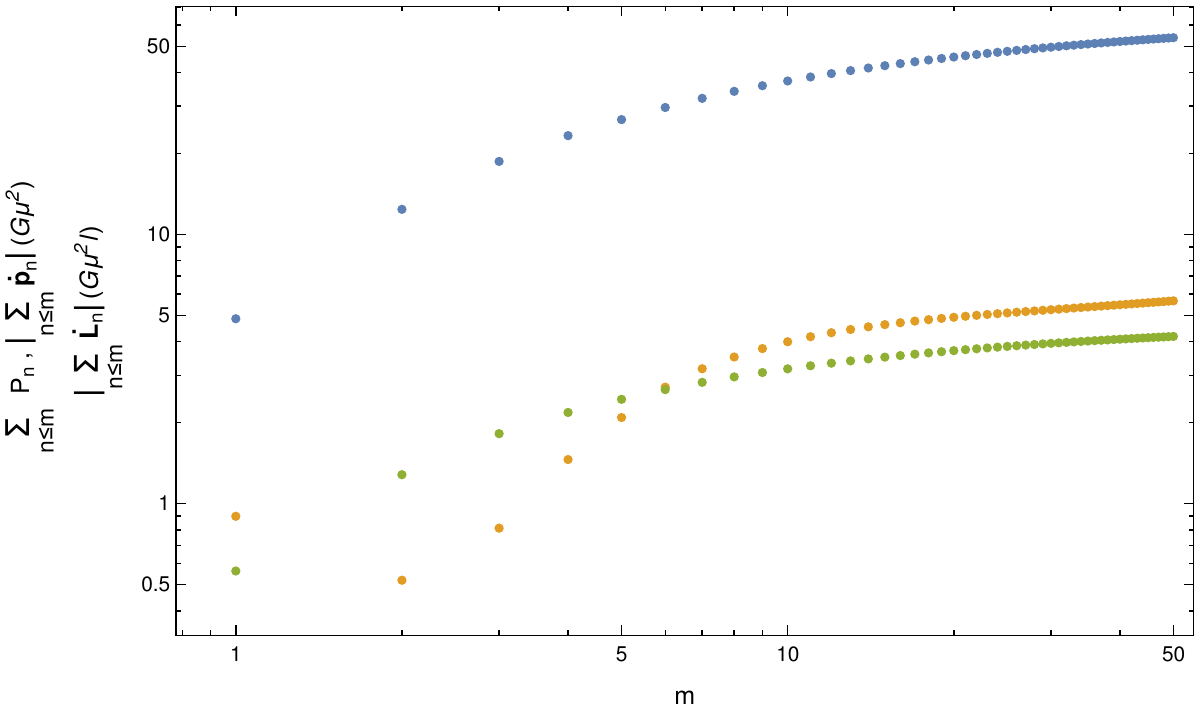}

\caption{Plot showing the radiated power, rate of radiation of momentum and angular momentum in each mode $m$ on the left and plot showing the cumulative quantities up to mode $m$ on the right, for the loop Eq. \eqref{eq:exampleloop}. The blue markers correspond to the power radiated, the orange markers correspond to the momentum and the green markers correspond to the angular momentum.}
\label{fig:example}
\end{figure*}

For the purposes of demonstration, let us select $\alpha=1/2$ and $\Phi=\pi/4$.
The integrals Eq. \eqref{Imu} and Eq. \eqref{Mmunu} are performed numerically and combined to calculate $\frac{dP_{m}}{d\Omega},\frac{d|\mathbf{\dot{p}}_{m}|}{d\Omega}$ and $\frac{d|\mathbf{\dot{L}}_{m}|}{d\Omega}$. The differential quantities are integrated over the celestial sphere numerically to give $P_{m}$, $|\mathbf{\dot{p}}_{m}|$ and $|\mathbf{\dot{L}}_{m}|$ according to Eqs. \eqref{eq:integratedEconservedquantities}, \eqref{eq:integratedPconservedquantities} and \eqref{eq:integratedLconservedquantities}. The left hand panel of \autoref{fig:example} shows the radiated power and the rate of radiated momentum and angular momentum per mode number for this loop up to $m=50$.
The calculated power, rate of radiation of momentum and rate of radiation of angular momentum from the first $50$ modes are $53.8$ $G\mu^{2}$, $5.67 G\mu^{2}$ and $4.18 G\mu^{2}l$, respectively, in general agreement with \cite{DURRER1989238}.
The right hand panel shows the cumulative quantities up to and including mode $m$. Evidently, most of the total radiated power,
momentum and angular momentum originates from the lower modes. These will be responsible for the dominant secular change in the loop's invariant length, center of mass acceleration and spin changes. This particular loop has zero
x-components of both $\mathbf{\dot{p}}_{m}$ and $\mathbf{\dot{L}}_{m}$.

We used numerical results to estimate the total power, momentum and
angular momentum emitted by the loop from all the modes. We performed
quadratures using a regular grid on the celestial sphere for all modes
with $m \le 200$ and using Monte Carlo sampling on the celestial
sphere for modes $m=100-200$ and $207$, $248$, $298$, $358$, $429$,
$515$, $619$, $743$ and $891$ (successive values increase by $\sim
1.2$).  We used linear interpolation of the Monte Carlo results to
infer the quadratures for $200 < m \le 891$.  We extrapolated the
results for $m>891$ based on the asymptotic scaling of the integrals
for a dominant cusp (large $m$ quadratures $\propto
m^{-4/3}$).  We observed and verified the power law scaling for the
numerically calculated quadratures with $300 \lesssim m \le 891$. Summing
these three contributions (explicit, interpolated and extrapolated
quadratures) allowed us to estimate the loop's total power, total rate
of radiation of momentum and total rate of radiation of angular
momentum: $P=(74.0-74.2) G\mu^{2}$, $|\mathbf{\dot{p}}|=(9.6-9.7)
G\mu^{2}$ and $|\mathbf{\dot{L}}|=(5.32-5.33) G\mu^{2}l$,
respectively. The quoted range is derived by comparing
regular versus Monte Carlo quadrature estimates in the overlap region
$101 \le m \le 200$.  Approximately 80\% of the total power, momentum
magnitude and angular momentum magnitude radiated is sourced by modes
with $m<121-122$, $527-549$ and $60$, respectively, for this
particular loop (the range is as above). 
Although the upper limits for $m$ may appear variable they can be
understood in terms of 3 qualitative factors: the intrinsic
magnitude of radiated quantities, the possibility of sign cancellations
in radiated quantities and the onset of asymptotic variation.
Details are quantitatively discussed in Appendix \ref{sec:LoopEmission}.

We find at least a factor of 2 variation among different Turok and
Vachaspati-Vilenkin loops in terms of the total rates of emission but
the qualitative conclusion that lower modes contribute the most
remains true.

\section{\label{sec:Techniques} Techniques to evaluate \texorpdfstring{$\mathbf{I_{\pm}^{\mu}}$}{Ipmmu}}

The fundamental step in the calculation of energy, momentum and angular
momentum radiated from a cosmic string loop is the estimation of the
integrals Eq. \eqref{Imu} and Eq. \eqref{Mmunu}. These are oscillatory
integrals with phase
$\frac{1}{2}\omega_{m}k.X_{\pm}=\frac{2\pi m}{l}k.X_{\pm}$. Here we describe two exact and two approximate methods of calculation.

\subsection {\label{sec:directreal} Direct Evaluation of the Real integral}

In general, the integrals $I^{\mu}_{\pm}$ cannot be reduced to an
explicit closed form (except for a few special cases
\cite{Vachaspati_Vilenkin_1985,DURRER1989238,BURDEN1985277}).
At low $m$ it is straightforward to use a
direct numerical evaluation to derive an ``exact'' numerical result. The simple trapezoidal rule is
exponentially convergent \cite{Trefethen2014TheEC} for smooth periodic
functions but there are two practical difficulties as $m$ grows large.  First, the integrand oscillates more
and more rapidly and the number of
function evaluations needed scales $\propto m$. Second, the
cancellation of signed quantities of order unity in the sum
accumulates errors that may exceed the exact, small final result.
Eventually, higher precision arithmetic must be used in the direct
evaluation because the residual answer is so small. Numerical schemes exist
that can integrate rapidly oscillating functions
\cite{Levin1996FastIO,Levin1997AnalysisOA} but in the next section we
will take advantage of the special features of the integrand and
develop a customized approach.

Direct numerical schemes related to the Fast Fourier Transform (FFT)
were explored in \cite{Allen_Shellard_1992, Blanco_Pillado_2017}. The
methodology may be succinctly summarized as follows: Assume that the
integrand, a periodic function in $\sigma_\pm$, is represented by $N$
discrete samples. A transformation of the original independent
variable to a new form absorbs part of the periodic function leaving a
pure harmonic (dependent on $m$) times a weight at the expense of
introducing a new grid with uneven spacing.  Adopt an interpolation
scheme for the weight. Let the new uniform grid have dimension $M = c
N$ with $c >> 1$ (values $c=8-16$ were typical in
\cite{Allen_Shellard_1992}).  Finally, use an FFT of length $M$ to
perform the quadrature sum. FFT-related methods have the great
advantage of computational speed for evaluating a range of mode
numbers but suffer from aliasing effects because power at high
frequencies is redistributed to low frequencies. As a practical
matter, the contamination at low $m$ is small because the true power
at high frequencies is small compared to the true power at low
frequencies, i.e. the induced relative error is small at low
frequencies; on the other hand, large $m$ results are significantly
impacted in terms of relative error. We will see by direct calculation
that the accuracy degrades as $m$ increases and, in general, the
largest mode $m$ at which good accuracy can be achieved will satisfy
$m < M/c$.

\subsection {Direct Evaluation by Deformed Complex Contour}

Let us focus on the complex analytic properties of the
integration of  $I^{\mu}_{-}$. We simplify the notation and consider one spacetime
component $\mu$ and one harmonic $m \in \mathbb{Z}^{+}$ of the
fundamental frequency.  Let $h = I^{\mu}_{-}$, $z=\sigma_-/l$,
$g=\Xdot^{\mu}_{-}$ and $f=\omega_1 k \cdot X_{-}/2$. The integral has the
form
\begin{equation}
h = \int_{-1/2}^{1/2} dz \  g(z) e^{-i m f(z)}
\label{eq:hz}
\end{equation}
where $f$ and $g$ are periodic functions of $z$. For convenience,
let $F = i f$ so that the exponential above is $e^{-m F}$.

If we were interested in finding an asymptotic approximation to the
integral at large $m$ we might follow the method of steepest descent
\cite{stein2010complex}.  The integrand is a smooth complex function
without poles and with various critical points $z_{c}$ in the complex
plane where $F'(z_c)=0$. The method of steepest descent proceeds by
deforming the integration contour to pass through critical point(s)
and also requiring that the imaginary part of $F(z)$ be constant over
the deformed contour. The latter condition suppresses the oscillations of the
integrand. Laplace's method may be used to provide an approximation
for the integral along a path for which the integrand is a {\it maximum} at
the critical point. The technique is commonly used to generate
asymptotic approximations
(e.g. \cite{stein2010complex,de1981asymptotic}).

We apply the same ideas here to deform the integration path in the
complex plane. A numerical calculation of $h$ along the new path
will {\it exactly} match the direct real calculation along the old path.
The new path also provides an asymptotic approximation to $h$ by way of the
contributions at each of the critical points visited.

\begin{figure}[h]
    \centering
    \includegraphics[width=\linewidth, height=8 cm]{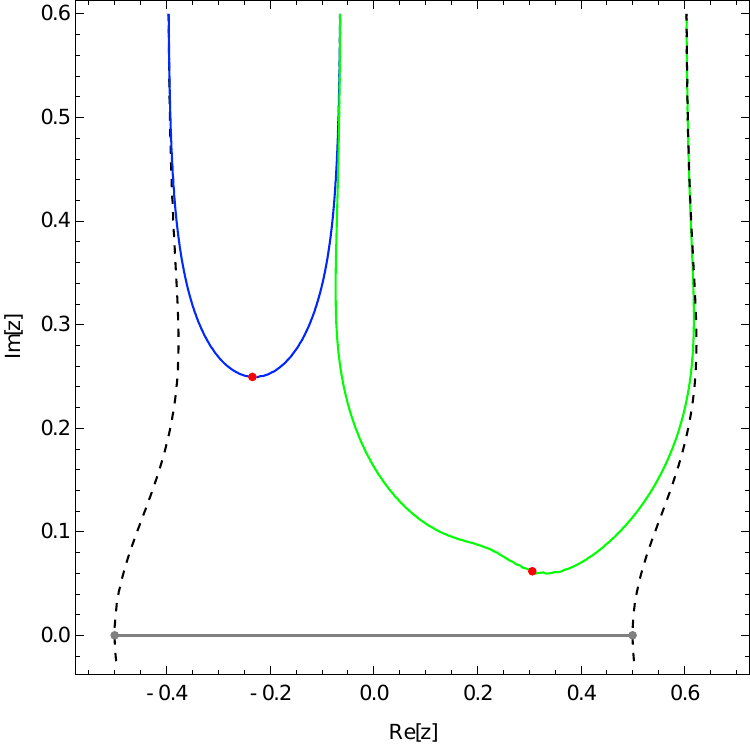}
    \caption{An example illustration of the integration contours of $I_{-}$. The light gray line shows the integration path along the real axis. The blue and green lines give the deformed contour passing through the critical points marked in red (a subset of all the critical points that are present). The path starts at $z=-1/2$, extends to positive imaginary infinity, the first critical point, positive imaginary infinity, the second critical point, positive imaginary infinity and ends at $z=1/2$.}
    \label{fig:complexphaseplanewitharrows}
\end{figure}

Figure \ref{fig:complexphaseplanewitharrows} illustrates part of the
complex $z$ plane for an example \footnote{The particulars are not crucial
at this point. This example is discussed in greater detail in
a following section. It is
Case 1 of the $X_-(\sigma_-)$
mode of a Turok loop with $\alpha=1/10$ and with $\hat k$ direction implied
by $\theta=7/10$ and $\phi=6/5$.}. The original integration path over
real $z$ is the light gray line that extends from one gray dot at
$z=-1/2$ to the other gray dot at $z=1/2$. The dashed black, blue and green lines are contours for different fixed
$\Im[F(z)]$. The red dots are critical points with $F'(z_c)=0$, a
subset of all the critical points in the plane for this particular
example. The original path is deformed to a connected set of 4 particular
segments. The first segment starts at $z=-1/2$, proceeds to large
imaginary $z$ (near vertical), then links to colored paths that
begin and end at $\Im[z]=+\infty$ and the last segment returns to the real
axis at $z=1/2$. The first segment (dashed) has
$\Im[F(z)]=\Im[F(z=-1/2)]$, ascending in the direction $\Im[z] \to
+\infty$ with large positive $\Re[F]$ (the integrand is zero). The last segment (also dashed)
descends from infinite $\Im[z]$ and large positive $\Re[F]$ to $z=1/2$ along the
contour with $\Im[F]=\Im[F(z=1/2)]$. These two segments are identical
except for a shift in $z$ by $1$. Since $f$ and $g$ are periodic in $z$
the two integrations, one up and one down, exactly cancel each
other. The remaining two segments form two looping jumps,
from and to $\Im[z]=+\infty$ with large positive $\Re[F]$ at each end; each
segment passes through a critical point at finite $z=z_c$ with $F'(z_c)=0$ and
$F''(z_c) > 0$. 

Carrying out an exact evaluation in the complex plane involves three
tasks: locating the critical points, finding the complete closed path with
piecewise-constant $\Im[F]$ and integrating along the path.
The deformed contour need not be found
exactly (in the sense that $\Im[F]$ is precisely constant) because {\it
  any} closed contour gives the same integrated result. The advantage
of locating a contour with constant $\Im[F]$ is most significant when
$m$ is large.  We have compared numerical results for
complex integration along the
deformed path to the real integration along the original contour
for $1 \le m \le 10^4$ and achieved agreement at the level of
machine precision \footnote{We used
the simple trapezoidal rule and also NIntegrate (the generic
Mathematica integration recipes) for the real direct method. We
used NIntegrate along a numerically determined path for
the complex direct integration.}.

It is worth pointing out a few practical aspects of the numerical
integration over the deformed path.  The path
is independent of $m$ so once the path is determined then large batches of $m$
can be done efficiently. The integration itself is
relatively easy to carry out because, by design, the phase of $e^{-m
  F}$ doesn't oscillate (the prefactor $g$ varies slowly). The maximum
of the integrand for each segment occurs near the associated critical
point $z_c$. The concentration of the integrand about the peak is
related to $\Re[F''(z_c)]$ -- large, positive values imply sharper
peaks near the critical point.  The peak size $e^{-m F(z_c)}$
sets the scale for additive arithmetic for any integration scheme. If
the relative error is $\epsilon$ (stemming from the
truncation error of the integration scheme) then the inferred absolute
error $\sim \epsilon e^{-m F}$. Other than its dependence on
$\epsilon$ the minimum absolute error that can be achieved numerically is constrained
by the smallest number that can be represented \footnote{For IEEE-754
double precision normalized floating point we estimate the absolute
error $\max \{ \epsilon e^{-m F}, 10^{-308} \}$.}. This is quite
different than the residual errors seen in the direct real
calculations which involves cancellation of signed quantities
of order unity.

\subsection{\label{sec:MultipointMethod} Approximate Real Multipoint Method}

The direct numerical schemes for real integration are suitable at
small $m$. Here we describe a new method, which turns out
to be appropriate for intermediate mode numbers $m$ where intermediate
means larger than those needing direct methods and smaller than those
most accurately evaluated by asymptotic means.

All previous approximations \cite{Vachaspati_Vilenkin_1985, Damour_2001, Blanco_Pillado_2017} to high mode number integrals exploit the existence of small neighborhoods where the integral builds up because the complex phase doesn't oscillate too rapidly. These methods are ideal for describing the emission from cusps where the derivatives of both the left and right-moving oscillatory phases $k.X_{\pm}$ vanish. Write the mode conformal coordinates at the cusp as $\sigma_{\pm}=\sigma_{\pm}^{(c)}$. At the cusp the two tangent vectors $\mathbf{\Xdot}_{\pm}(\sigma_{\pm}^{(c)})$ are identical. These pick out a special direction which becomes the center of expansion for the approximation. At large $m$ the emission is strongest along $\mathbf{\Xdot}_{\pm}(\sigma_{\pm}^{(c)})$ and falls rapidly away from this special direction on the celestial sphere.

This approach can be modified to yield an improved description in two senses: it will treat emission even when the special direction associated with a cusp does not exist or is irrelevant and it will improve the fidelity of the beam shape. As we will see the improvements also will enable us to extend the description to somewhat lower mode frequencies than previous methods.

For each mode the integral Eq. \eqref{Imu} gets its maximum contribution if $\hat{k}$ aligns with some $\mathbf{\Xdot}^{*}_{\pm}=\mathbf{\Xdot}_{\pm}(\sigma_{\pm}^{*})$. But this situation does not hold for either mode for most emission directions. In general, $\hat{k}$ will point away from $\mathbf{\Xdot}^{*}_{\pm}$,
\begin{equation}
    k^{\mu}(\theta,\phi)=\Xdot^{*\mu}_{\pm}+\delta_{\pm}^{\mu}
\end{equation}
where $\delta_{\pm}^{\mu}$ corresponds to the deviation for each mode. With
our definitions $k^\mu=(1,{\hat k})$ and $\Xdot^{*\mu}=(1,{\hat n})$ for
unit vectors ${\hat k}$ and ${\hat n}$ so that
$\delta^\mu=(0,{\hat k}-{\hat n})$. For any $\sigma_{\pm}^{*}$, there is a corresponding $\delta_{\pm}^{\mu}$, which is fixed by the choice of $\theta$, $\phi$ and $\sigma_{\pm}^{*}$. Previous authors have generally worked in the limit of small $\delta_{\pm}^{\mu}$, but we will retain $\delta_{\pm}^{\mu}$ exactly for the moment.
The four-vectors $X_{\pm}$ and $\Xdot_{\pm}$ can be Taylor-expanded in a small region around $\sigma_{\pm}^{*}$. The largest contribution to the variation of the phase $k.X_{\pm}$ comes from the first three powers of $\sigma_{\pm}$. At the cusp, the leading contribution comes from the cubic term $\left(\mathcal{O}\left(\sigma_{\pm}^{3}\right)\right)$ while the linear $\left(\mathcal{O}\left(\sigma_{\pm}\right)\right)$ and quadratic terms $\left(\mathcal{O}\left(\sigma_{\pm}^{2}\right)\right)$ are zero. So, for the description of the phase, we shall retain terms up to cubic order in $\sigma_{\pm}$ whereas for the prefactor $\Xdot_{\pm}$, we will retain the linear term only. With these assumptions, the Taylor expansion of $X_{\pm}, k.X_{\pm}$ and $\Xdot_{\pm}$ are
\begin{align}
    X_{\pm}(\sigma_{\pm})&= X_{\pm}^{*}+\Xdot_{\pm}^{*}(\sigma_{\pm}-\sigstar_{\pm})\nonumber\\&+\frac{1}{2}\Xddot_{\pm}^{*}(\sigma_{\pm}-\sigstar_{\pm})^{2}\nonumber\\&+\frac{1}{6}X^{(3)*}_{\pm}(\sigma_{\pm}-\sigstar_{\pm})^{3}\\
    k_{\mu}X^{\mu}_{\pm}(\sigma_{\pm})&=\left(\Xdot^{*}_{\pm,\mu}+\delta_{\pm,\mu}\right)X^{\mu}_{\pm}(\sigma_{\pm})\nonumber\\&=\Xdot_{\pm,\mu}^{*}X^{*\mu}_{\pm}-\frac{1}{6}|\Xddot_{\pm}^{*}|^{2}(\sigma_{\pm}-\sigma_{\pm}^{*})^{3}\nonumber\\&+\delta_{\pm,\mu}X^{*\mu}_{\pm}+\delta_{\pm,\mu}\Xdot^{*\mu}_{\pm}(\sigma_{\pm}-\sigma_{\pm}^{*})\nonumber\\
    &+\frac{1}{2}\delta_{\pm,\mu}\Xddot^{*\mu}_{\pm}(\sigma_{\pm}-\sigma_{\pm}^{*})^{2}\nonumber\\
    &+\frac{1}{6}\delta_{\pm,\mu}X^{(3)*\mu}_{\pm}(\sigma_{\pm}-\sigma_{\pm}^{*})^{3},\quad\text{to cubic order}
    \label{kmuXmu}\\
    \Xdot_{\pm}(\sigma_{\pm})&= \Xdot_{\pm}^{*}+\Xddot_{\pm}^{*}(\sigma_{\pm}-\sigstar_{\pm}), \quad \text{to linear order}
\end{align}
where the terms in the expansion of the phase, Eq. \eqref{kmuXmu}, have been rewritten using Eq. \eqref{eq:diff Virasoro}. We define the coefficients,
\begin{align}
    A_{m} &\equiv -\frac{2\pi m}{l}\left(-\frac{1}{6}|\Xddot^{*}_{\pm}|^{2}+\frac{1}{6}\delta_{\pm,\mu}X^{(3)*\mu}_{\pm}\right),\label{Am}\\
    B_{m} &\equiv -\frac{2\pi m}{l}\frac{1}{2}\delta_{\pm,\mu}\Xddot^{*\mu}_{\pm},\label{Bm}\\
    C_{m} &\equiv -\frac{2\pi m}{l}\delta_{\pm,\mu}\Xdot^{*\mu}_{\pm}, \label{Cm}\\
    D_{m} &\equiv -\frac{2\pi m}{l}\left(\Xdot^{*}_{\pm,\mu}X^{*\mu}_{\pm}+\delta_{\pm,\mu}X^{*\mu}_{\pm}\right), \label{Dm}
\end{align}
and make a change of variables,
\begin{align}
    \sigma_{\pm} &\rightarrow t-\frac{B_{m}}{3 A_{m}},\label{sigmasub}\\
    p &\rightarrow \frac{3 A_{m}C_{m}-B_{m}^{2}}{3A_{m}^{2}},\label{pm}\\
    q &\rightarrow \frac{2 B_{m}^{3}-9A_{m}B_{m}C_{m}+27A_{m}^{2}D_{m}}{27A_{m}^{3}}.\label{qm}
\end{align}
Substituting for these quantities in the integral Eq. \eqref{Imu}, shifting the origin $(\sigma_{\pm}-\sigstar_{\pm}) \rightarrow \sigma_{\pm}$ and extending the integration limits to $\pm \infty$ yields
\begin{align}
    I^{\mu}_{\pm}&= \left(\Xdot^{*\mu}_{\pm}-\frac{B_{m}}{3 A_{m}}\Xddot^{*\mu}_{\pm}\right)\int_{-\infty}^{\infty}dt \;e^{i A_{m}(t^{3}+p t+q)}\nonumber\\
    &+\Xddot^{*\mu}_{\pm}\int_{-\infty}^{\infty}dt\;t\;e^{i A_{m}(t^{3}+p t+q)}.
    \label{Iminus final}
\end{align}
The integrals 
\begin{subequations}
\label{eq:analytic_integrals}
\begin{align}
    I_{1} &\equiv \int_{-\infty}^{\infty}dt \;e^{i A_{m}(t^{3}+p t+q)},\\
    I_{2} &\equiv \int_{-\infty}^{\infty}dt\;t\;e^{i A_{m}(t^{3}+p t+q)}
\end{align}
\end{subequations}
can be evaluated analytically. See Appendix \ref{sec:AnalyticIntegrals} for the explicit expressions.

If ${\hat k}$ aligns exactly with one of the tangent vectors $\mathbf{\Xdot}_{\pm}(\sigma_{\pm})$ then $\delta_\pm^\mu = 0$ and the results for that mode simplify as follows:
\begin{align}
    A_{m} &\rightarrow -\frac{2\pi m}{l}\left(-\frac{1}{6}|\Xddot^{*}_{\pm}|^{2}\right),\\
    B_{m} &\rightarrow 0\\
    C_{m} &\rightarrow 0\\
    D_{m} &\rightarrow -\frac{2\pi m}{l}\left(\Xdot^{*}_{\pm,\mu}X^{*\mu}_{\pm}
    \right),\\
    \sigma_{\pm} &\rightarrow t\\
    p &\rightarrow 0\\
    q &\rightarrow \frac{D_{m}}{A_{m}}\\
    I^{\mu}_{\pm}&\rightarrow \Xdot^{*\mu}_{\pm} I_1 + \Xddot^{*\mu}_{\pm} I_2 \label{eq:Imu_cusp}\\
    I_{1} &\rightarrow \frac{-2 \pi e^{i D_m}}{A_m^{1/3} \Gamma(-1/3)}\\
    I_{2} &\rightarrow \frac{-i \pi e^{i D_m}}{A_m^{2/3} \Gamma(-2/3)} .
\end{align}
In this case $I_{1} \propto m^{-1/3}$ and $I_{2} \propto
m^{-2/3}$.

In the case that the two tangent vectors coincide forming a cusp write $l^\mu
\equiv \Xdot_{+}^{\mu}= \Xdot_{-}^{\mu}$. Exact alignment of viewing direction is
$k^\mu = l^\mu$. Now dropping constants for clarity $I^{\mu}_{\pm} \sim
l^{\mu}I_{1} + \Xddot^{*\mu}_{\pm}I_{2}$.  From above $I^{\mu}_\pm$
contain terms $\propto m^{-1/3}$ and $m^{-2/3}$. The stress energy
tensor involves symmetrized products like $I_+^{( \mu } I_-^{\nu )}$
and such expressions involve powers $m^{-2/3}$, $m^{-1}$ and
$m^{-4/3}$.  Now \cite{Damour_2001} showed that all but the last were
gauge dependent and could be removed from the stress tensor
by a suitable coordinate
transformation. In our treatment we do not make any explicit gauge
transformations but rely on the fact that we are calculating physical
observables even though they are written in terms of
gauge dependent quantities. Such observables depend
on the appropriate symmetrized combination of the one-dimensional
integrals, e.g. the differential power in Eq. \eqref{dPdOmegaIpm}.
Using the Virasoro conditions in Section \ref{sec:formalism} we find all the
gauge-dependent terms identified by \cite{Damour_2001} explicitly
vanish when we substitute $I^{\mu}_{\pm} \propto l^{\mu}I_{1} +
\Xddot^{*\mu}_{\pm}I_{2}$. The final differential cusp power is
$dP_{m}/d\Omega \propto m^2 (m^{-4/3})^2 \propto m^{-2/3}$.

For the cases where $\hat{k}$ aligns with only one of or neither of
the tangent vectors the cross terms $\dot{X}_{-}.\dot{X}_{+}$ do not
cancel out. Then $dP_{m}/d\Omega$ includes contributions from the
lower powers of $m$. The gauge argument in \cite{Damour_2001} is valid
only for the case of a cusp and in the direction of the cusp. It cannot
be applied to the more general case. It is important to recognize that
the asymptotic behavior as $m \to \infty$ will involve both
exponentials and powers of $m$, i.e. one should not conclude that the drop
off with $m$ is necessarily slower than $m^{-2/3}$. We will see, however, that
at intermediate $m$ the emission in the exact cusp direction need not be
dominant.

A similar procedure can be employed to estimate $M_{\pm}^{\mu\nu}$, with the prefactor $\Xdot_{\pm}^{\mu}X_{\pm}^{\nu}$ expanded to linear order and the phase expanded to cubic order.

The cubic expansion Eq. \eqref{kmuXmu} obviously ignores higher order terms. Consider the expansion of the phase up to the 5$^{\text{th}}$ order, i.e.,
\begin{align}
   -\frac{2\pi m}{l} k_{\mu}X_{\pm}^{\mu}\left(\sigma_{\pm}\right)&=D_{m}+C_{m}(\sigma_{\pm}-\sigma^{*}_{\pm})+B_{m}(\sigma_{\pm}-\sigma^{*}_{\pm})^{2}\nonumber\\
   &+A_{m}(\sigma_{\pm}-\sigma^{*}_{\pm})^{3}+G_{m}(\sigma_{\pm}-\sigma^{*}_{\pm})^{4}\nonumber\\
   &+H_{m}(\sigma_{\pm}-\sigma^{*}_{\pm})^{5}
\end{align}
where $A_{m}, B_{m}, C_{m}, D_{m}$ are defined in Eq. \eqref{Am} - Eq. \eqref{Dm} and 
\begin{align}
    G_{m} &= -\frac{2\pi m}{l}\frac{1}{24}k_{\mu}X^{(4)*\mu}_{\pm},\\
    H_{m} &= -\frac{2\pi m}{l}\frac{1}{120}k_{\mu}X^{(5)*\mu}_{\pm}.
\end{align}
Assume that the main contribution for the cubic phase occurs over width $|\sigma_{\pm}-\sigma^{*}_{\pm}| \sim 1/|A_m|^{1/3}$. Then the
cubic term is dominant if both
\begin{equation}
    \left(\frac{|A_{m}|}{|G_{m}|^{3/4}},\frac{|A_{m}|}{|H_{m}|^{3/5}}\right) \gtrsim \mathcal{O}(1).
    \label{AmGmHm}
\end{equation}
The above condition sets an approximate lower bound on $m$ given by, 
\begin{align}
  m &\gtrsim \frac{3l}{64\pi} \text{max}\left(\frac{|k.X^{(4)*}|^{3}}{|k.X^{(3)*}|^{4}},\frac{8}{5^{3/2}}\frac{|k.X^{(5)*}|^{3/2}}{|k.X^{(3)*}|^{5/2}}\right)  \label{eqn:conditiononm}  \\
  & \equiv m_{low,4+5}
\end{align}

Whenever any of the aforementioned conditions is violated there is no
good reason to assume the cubic expansion will provide reliable
results.  More generally {\it any} Taylor series expansion
may have a limited radius of convergence so that even
if one wanted to extend the method beyond cubic order it might not
yield more accurate results.

There are other considerations as well. The Taylor expansion of the prefactor, $\vecXdotpm$ at linear order must be a good approximation over the contributing width $|\sigma_{\pm}-\sigma^{*}_{\pm}|\sim |A_{m}|^{-1/3}$. A necessary condition
is
\begin{equation}
\left|\frac{1}{2}\mathbf{\Xdddot}^{*}_{\pm}\left(\sigma_{\pm}-\sigma^{*}_{\pm}\right)^{2}\right|<\left| \mathbf{\Xdot}^{*}_{\pm} +\mathbf{\Xddot}^{*}_{\pm}\left(\sigma_{\pm}-\sigma^{*}_{\pm}\right)\right|.
\label{eq:prefactor}
\end{equation}
In addition, the width of the peak should
be small compared to the size of the fundamental domain ($|A_m| >> 1$)
and small compared to the distance to nearby peaks.

Collectively, these inequalities imply the existence of a
lower mode $m$ for which Eq. \eqref{Iminus final} is a good approximation.
We summarize this by $m > m_{low} \equiv \max \left( m_{low,4+5},
... \right) $.

We will work exclusively at cubic order for the phase and at linear order
for the prefactor and quantify the errors of
that approach with respect to exactly known answers.

\subsubsection{Finding \texorpdfstring{$\mathbf{\sigma_{\pm}^{*}}$}{sigma*}}
For any $\hat{k}$ direction, each value of $\sigma_{\pm}^{*}$ identifies one center that is responsible for one contribution to the full integral in Eq. \eqref{Imu}. In general there can be several distinct centers, i.e. several values of $\sigma^{*}_{\pm}$. The principle is to identify all the distinct stationary or nearly-stationary points of the phase responsible for making localized contributions to Eq. \eqref{Imu} and sum these contributions.

The centers of expansion are points where $\dot{X}_{\pm}$ most closely aligns with $\hat{k}$, i.e. where the derivative of the phase $k.\dot{X}_{\pm}$ is closest to zero. The quantity of interest has the form
\begin{equation}
    k_{\mu}\Xdot^{\mu}_{\pm}=-1+\hat{k}.\vecXdotpm.
\end{equation}

Both $\hat{k}$ and $\vecXdotpm$ are unit vectors implying $-2 \leq k.\dot{X}_{\pm} \leq 0$. For the case of a cusp, the propagation direction aligns exactly with $\vecXdotpm$ i.e. $k.\dot{X}_{\pm}= 0$. If $k.\dot{X}_{\pm}<0$ (no exact alignment) then $k.\dot{X}_{\pm}$ has to be as close to zero as possible, i.e. $\sigma_{\pm}^{*}$ should be an isolated point with $k.\ddot{X}^{*}_{\pm}=0$ and a local maximum i.e. $k.X^{(3)*}_{\pm} < 0$. This is what we mean by a \textit{near-stationary point}. Note that the integral contribution will not be localized at third order if the requirement of a local maximum is dropped. This can be seen heuristically from the third order fit (with $k.X^{(3)*}_{\pm} >0$) about a putative center -- the cubic fit inevitably gives rise to a distant and distinct stationary point with $k.\dot{X}_{\pm}=0$. All such stationary points are supposed to be found as separate expansion centers. To avoid confusion with the negative signs, we will use $|k.\Xdot_{\pm}|$ as the quantity of interest and so the condition for maxima of $k.\Xdot_{\pm}$ translates to condition for minima of $|k.\Xdot_{\pm}|$. In the following discussions, ``minimum (minima)" refers to the minimum (minima) of $|k.\Xdot_{\pm}|$. In summary, we find all the local minima of $|k.\dot{X}_{\pm}|$ for each $I_{\pm}$ and sum the individual contributions to approximate the full integral given by Eq. \eqref{Imu}. The same recipe may be used to approximate $M^{\mu\nu}_{\pm}$ in Eq. \eqref{Mmunu}.
\\\\
Given a loop and propagation direction $\hat{k}$ the contributions to the functions $I_{\pm}$ are calculated according to the following general procedure:

\begin{itemize}
    \item Select $\sigstar_{\pm}$ associated with the direction $\hat{k}$ by finding the minima of $|k.\Xdot_{\pm}|$.
    
    \item Expand the phase $k.X_{\pm}$ to cubic order and the prefactor $\dot{X}_{\pm}$ to linear order in $\sigma_{\pm}$ about $\sigstar_{\pm}$.
    
    \item Make the substitutions Eq. \eqref{Am} - Eq. \eqref{qm} and estimate the integral $I_{\pm}$ given by Eq. \eqref{Iminus final}.

    \item Add up the separate contributions from all the distinct stationary or near-stationary points.
\end{itemize}

    Once the integrals $I_{\pm}$ are estimated, they can be combined using Eq. \eqref{dPdOmegaIpm} to give $dP_{m}/d\Omega$ for that direction on the celestial sphere.

The key difference between the multipoint method and previous approaches \cite{Damour_2001,Blanco_Pillado_2017} is in the treatment of different points on the string loop. The existing approaches start with the assumption that the cusps dominate the emission at high modes, since they correspond to stationary points in both the phases $k.X_{\pm}$. To find the emission in regions surrounding the cusp, one expands the phase around the cusp up to a cutoff angle set by the mode number $\theta_{cutoff} \sim m^{-1/3}$. Beyond this cutoff angle, the emission is exponentially small \cite{Damour_2001} or calculated using numerical methods \cite{Blanco_Pillado_2017}. These approaches give accurate results for the emission from a cusp as $m \to \infty$. In contrast, for each left-moving and right-moving mode the multipoint method finds the expansion point(s) on the loop which contribute the most for a chosen direction on the sphere. These points are not fixed. They need not correspond to a cusp direction and, in addition, a cusp may not even be present. For power calculations the method works separately at the level of the individual one-dimensional integrals $I_{+}$ and $I_{-}$, not the products that define $dP_m/d\Omega$. It handles contributions from both first order (cusps) and second order (non-cusps) stationary points of the phase.

\subsection{\label{sec:ComplexMethod} Asymptotic Complex Integration: Steepest Descent }

We make use of the techniques of asymptotic expansion \cite{erdelyi1956asymptotic} to construct a simple algebraic expression for $I_{\pm}^{\mu}$ for the asymptotic limit $m \to \infty$. Consider the integral $h(z)$ in Eq. \eqref{eq:hz}. The first step in the steepest descent method is to determine the critical points $F'(z_{c})=0$ on the deformed integration path. At the $i$-th
critical point $z_{c,i}$ where $g=g(z_{c,i})$, $F=F(z_{c,i})$ and
$F^{(2)}=(d^2F/dz^2)(z_{c,i})$ Laplace's method gives
\begin{equation}
    h^{(0)}_{asym,i} = e^{-m F + i \psi} \sqrt{\frac{2 \pi}{|F^{(2)}|}} \frac{g}{m^{1/2}} .
    \label{eq:asymp_h}
\end{equation}
Here, $\psi$ is the angle that the constant $\Im[F(z)]$ path makes at $z=z_{c, i}$.
It is straightforward to derive higher order approximations as an inverse expansion in powers of $m$:
\begin{eqnarray}
  h^{(1)}_{asym,i} & = & h^{(0)}_{asym,i} \left( 1 + \frac{a}{m} \right) \\
  h^{(2)}_{asym,i} & = & h^{(0)}_{asym,i} \left( 1 + \frac{a}{m} + \frac{b}{m^2} \right)
\end{eqnarray}
where $a$, $b$, etc. are functions of $g$, $F$ and higher
derivatives at the $i$-th critical point. Explicit
expressions for $a$, $b$, etc. are given in Appendix \ref{sec:Asymptotics}. The
numerical evaluation of the asymptotic values is very fast. 

\subsection{\label{sec:AlignmentLimit}Alignment Limit for the Approximate Methods}

\begin{figure}[h]
     \centering
    \includegraphics[width=\linewidth, height=6 cm]{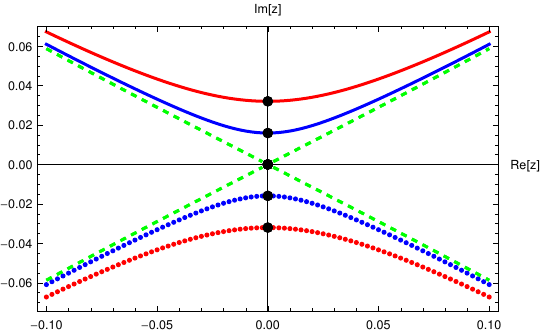}
    \caption{The complex plane for a great circle tangent vector path
      for exact and near alignment of the tangent vector and viewing
      direction.
      The large black dots are the critical points $z_c$ such that
      $F'(z_c)=0$, the lines are the contours of constant $\Im[F(z)]$
      that pass through the critical points (a common vertical contour
      with $\Re[z]=0$ has been omitted). Solid lines have
      $\Re[F''(z_c)]>0$, dotted lines has $\Re[F''(z_c)]<0$ and dashed
      lines have $\Re[F''(z_c)]=0$.  The deviation from alignment is
      indicated by the color: red lines have $|\epsilon| = 1/5$, blue
      lines $|\epsilon|=1/10$ and green lines $\epsilon=0$.}
   \label{fig:alignment}
\end{figure}

Let's examine the limiting behavior of the real and complex approximations
in a particular example in which
the beam direction approaches the locus of a tangent vector that
traces a great circle on the tangent sphere. 
Assume the tangent vector lies in the $x-y$ plane $\vecXdot_{+} = \left(l/2\pi\right)\{\cos
  \left( 2 \pi \sigma_+/l \right), -\sin \left( 2 \pi \sigma_+/l \right),
  0\}$. This example is simple enough that we can straightforwardly find the location of the critical points $z_{c,i}$, the integration path
and $F''(z_{c,i})$. Let the beam direction ${\hat k}$ be parameterized by spherical polar angles
$\theta$ and $\phi$. Beam directions corresponding to $\theta = \pi/2 + \epsilon$ for
small $|\epsilon|$ imply near alignment. 

The critical points of $F(z)$ are given by the complex $z$ such that
$\cos \epsilon \cos \left( 2 \pi z \right) = 1$ and the constant
$\Im[F]$ curves that pass through the critical point satisfy $\cos
\epsilon \cosh \left( 2 \pi y \right) \sin \left( 2 \pi x\right) = 2
\pi x$ for $z=x + i y$. These are illustrated in
\autoref{fig:alignment}.

For slight misalignment there are two distinct critical points, one
above and one below the real axis.  Note that the solutions are
invariant under the change of sign of $\epsilon$. The critical points
and constant $\Im[F]$ curves above and below the real axis are
different solutions, they do not represent positive and negative
$\epsilon$.  As $\epsilon \to 0$ exact alignment of the tangent
vector and ${\hat k}$ occurs. Now, the two separate critical points
merge on the real axis.

The second derivative of $F(z)$ along the path of constant $\Im[F(z)]$
controls whether the critical point is maximum or
minimum. $\Re[F''(z_c)]>0$ implies a peak and $\Re[F''(z_c)]<0$ implies
a bowl. In the figure the critical points with positive imaginary part
are peaks, those with negative imaginary parts are bowls. When exact
alignment occurs, $F''(z_c)=0$ and the Laplace treatment which relies on the integrand
being peaked near the critical point fails. A
numerical integration along the constant $\Im[F]$ path will still yield the
exact answer but the approximation that the bulk of the integral is
contributed near the critical point is incorrect.

Roughly speaking the argument of the exponential that is being
integrated has the form $-m \Re[F''(z_c)] \Delta z^2$ and the maximum range of
integration is $\Delta z \sim 1$. We would anticipate that $m$
must exceed $1/\Re[F''(z_c)]$ to use the complex asymptotic
approach.

\begin{figure}[h]
    \centering
    \includegraphics[width=\linewidth]{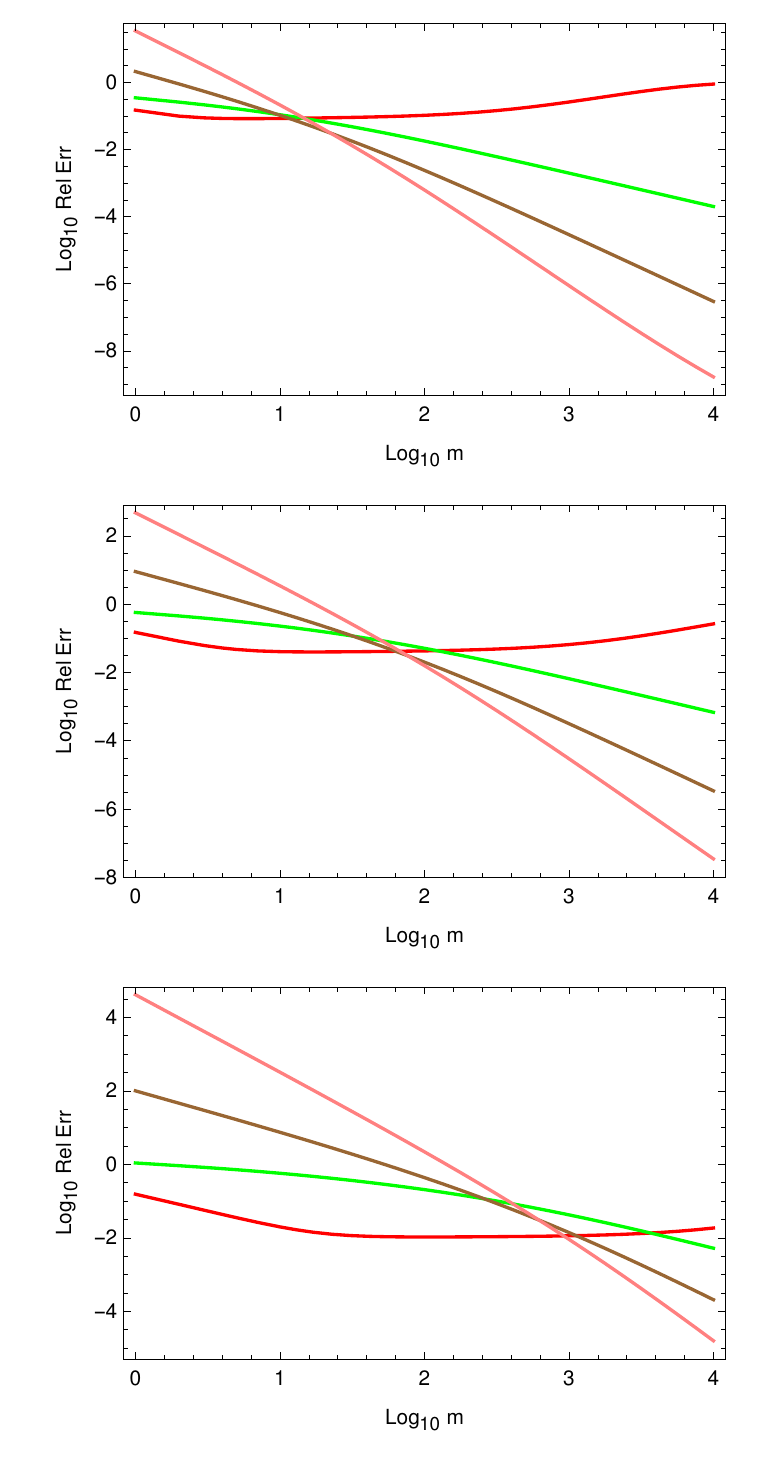}
    \caption{The relative errors for the three asymptotic approximations and the multipoint method for three different directions $\hat{k}$ on the celestial sphere for the circular tangent loop. The three directions correspond to $\theta = 0.35\pi, 0.4\pi, 0.45\pi$ and $\phi=0$. The green, brown and pink plots correspond to $\sum_{i}h^{(0)}_{asym,i}, \sum_{i}h^{(1)}_{asym,i}$ and $\sum_{i}h^{(2)}_{asym,i}$ respectively and the red plot corresponds to the multipoint method.}
    \label{fig:sequence}
\end{figure}

Figure \ref{fig:sequence} depicts the relative errors (with respect
to the exact answer) as a function of the mode number for the real
multipoint (red) and complex asymptotic
(green, brown and pink corresponding to three
separate orders $h^{(0)}$, $h^{(1)}$, $h^{(2)}$) methods. The
different panels show different misalignments between the tangent
vector and ${\hat k}$.  From top to bottom $\theta$
varies from $0.35\pi$ to $0.45\pi$ in increments of $0.05\pi$ ($\phi=0$
in all cases) -- exact alignment is $\theta=0.5\pi$. We empirically
determine the crossover $m$ (hereafter $m_{cross}$) such that the real
multipoint method yields lower relative errors for $m<m_{cross}$ and,
conversely, that complex asymptotic method yield lower relative errors
for $m>m_{cross}$.

We can infer from the figure that the size of $m_{cross}$ increases as
the angle between the beam direction and the tangent vector decreases
and that the crossover point is roughly independent of the order of
the complex asymptotic method.

The practical consequences are that for $m < m_{cross}$ we should
adopt a direct or multipoint method while for $m > m_{cross}$ it will
always be advantageous to use the asymptotic methods. In the latter
case one would adopt the highest order asymptotic scheme available.

In summary, to calculate the beam shape on the full celestial sphere
for $m>m_{low}$ we will need two approximate methods, one for near
alignment and/or small $m$, the other for misalignment and/or large
$m$.  This need for multiple methods, somewhat reminiscent of Stokes
phenomena, might have been formally anticipated from the qualitatively
different mathematical forms that the real multipoint and complex asymptotic
methods
generate. The real method gives a power law $I \propto m^{-1/3}$ for
exact alignment whereas the complex asymptotic method produces a
leading, exponential $I \propto e^{-m F}$. Consider a sequence of
misaligned beams that approach alignment. Evidently one would need
many correction terms beyond those of $h^{(0)}$, $h^{(1)}$,
etc. multiplying the exponential to reproduce the exact, aligned
results.

\section{\label{sec:Turokloop}Turok Loop -- Multipoint Examples}
In this section we will systematically
explore the qualitatively different cases that emerge in a practical
application of the multipoint calculation.
We choose a set of loops introduced by Kibble and Turok in \cite{KIBBLE1982141}. These loops (which we shall refer to as ``Turok loops") are distinguished from the Vachaspati-Vilenkin loops (Eq. \eqref{eq:exampleloop}) introduced earlier in their description of $X_{-}$. Turok loops involve two frequencies and are characterized by two parameters $\alpha \in [0,1]$ and $\Phi \in [-\pi,\pi]$,
\begin{subequations}
\label{eq:Turokloop}
\begin{align}
    X_{-}(\sigma_{-})&=\frac{l}{2\pi}\left\{\frac{2\pi}{l}\sigma_{-},\right.\nonumber\\
    &\left.-(1-\alpha)\sin \left(\frac{2\pi\sigma_{-}}{l}\right)-\frac{\alpha}{3}\sin\left(\frac{6\pi\sigma_{-}}{l}\right),\right.\nonumber\\
    &\left.-(1-\alpha)\cos\left(\frac{2\pi\sigma_{-}}{l}\right)-\frac{\alpha}{3}\cos\left(\frac{6\pi\sigma_{-}}{l}\right),\right.\nonumber\\&
    \left.-2\sqrt{\alpha(1-\alpha)}\cos\left(\frac{2\pi\sigma_{-}}{l}\right)\right\},
    \label{Xminus Turok}
    \end{align}
    \begin{align}
    X_{+}(\sigma_{+})&=\frac{l}{2\pi}\left\{\frac{2\pi}{l}\sigma_{+},\sin\left(\frac{2\pi\sigma_{+}}{l}\right),\right.\nonumber\\&\left.-\cos\left(\frac{2\pi\sigma_{+}}{l}\right)\cos\Phi,-\cos\left(\frac{2\pi\sigma_{+}}{l}\right)\sin\Phi\right\}.
    \label{Xplus Turok}
    \end{align}
\end{subequations}    

\begin{figure}[h!]
    \centering
    \includegraphics[width=\linewidth, height=8 cm]{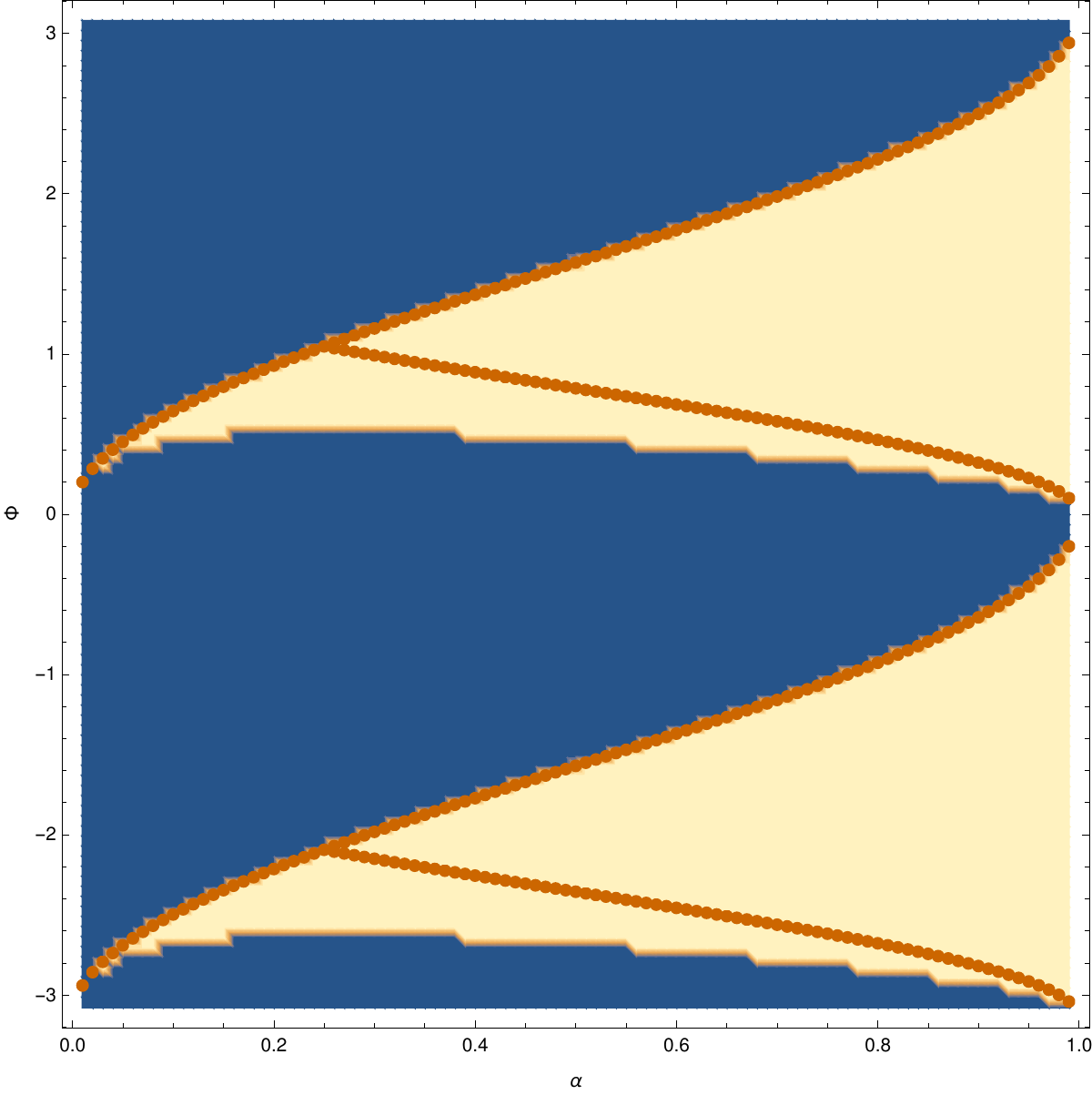}
    \caption{Plot showing number of cusps formed in different regions of parameter space spanned by $(\alpha,\Phi)$ for the Turok loop described by Eq. \eqref{eq:Turokloop}. The blue region corresponds to two cusps and the beige corresponds to six cusps. The orange dotted line traces a one-dimensional set of points in the parameter space that produces four cusps.}
    \label{fig:paramspace}
\end{figure}

Both $X_{-}$ and $X_{+}$ satisfy the Virasoro condition Eq. \eqref{VirasoroCOM}, and are periodic with a period of $l$. The loop described by Eq. \eqref{eq:Turokloop} can have two, four or six cusps depending on the values of $\alpha$ and $\Phi$ as shown in \autoref{fig:paramspace}.

\subsection{Calculation of \texorpdfstring{$I_{\pm}$}{I+-}}

The integral $I_{+}$ is calculated in closed form for a given $m$ in
Appendix \ref{sec:IplusforTurok} whereas $I_{-}$ cannot be done in an
equivalent manner. We compute numerically exact values of $I_{-}$ with
direct methods and approximate values using the multipoint and
asymptotic methods. We focus on the approximate methods here.

\subsubsection{Multipoint Method}
Here, we provide a framework for the calculation of $I_{-}$ using the multipoint method and discuss different scenarios which might arise.

For each direction $\hat{k}$ on the celestial sphere, the first step is determining the value(s) of $\sigma^{*}_{-}$. Depending on the value of $\theta,\phi$ and $\alpha$, the quantity $|k_{\mu}(\theta,\phi)\Xdot^{\mu}_{-}|$ can have either one minimum or three local minima as a function of $\sigma_-$. For each minimum the value $|k.\dot{X}_{-}|$ and the mode number $m$ determine the local contribution to $I_{-}$.
In general, we sum over the individual contributions
\begin{equation}
  I_{-}^{\mu}=\sum_{i}I^{\mu}_{-}(\sigma_{-,i}^{*})  
\end{equation}
where $\sigma_{-,i}^*$ denotes the $i$-th minimum. Each individual
contribution is based on a cubic fit to the phase derivative $k.{\dot X}_{-,i}^*$ in which the integration has been formally extended
to $\sigma_- = \pm \infty$. We heuristically refer to the phase
derivative as a function of $\sigma_-$ as the ``phase curve". The phase
curves are independent of the mode $m$ while the integrated quantity which involves the exponential of the phase does depend upon
$m$.

We will compare the results of an exact calculation of $I_-$ to the approximate multipoint method for selected values of $\alpha$,
$\theta$ and $\phi$. These choices indirectly control the number of
minima, the relative depths and spacings of the minima, the relative
size of the cubic and higher order terms near the minima, etc. We
present a set of phase curves that covers several qualitatively different
situations that arise. 

\begin{table*}[h!]
    \centering
    \begin{tblr}{colspec={Q[c,m] Q[c,m]  Q[c,m]  Q[c,m] Q[c,m]}}
    \hline \hline
        % \SetCell[r=2]{m}Case & \SetCell[r=2]{m}($\alpha,\theta,\phi$) & \SetCell[r=2]{m} Numerical $\mathbf{I_{-}}$ & \SetCell[r=2,c=2]{m} Cumulative RMS difference \\ $\left|\text{Approximate}\; \mathbf{I}_{-,i} - \text{Numerical}\; \mathbf{I}_{-}\right|$ & \SetCell[r=2]{m} m \\
        Case & ($\alpha,\theta,\phi$) & \centering Numerical $\mathbf{I_{-}}$ & Absolute Error, Relative Error & $m_{low}$, $m_{cross}$\\ 
    \hline 
        \SetCell[r=3]{m}\hypertarget{case1}{1} & \SetCell[r=3]{m}($\frac{1}{10},\frac{7}{10},\frac{6}{5}$) & \SetCell[r=3]{m}{\{$-0.00881401 - 0.0064074\; i$,\\$-0.024022 + 0.0104084 \;i$,\\$-0.0169884 + 
        0.00986654 \;i$\}} &  \SetCell[r=3]{m} 0.0026, 0.076 &\SetCell[r=3]{m} 2, 100\\
        &&&&\\
        &&&&\\
    
        \hypertarget{case2}{2} & ($\frac{3}{20}, \frac{9}{10}, \frac{\pi}{2}$) & \{$0.0238659 \;i, 0.102931,
        0.101559$\}  & 0.026, 0.18 & 2600, 200\\
    
        \hypertarget{case3}{3} &($\frac{1}{5}, \frac{4}{5}, \frac{\pi}{2}$) & \{$0.0243171\; i,0.107074,
        0.130373$\} & 0.039, 0.23 & 200, 400\\
    
        \SetCell[r=3]{m}\hypertarget{case4}{4} & \SetCell[r=3]{m}($\frac{3}{10}, 1, \frac{7}{5}$) & \SetCell[r=3]{m}{\{$-0.00730444 - 0.015009 \;i$,\\ $-0.0242681 - 0.0130257\; i$,\\ $-0.0309998 - 
        0.00183296 \;i$\}} &\SetCell[r=3]{m} 0.005, 0.11 &\SetCell[r=3]{m} 2, 300\\
        &&&&\\
        &&&&\\
    
        \SetCell[r=3]{m}\hypertarget{case5}{5} & \SetCell[r=3]{m}($\frac{3}{5}, \frac{27}{10}, \frac{3}{2}$) & \SetCell[r=3]{m}{\{$0.00643152 + 0.00844617 \;i$,\\ 
        $0.00834495 - 0.00592393\; i$, \\$-0.0346817 + 0.0221558\; i$\}} &\SetCell[r=3]{m} 0.0019, 0.043 &\SetCell[r=3]{m} 5, 200\\
        &&&&\\
        &&&&\\
    
        \hypertarget{case6}{6} & ($\frac{19}{20}, \frac{\pi}{2}, \frac{\pi}{2}$) &{\{$0.0294055 \;i, 0.0827689,
        0.00990778$\}}& 0.0026, 0.03 & 2, 400\\
    \hline \hline
    \end{tblr}
    \caption{Different cases for the derivative of the phase $|k.\Xdot_{-}|$ and the difference between the values of $\mathbf{I_{-}}$ computed numerically and using the multipoint method, all for $m=100$. Column 3 gives the numerically computed value of $\mathbf{I_{-}}$. Column 4 shows the absolute and relative errors between the values of $\mathbf{I_{-}}$ estimated numerically and that computed using the multipoint method, where the Absolute Error = $\left|\text{Approximate}\;\mathbf{I}_{-}-\text{Numerical}\;\mathbf{I}_{-}\right|$ and the Relative Error = $\text{Absolute Error}/\left|\text{Numerical}\;\mathbf{I}_{-}\right|$. Column 5 gives the values of the mode number $m_{low}$ and $m_{cross}$; $m>m_{low}$ for the polynomial fits to apply and $m>m_{cross}$ to prefer the asymptotic method.
The cases have been selected to illustrate phase curves with qualitatively different features that are relevant to the multipoint method, in particular, the number of inflection points, the stationarity of such points, the adequacy of the fit in the point's vicinity, the point separation, etc.
    }
\label{tab:phaseder_cases}
\end{table*}

Table \ref{tab:phaseder_cases} gives a list of six different cases of
phase curves which are discussed in more detail below. For this survey
we calculate at fixed mode $m=100$. In all cases the contribution to
the (cubic) integral about each peak is narrow in the sense
$\Delta\sigma_- \sim 1/|A_m|^{1/3}$ is small compared to the available
extent of $\sigma_-$ and the multipoint method is favored over the
complex asymptotic method ($m_{cross} \gtrsim 100$). We briefly describe
the cases:
\begin{itemize}
    \item Case \hyperlink{case1}{1}: There is one minimum (see phase
      curve in \autoref{fig:phase_der_1}), the cubic expansion is good
      (see the orange dashed line in the plot; $m_{low} \approx 2$ set
      by the 5th order term). Since $|k.{\dot X}_{-}|$ is close to
      zero this is a nearly-stationary point. The relative error is
      $\sim 8$\%. This case is most similar to situations covered
      by traditional single point expansion techniques.
    \begin{figure}[htbp!]
        \centering
        \includegraphics[width=\columnwidth, height=6cm]{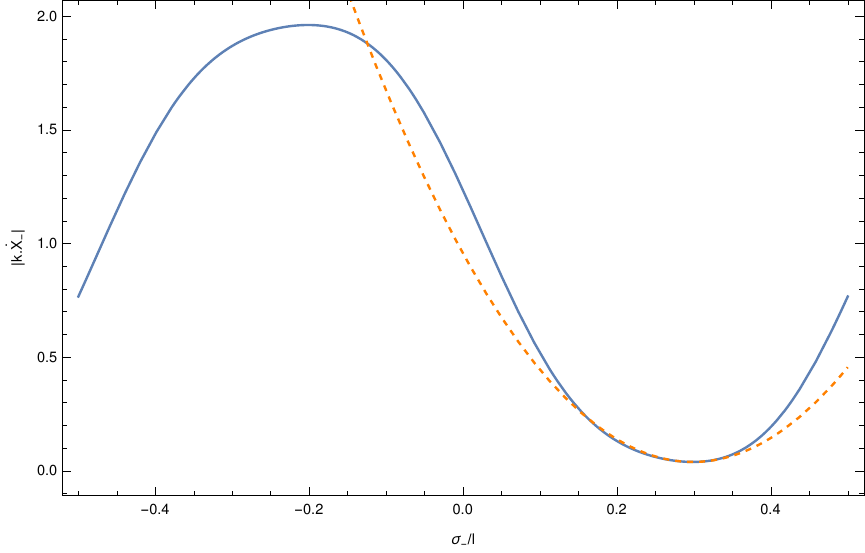}
        \caption{Case 1: plot of $|k.\Xdot_{-}|$ estimated for $\alpha=1/10$ at $(\theta,\phi)=(7/10,6/5)$, with a single minimum ($\sigstar_{-}$). The blue curve is the exact derivative of the phase and the orange dashed curve is the cubic fit to the derivative of the phase about the minimum.}
        \label{fig:phase_der_1}
    \end{figure}
    \item Case \hyperlink{case2}{2}: Like the previous case there is
      one minimum (see \autoref{fig:phase_der_2}) with small $|k.{\dot
        X}_{-}|$, however, the phase curve is {\it not} well fit by a
      cubic in the vicinity of the minimum (note that the orange
      dashed line in the plot misses the shape of the peak; $m_{low}
      \approx 2600$ set by 5th order term). The relative error is
      $\sim 18$\%, larger than the previous case.
    \begin{figure}[htbp!]
        \centering
        \includegraphics[width=\columnwidth, height=6cm]{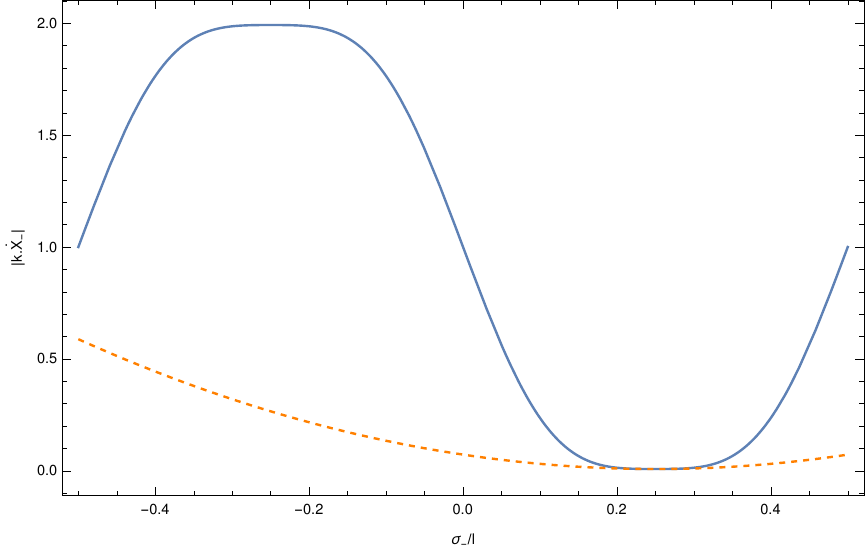}
        \caption{Case 2: plot of $|k.\Xdot_{-}|$ estimated for $\alpha=3/20$ at $(\theta,\phi)=(9/10,\pi/2)$ with a single minimum ($\sigstar_{-}$). The blue curve is the exact derivative of the phase and the orange dashed curve is the cubic fit about the minimum. The derivative of the phase varies very slowly in the neighborhood of the minimum.}
        \label{fig:phase_der_2}
    \end{figure} 
    \item Case \hyperlink{case3}{3}: Now there are three minima (see
      \autoref{fig:phase_der_3}). Two have similar magnitudes
      $|k.{\dot X}_{-}|$ close to zero and are expected to dominate
      the sum. They are difficult to distinguish and the cubic fit is {\it
        not} good ($m_{low} \approx 200$ set by the 4th order
      term near the dominant peaks). The relative error is $\sim
      23\%$.
    \begin{figure}[htbp!]
        \centering
        \includegraphics[width=1\columnwidth, height=6cm]{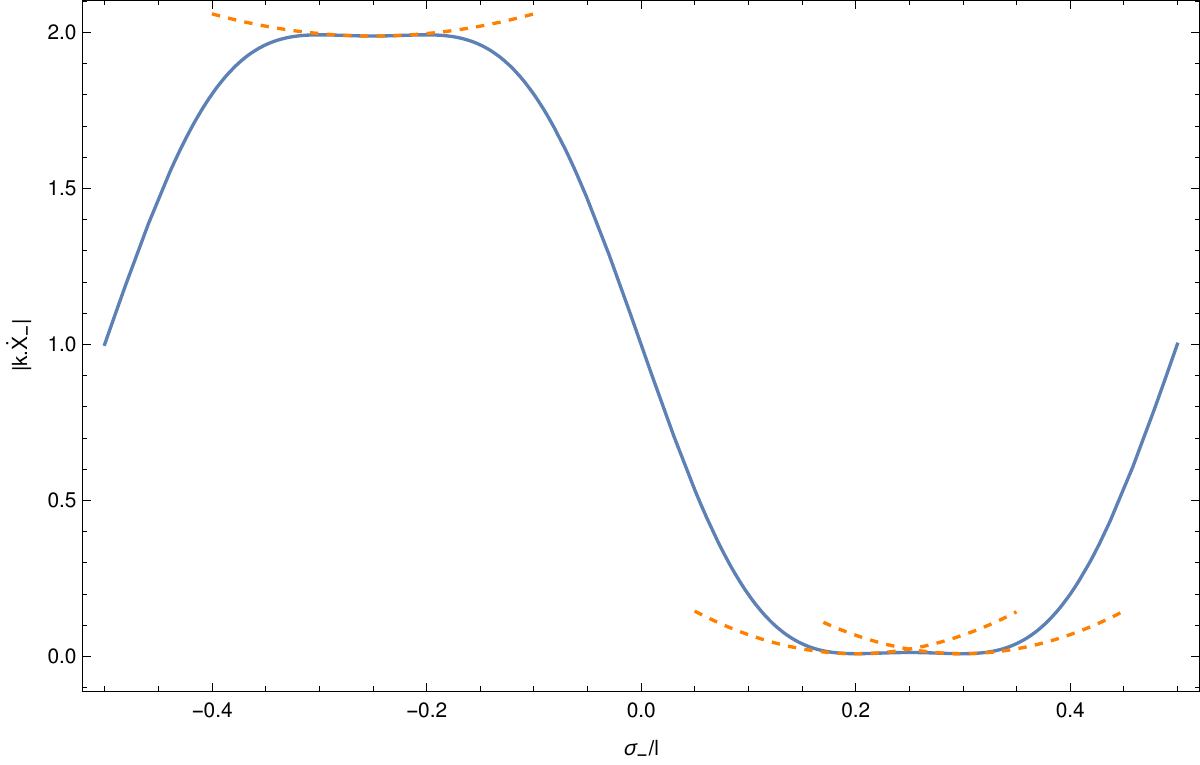}
        \caption{Case 3: plot of $|k.\Xdot_{-}|$ estimated for $\alpha=1/5$ at $(\theta,\phi)=(4/5,\pi/2)$ with three minima ($\sigstar_{-,i}$). Two of the minima are similar in magnitude and the derivative of the phase varies very slowly in between them.}
        \label{fig:phase_der_3}
    \end{figure}
    \item Case \hyperlink{case4}{4}: Like the previous case there are
      three minima (see \autoref{fig:phase_der_4}) and two have
      similar magnitudes $|k.{\dot X}_{-}|$ close to zero. Here,
      however, the cubic fits to the phase and the linear fit to the prefactor are good ($m_{low} \approx 2$, set by the prefactor), the
      peaks are better separated and the
      two dominate the sum.  The relative error is $\sim 11$\%.
    \begin{figure}[htbp!]
        \centering
        \includegraphics[width=1\columnwidth, height=6cm]{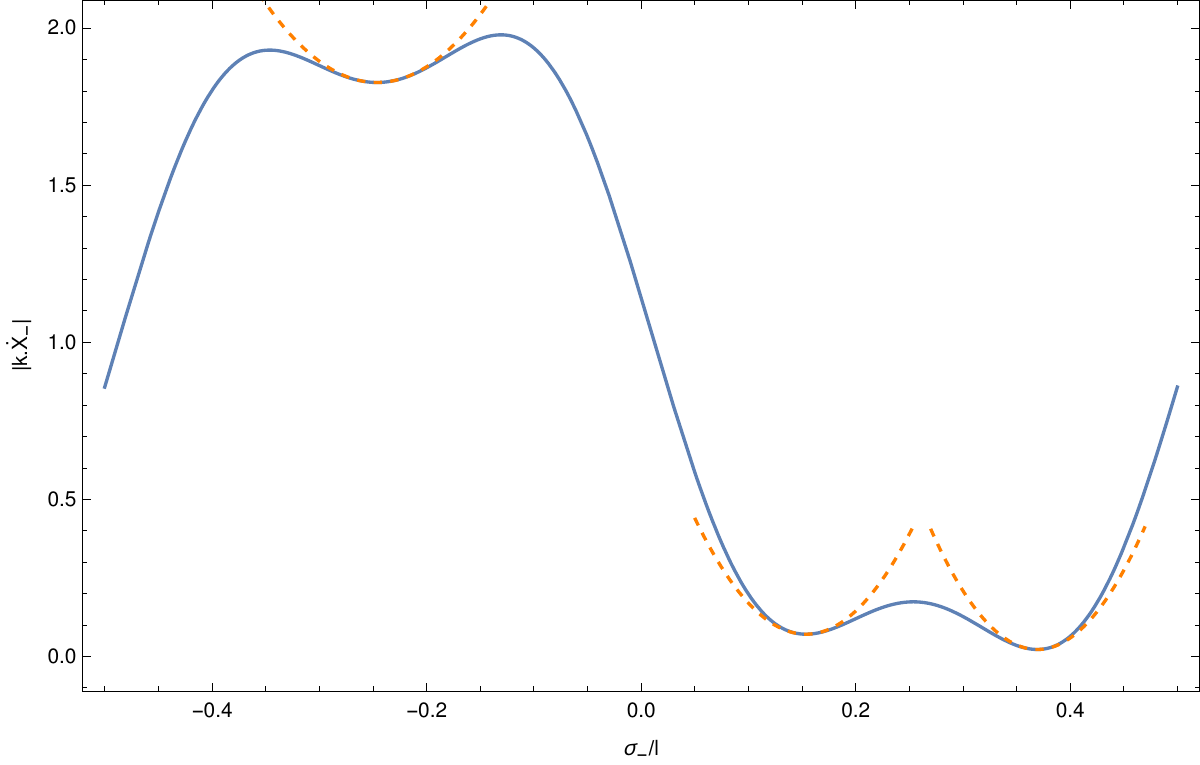}
        \caption{Case 4: plot of $|k.\Xdot_{-}|$ estimated for $\alpha=3/10$ at $(\theta,\phi)=(1,7/5)$ with three minima ($\sigstar_{-}$). Two minima are similar in magnitude but the derivative of the phase in between them is not flat.}
        \label{fig:phase_der_4}
    \end{figure}
    \item Case \hyperlink{case5}{5}: Again there are three minima (see
      \autoref{fig:phase_der_5}) but only one has $|k.{\dot X}_{-}|$
      close to zero, well fit by the cubic
      and is well separated from the others. The linear fit to the prefactor gives $m_{low} \approx 5$.
      The relative error is $\sim 4$\%.
    \begin{figure}[htbp!]
        \centering
        \includegraphics[width=1\columnwidth, height=6cm]{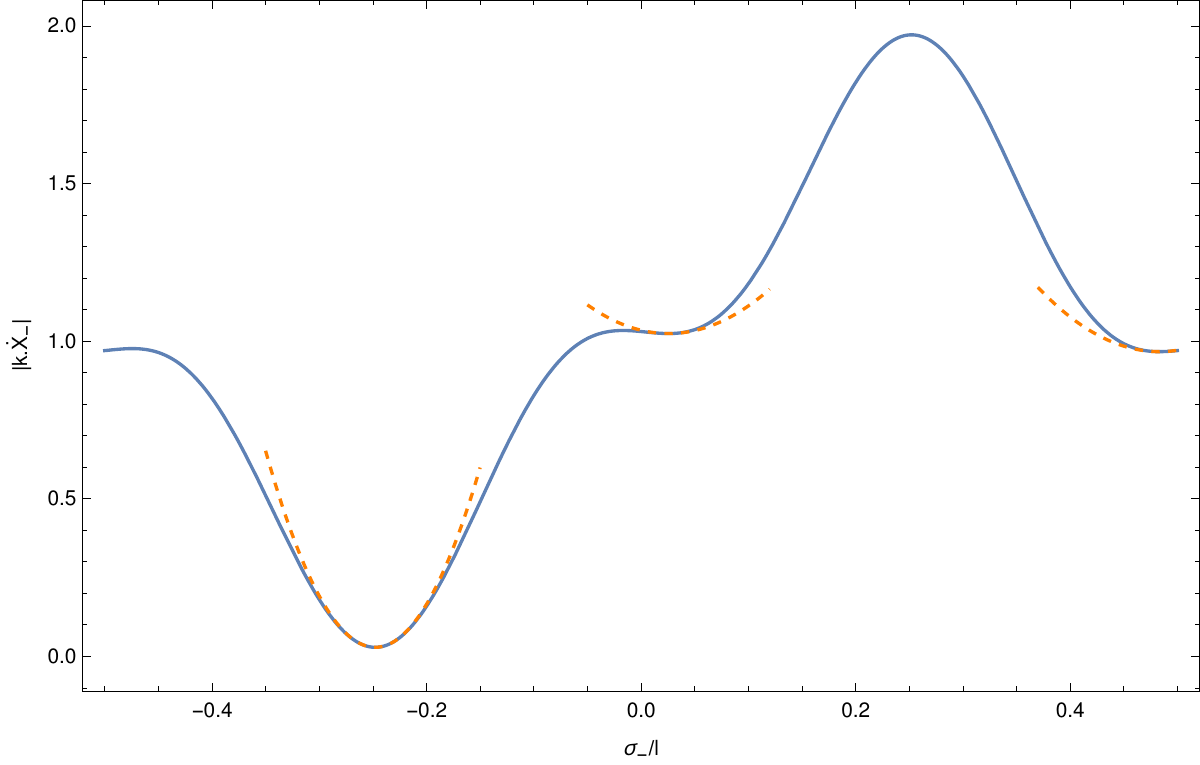}
        \caption{Case 5: plot of $|k.\Xdot_{-}|$ estimated for $\alpha=3/5$ at $(\theta,\phi)=(27/10,3/2)$ with three minima ($\sigstar_{-}$). There is a global minimum which is more prominent than the other two local minima.}
        \label{fig:phase_der_5}
    \end{figure}
    \item Case \hyperlink{case6}{6}: Figure \ref{fig:phase_der_6} shows
      an ideal case for the multipoint method. There are three minima,
      all with $|k.{\dot X}_{-}|$ close to zero. The cubic fit is good
      for each peak. All are well separated. The relative error $\sim
      3$\% and each peak improves the answer. The linear fit to the prefactor is also good ($m_{low} \approx 2$).
    \begin{figure}[htbp!]
        \centering
        \includegraphics[width=1\columnwidth, height=6cm]{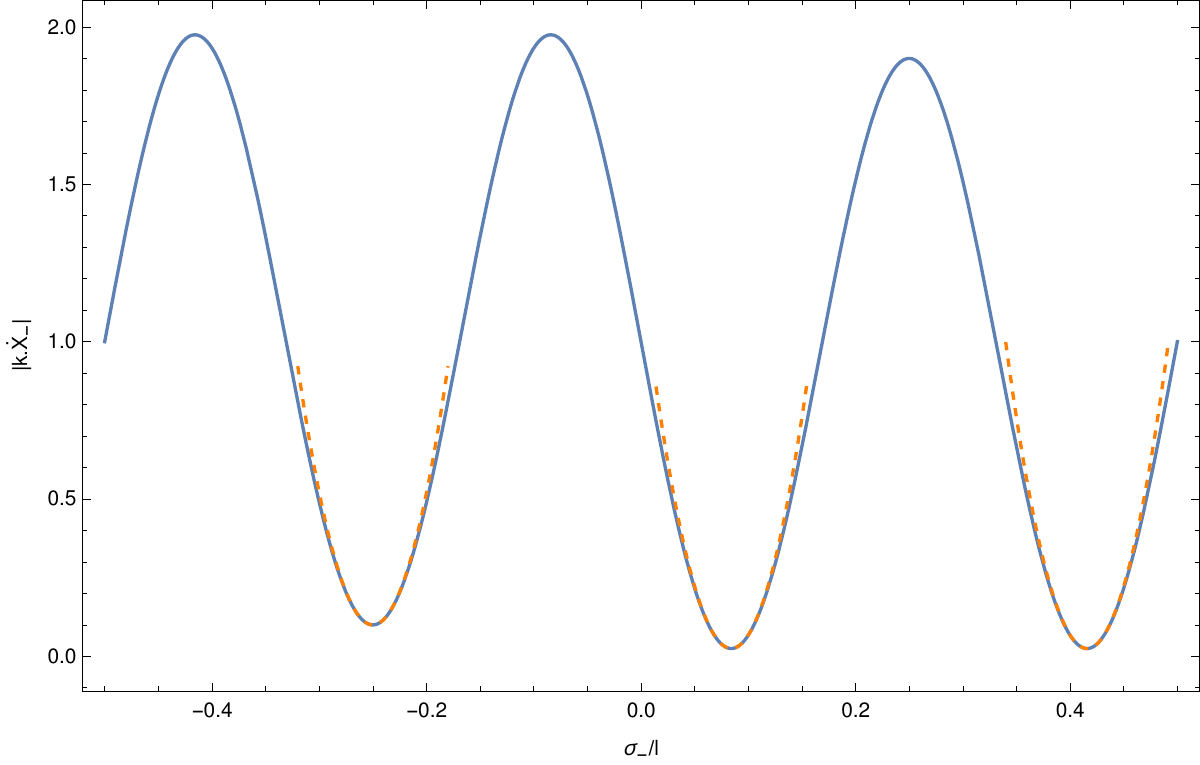}
        \caption{Case 6: plot of $|k.\Xdot_{-}|$ estimated for $\alpha=19/20$ at $(\theta,\phi)=(\pi/2,\pi/2)$ with three minima ($\sigstar_{-}$). The three minima are of comparable magnitude and all contribute, especially at low $m$.}
        \label{fig:phase_der_6}
    \end{figure}
    
\end{itemize}

These cases cover most of the qualitatively different phase curves
encountered. The trend of relative errors largely traces the
adequacy of the polynomial fits and the independence (degree of separation) of the
peaks. The method achieves
relative errors $\lesssim 10\%$ when the fit is good and the peaks are not too close to each other.

\begin{figure}[h]
    \centering
    \includegraphics[width=\linewidth, height=8 cm]{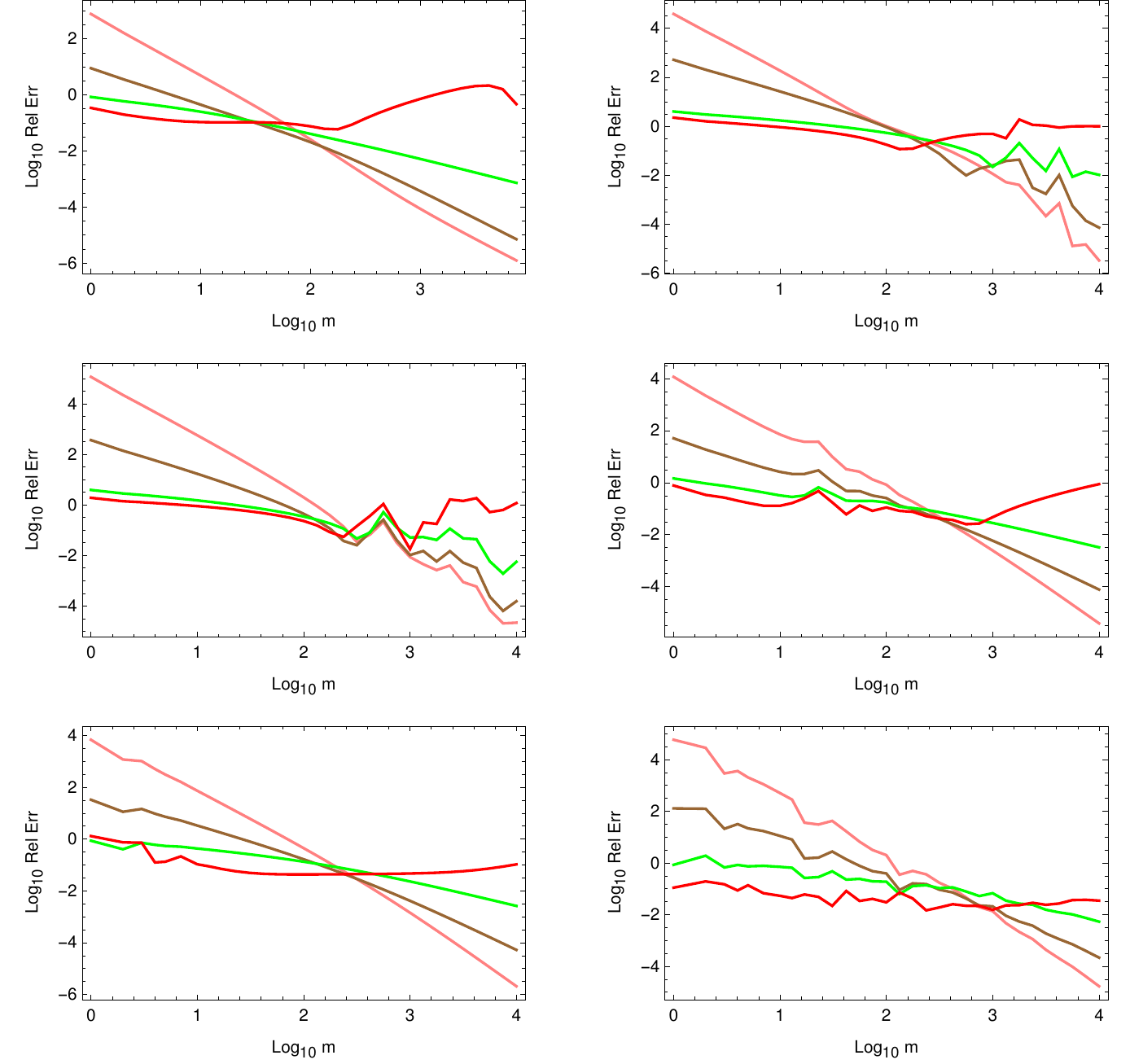}
    \caption{The relative errors for the three asymptotic approximations and the multipoint method as a function of $m$ for the six different cases. The upper two panels
are cases 1 and 2, the middle two are cases 3 and 4, the lower two are
cases 5 and 6. The red line corresponds to the multipoint method and the green,brown and pink lines correspond to $h^{(0)}_{asym}, h^{(1)}_{asym}$ and $h^{(2)}_{asym}$ respectively.}
    \label{fig:cases1-6}
\end{figure}

\subsubsection{Errors as functions of $m$}

It is useful to compare the errors of the real multipoint treatment
to those of the complex asymptotic treatment.
Figure \ref{fig:cases1-6} presents results for a selected set of modes $1 \le m
\le 10^4$ for the six cases. Each solution is compared to the exact solution in terms of
norm difference divided by norm of exact result. 
The trends are clear: the relative error for the multipoint method is
roughly constant with $m$ whereas the relative error of the asymptotic
methods decreases with $m$. The multipoint method is superior at small
$m$ while the asymptotic methods are better at large $m$. The relative
error of the highest order asymptotic approximation decreases most
rapidly.

In all the illustrated cases the relative error of the multipoint
results does not systematically decrease as $m$
increases. Nonetheless, the magnitude of the exact, multipoint and
asymptotic results all decrease together. In other words, the
multipoint answer falls as rapidly as the numerical answer and for
many purposes a fixed relative error in an exponentially shrinking
quantity is adequate.

\subsubsection{Errors on the celestial sphere}

\begin{figure}[h]
    \centering
    \includegraphics[width=\linewidth, height=8 cm]{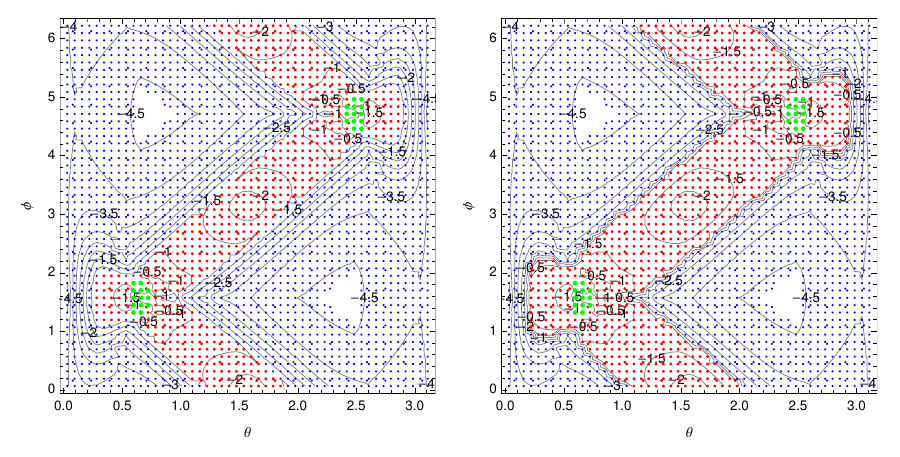}
    \caption{Regions of the celestial sphere where the multipoint method and the asymptotic method have been selected for a Turok loop with $\alpha=1/5$ and for $m=100$. The red dots correspond to the points where the multipoint method has been selected (i.e. where the multipoint method yields lower relative errors) and the blue dots correspond to the points where the asymptotic method is selected. The green dots show where the fit is poor because $m<m_{low}$ (see
Eq. \ref{eqn:conditiononm}).}
    \label{fig:eps}
\end{figure}

In Section \ref{sec:AlignmentLimit} we showed that
for directions on the celestial sphere aligning (or nearly aligning) with tangent curves, the multipoint method must be used.
Let's consider a Turok loop with $\alpha=1/5$ (same as Case 3) and harmonic $m = 100$
and quantify the errors in $\mathbf{I_{-}}$ as a function of position on the celestial
sphere. The logarithmic relative errors with respect to
an exact numerical answer are shown in a contour
representation in \autoref{fig:eps}. The figure axes are the
angles $\theta$ and $\phi$. Each point represents a direction of ${\hat
  k}$. The graph on the left displays contours of the minimum relative
error selected from two different approximations: the multipoint and asymptotic ($h^{(2)}$) methods. The red dots show where the multipoint treatment
is selected, the blue dots are where the asymptotic treatment is.
The graph on the right is similar (it contours the logarithmic relative
$\mathbf{I_{-}}$ with respect to
an exact numerical answer)
but takes the multipoint results for $m <
m_{cross}$ and the asymptotic results for $m > m_{cross}$. The color
coding is identical. Here, we assumed $m_{cross}$ to be the approximate value of $m$ where the three asymptotic terms were comparable.

Both plots show that the peak errors lie
along the direction of the tangent vector sweep. The asymptotic errors
smoothly decrease on both sides away from the peak.
The beam is described to roughly 1-10\% near the peak (red areas) and with
increasing relative (and absolute) accuracy in the asymptotic regime (blue areas).

The source of the error near the tangent vector sweep can be traced back to the
fit to the phase. Figure \ref{fig:eps} shows with green dots the regions of
the celestial sphere where the multipoint approximation was used
for $m<m_{low}$ (i.e. points that violate
Eq. \ref{eqn:conditiononm} because 4th or 5th order terms are
important). Note that these points coincide with the largest
errors.

\subsubsection{A Schematic Methodological Partition}

Finally we want to describe schematically how one might select amongst
the different methods mentioned so far.  Choice of methodology
revolves around considerations of error size and computational
timing. Quadrature methods that use increasingly dense point spacing
can potentially achieve arbitrarily small errors. Quadratures can be
done for individual $m$ with adjustable tolerances.  In general,
point-by-point quadratures are the most flexible in terms of accuracy
and also the most time consuming per point. Sets of $m$ values can be
evaluated en masse by quadrature (e.g. FFT method) for efficiency but
then the inaccuracies are linked together and group timings must be
considered.

On the other hand, the multipoint and complex asymptotic methodologies
provide approximate answers. Mathematically, both are asymptotic
expansions. The multipoint method utilizes polynomials to describe the
region of the phase function that contributes to the integral; the
complex asymptotic analysis approximates the complex phase function by
Taylor series expansion of given order. In a practical sense both
methods are limited by the finite expansion orders utilized and, in a
more fundamental sense, by radius of convergence issues. The methods
are not very flexible in terms of accuracy but both are quite fast.

So an elementary consideration is what sort of accuracy is required
for a particular application? If very high accuracy is needed then
numerical quadrature must be done. Otherwise, multipoint and complex
asymptotic may suffice.

A related consideration is whether small absolute errors or small
relative errors are needed for the particular application. Note that
away from a tangent vector curve the magnitude of the integral answer
decreases exponentially as $m$ increases. The multipoint method
generally achieves constant relative errors as $m$ increases whereas
the complex asymptotic method gives decreasing relative errors in that
limit.  One or the other may be sufficient for practical purposes
since the magnitude of the answer in that limit is so small.

\begin{figure}[H]
    \centering
    \includegraphics[width=\linewidth, height=6cm]{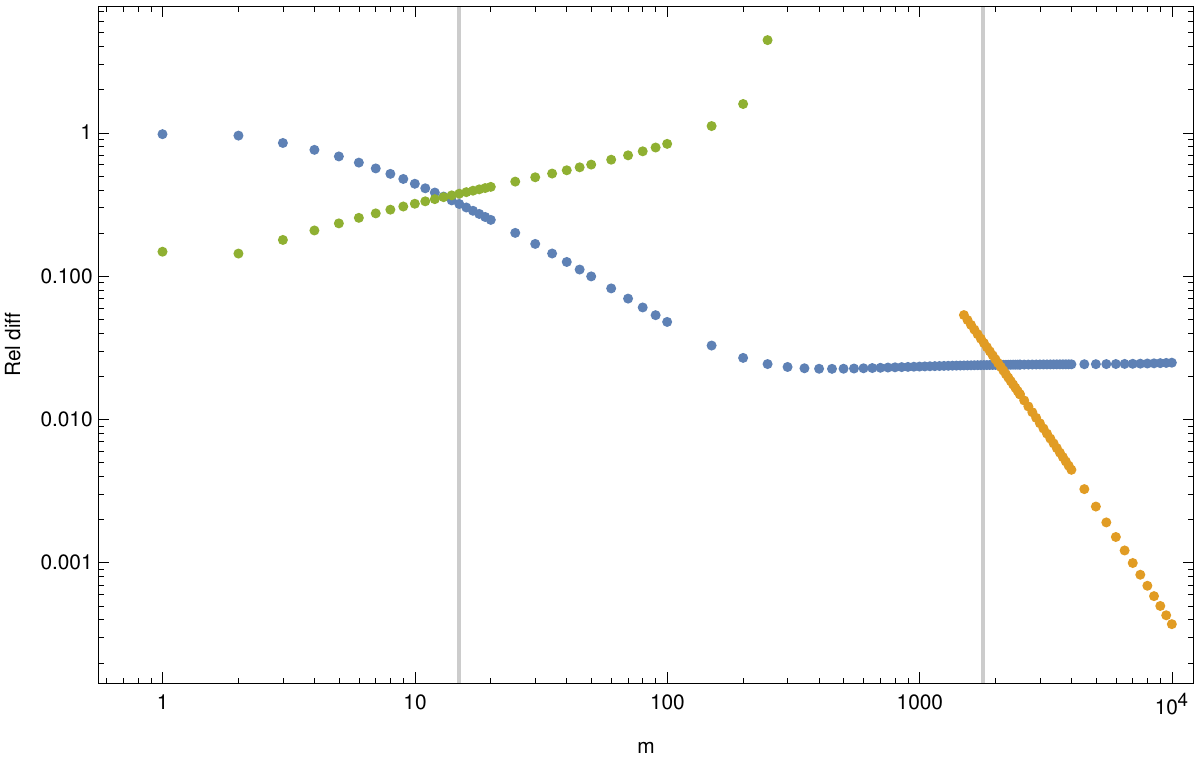}
    \caption{The relative errors of $I_{-}$ with respect to a
      numerically exact treatment for three different methods -- FFT
      (green dots), multipoint (blue) and complex asymptotic
      (orange). The source is a Turok loop with $\alpha=2/5$ and
      emission direction $\theta=0.1, \phi=1.5$. The FFT was computed
      with $N=2^{8}$ points. The vertical lines are $m_{low} \sim 15$
      (on the left) and $m_{cross} \sim 1780$ (on the right).}
    \label{fig:3methods}
\end{figure}

To illustrate we now pick a point on the celestial sphere for the beam
direction and fix the loop parameters. We have selected a direction
close to the location of a tangent vector.  Figure \ref{fig:3methods}
displays the relative errors of $I_-$ as a function of mode $m$ for
three methods -- the direct evaluation of the real integral using FFT,
the multipoint method and the asymptotic method (all with respect to a
numerically exact treatment). Two vertical lines $m=m_{low}$ and
$m=m_{cross}$ partition the mode space into three ranges. For the case
depicted, $m_{low} \approx 15$ (due to the accuracy of the linear prefactor)
and $m_{cross} \approx 1780$ (inferred by comparing $h_{asym}^{(1)}$ and
$h_{asym}^{(2)}$). The multipoint method is applicable for
$m_{low}<m<m_{cross}$ and the complex asymptotic method for
$m>m_{cross}$.  The FFT method is not apriori restricted.  The mode-by-mode
errors are displayed.  As $m$ increases the lowest
errors are provided first by the FFT, then the multipoint and finally
the asymptotic method.  For the specific numerical choices made the
FFT relative errors equal those of the multipoint method at $m \sim
10$ when both are $\sim 0.3$.

We now consider adjustments that might be made.  As the figure
demonstrates, the accuracy of the FFT method with a transform of fixed
length degrades as $m$ increases, an aliasing effect related to
evaluation of the integrand at unequally spaced points. The FFT may be
oversampled by the factor $c$ (see Section \ref{sec:directreal})
whence errors at fixed $m$ decrease exponentially as $c$
increases. So, by increasing $c$ the relative errors for a fixed range
of modes may be arbitrarily lowered.  If one is satisfied with the
typical multipoint error at a given $m=m_{OK}$ (implicitly in the range
$m_{cross}>m_{OK}>m_{low}$) then the simplest ansatz is to increase
$c$ until the FFT errors for $m<m_{OK}$ are less than or equal to that
error. Since the FFT relative errors typically increase with $m$ while
the multipoint errors decrease with $m$ the point $m=m_{OK}$ serves to
stitch the two methods together. The FFT method is used for
$m<m_{OK}$, the multipoint method for $m_{OK}<m<m_{cross}$ and the
complex asymptotic method for $m>m_{cross}$. (In the figure
$m_{OK}\sim 10$ which is less than $m_{low}$. Increasing $c$ decreases
the error at all $m$ and $m_{OK}$ increases.)  In patching together
the coverage of the three methods we are implicitly assuming that the
multipoint method is more efficient per point than the FFT method (see
Appendix \ref{sec:3methods_scenarios} for details on timing and other
efficiency considerations) and that the complex asymptotic method is
as efficient as the multipoint method and also more accurate for
$m>m_{cross}$. As $c$ increases (error level decreases) eventually one
transitions directly from the FFT to the asymptotic treatment at
$m_{OK} > m_{cross}$.

The case shown in Fig. \ref{fig:3methods} depicts a well-behaved and
consistent decrease in the relative errors for the multipoint and the
asymptotic methods as $m$ increases. We observe this smooth variation
when the beam direction is close to one and only one point on the
tangent vector curve. It is not difficult to choose directions that
have close encounters with multiple points on the tangent vector
curve. This is a common occurrence for directions in the vicinity of
self-intersecting points of the tangent vector curve. Interference
between multiple individual contributions may induce oscillating
relative errors in plots like Fig. \ref{fig:3methods}. Nonetheless, we
observe that the envelope of the oscillatory relative errors behaves
in a manner similar to those shown. In such cases the partitioning
values for $m$ should be based on the envelope's variation.

\section{\label{sec:Pseudocusp} A Particular Pseudocusp Example in the Turok loop}

Up to now we have concentrated on methods for calculating individual
one dimensional integrals like
$I_{-}$.  Now we turn our attention to using the methodology to
analyze the pseudocusp phenomena and the energy flux radiated.

\subsection{The Phenomena}
\begin{figure}[htbp]
    \centering
    \includegraphics[width=0.8\linewidth, height=7cm]{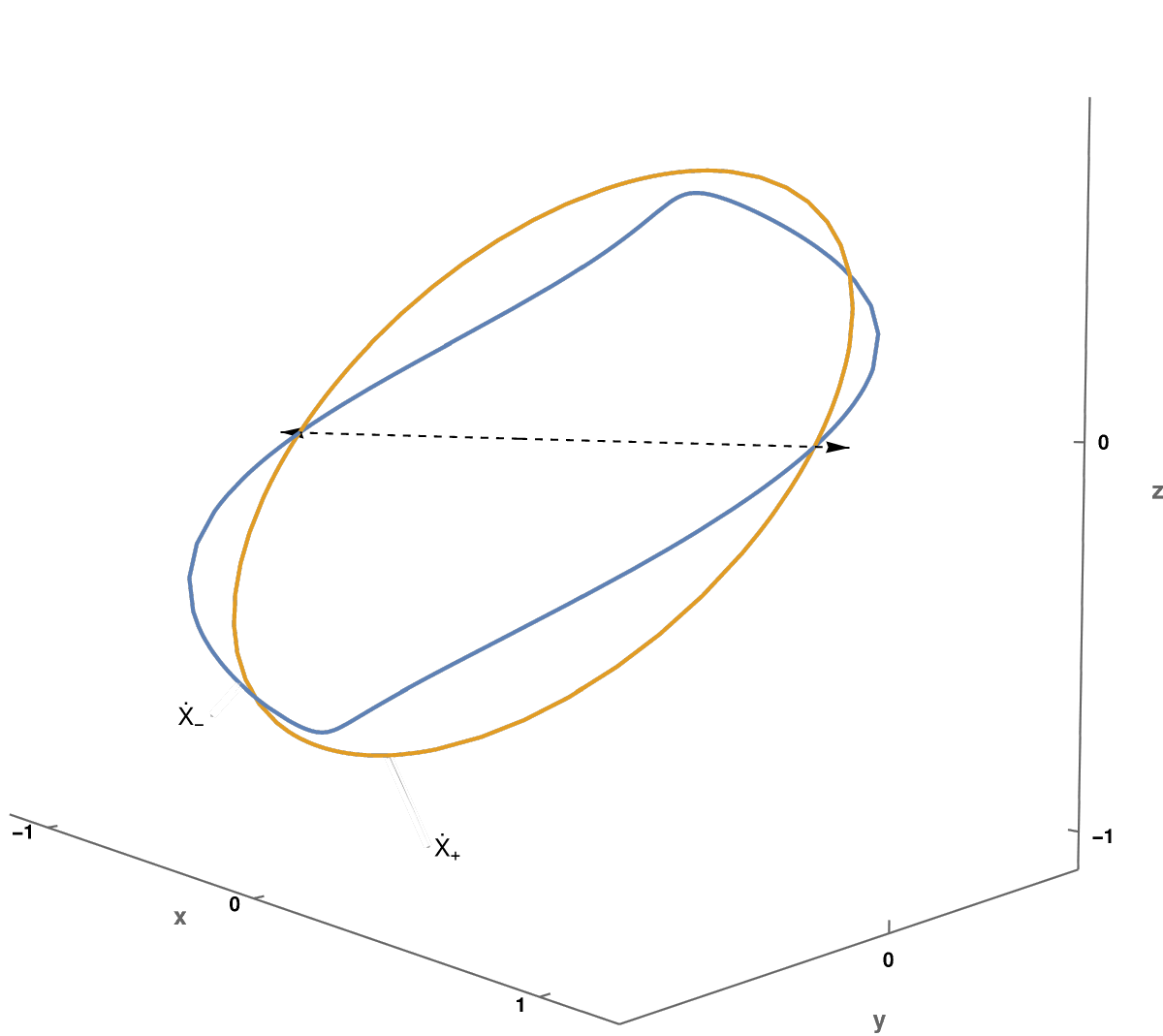}
    \caption{The tangent curves described by $\vecXdotpm$ for $(\alpha,\Phi)=(3/20,-18\pi/25).$ The curves intersect at two points to form two cusps depicted by black arrows. The curves come close together on the plane $x=0$, but do not intersect.}
    \label{fig:example_tangent_curves}
\end{figure}

Consider a Turok loop with parameters
$(\alpha,\Phi)=(3/20,-18\pi/25)$. Figure \ref{fig:example_tangent_curves}
shows the tangent curves for this loop on the unit sphere. There are
two cusps with cusp velocities along $\hat x$ and $-\hat x$ ($\phi=0$
and $\pi$ for $\theta=\pi/2$, respectively). In addition to the intersection
that creates the cusp
itself the tangent curves come close to each other in the $x=0$ plane
where the $\Xdot_-$ curve has a ``nub''.

\begin{figure}[htbp]

\includegraphics[width=1\linewidth, height=6cm]{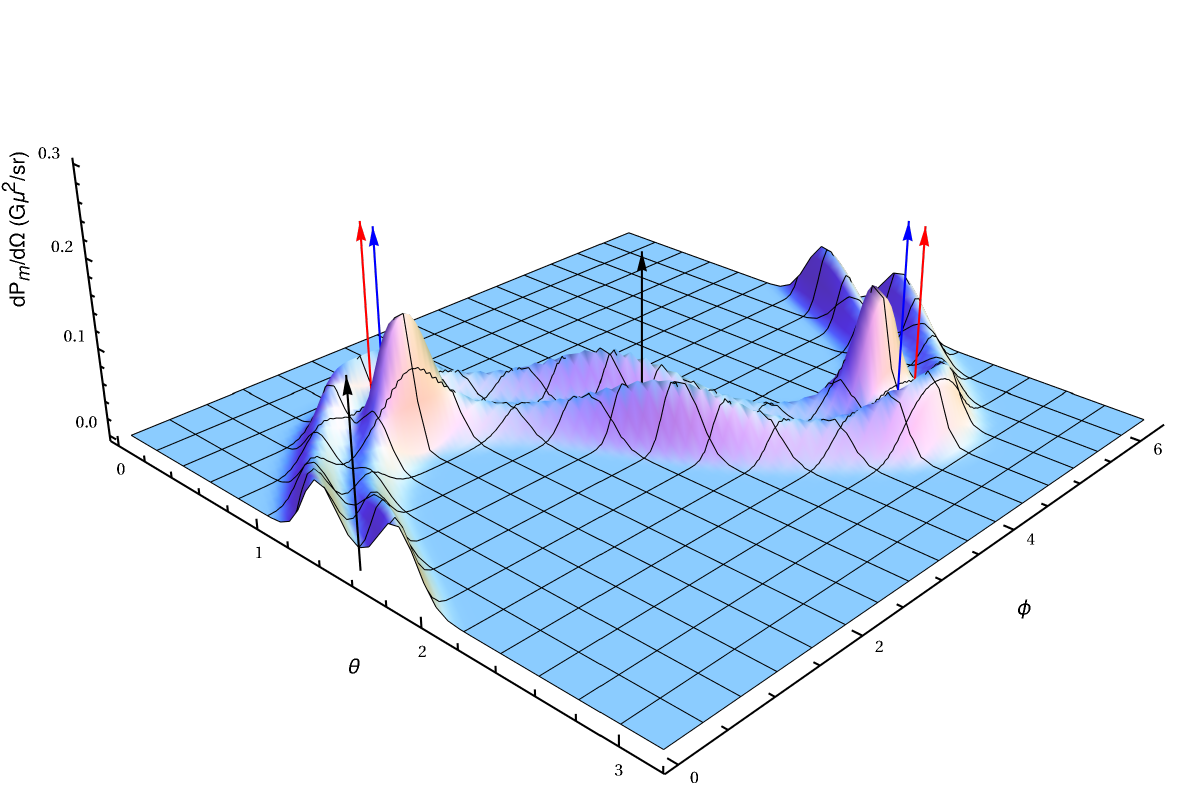} 

\caption{$dP_{m}/d\Omega$ vs. $(\theta,\phi)$ calculated numerically for $(\alpha,\Phi)=(3/20,-18\pi/25)$ for $m=100$. The black arrows depict the directions of velocities of the two cusps, the blue arrows depict directions where $k.\Xdot_{-}(\pm l/4)=0$ and the red arrows depict the directions where $k.\Xdot_{+}(\pm l/4)=0$. The plot is periodic in $\phi$ with a period of $2\pi$ and the peaks at $\phi=0$ and at $\phi=2\pi$ are identified to be part of the structure.}
\label{fig:example_dPdOmega_numerical}
\end{figure}

Figure \ref{fig:example_dPdOmega_numerical} displays $dP_{m}/d\Omega$ as a
function of $(\theta,\phi)$ for $m=100$ calculated by a direct numerical
method. The black arrows mark these cusp directions of emission,
i.e. $\hat{k}$ such that both $k.\Xdot_{\pm}=0$.  Note that the peaks
of the emission {\it don't} align with the cusp directions. The peaks are
the ``pseudocusps" discussed in \cite{Stott_2017}.

\subsection{Qualitative Considerations}

Qualitatively, the regions of maximum emission for low mode numbers
correspond to points on the unit sphere where the tangent curves come
close to each other but do not intersect.  The multipoint method
approximates $I_+$ and $I_-$ by separately selecting expansion centers
from each tangent vector for a given direction $\hat k$.  The
Euclidean separation between the tangent vectors 
$d(\sigma_-,\sigma_+) =
\left|\mathbf{\Xdot}_{-}(\sigma_{-})-\mathbf{\Xdot}_{+}(\sigma_{+})\right|$
provides a suggestive means for locating two nearby centers and a strip
between them.  At true
cusps, the tangent vectors intersect and $d=0$. If $d$ is small the
two tangent vectors are close to each other and if $d$ is a local
minimum we may expect, roughly speaking, that the two specific points (one
on each tangent vector)
might serve as expansion centers for the points along the
chord that connects them.

Note, it is merely a matter of convenience whether we specify that
expansion center in terms of the values of $\sigma_+$ and $\sigma_-$
for each mode or by means of the vector directions $\mathbf{\Xdot}_+$
and $\mathbf{\Xdot}_-$ to
the tangent curves or in terms of angular directions on the celestial
sphere $\{ \theta_+, \phi_+ \}$ and $\{ \theta_-, \phi_- \}$. We will
use all of these.

In the example, there are two minimal distance solutions:
$\left(\sigma_{-},\sigma_{+}\right)=\pm (l/4) \left(1,-1\right)$.  The
corresponding directions are antipodal on the celestial sphere. The
first solution has expansion center $\{ \theta_+, \phi_+ \} = \{0.69,
\pi/2\}$ and $\{ \theta_-, \phi_- \} = \{0.78, \pi/2\}$ for $I_+$ and
$I_-$, respectively; the second solution is $\{ \theta_+, \phi_+ \} =
\{2.45, 3\pi/2\}$ and $\{ \theta_-, \phi_- \} = \{2.37, 3\pi/2\}$.
Now, let us mark the directions $\mathbf{\Xdot}_-$ and
$\mathbf{\Xdot}_+$ with blue and red arrows respectively for
both solutions. The key observation is that the maximum emission
occurs in the vicinity of the blue and red arrows.

\subsection{Small Angle, Analytic Treatment }

We now present a more quantitative approach to this case based
on the multipoint method. On the tangent
sphere the tangent vectors are symmetric about the $x=0$ surface (the
x-plane). In \autoref{fig:example_tangent_curves} the ``nubs'' of the
blue curve lies in the x-plane. Construct a small arc in the x-plane
that stretches from the $\dot X_-$ to $\dot X_+$ tangent curves (blue
to orange along the short segment of a great circle). Consider
directions $\hat k$ that lie along the arc. Equivalently, in
\autoref{fig:example_dPdOmega_numerical} the arc is a constant $\phi$
slice through the peak. On account of the planar symmetry the centers
of the multipoint expansion for $I_+$ and $I_-$ are fixed at
$\sigma_-=-\sigma_+=(l/4)$ for direction along a part of the arc,
including the segment that lies between the two tangent curves.
This simplifies the application of the multipoint method.

We calculate $I_-$, $I_+$ and $dP_{m}/d\Omega$ by direct real and
multipoint methods.  We then provide a simplified analytic and
approximate version of the multipoint result to demonstrate the
scalings.

\begin{figure}[htbp]
    \centering
    \includegraphics[width=0.8\linewidth, height=7cm]{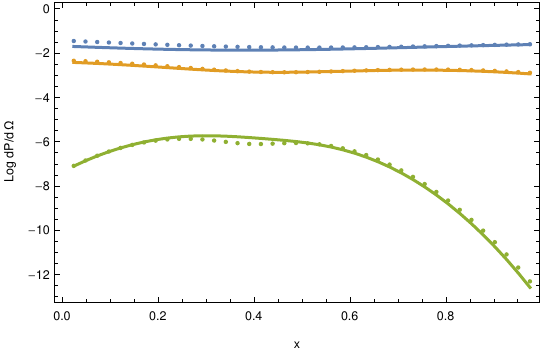}
    \caption{Energy flux $dP_{m}/d\Omega$ as a function of $x$ (where $x=0$ points along $\dot X_-$ and $x=1$ along $\dot X_+$) for $m=3 \times 10^2$, $3 \times 10^3$ and $3 \times 10^4$
      (blue, orange and green, respectively) by direct numerical evaluation (solid) and multipoint (dotted).}
    \label{fig:ExactvsMultidPdOmega}
\end{figure}
\begin{figure}[htbp]
    \centering
    \includegraphics[width=0.8\linewidth, height=7cm]{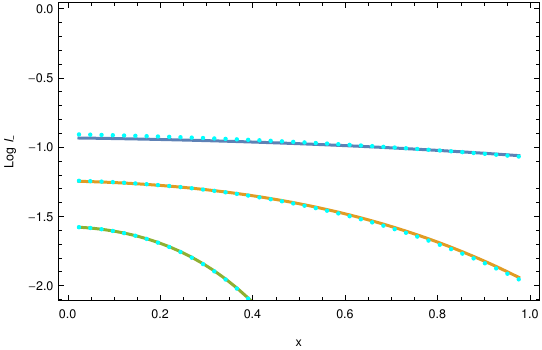}
    \caption{$I_-$ as a function of $x$ for $m=3 \times 10^2$, $3 \times 10^3$ and $3 \times 10^4$
      by exact numerical evaluation
      (blue, orange and green solid lines, respectively)
      and by multipoint (cyan dotted).}
    \label{fig:ExactvsMultidI}
\end{figure}
\begin{figure}[htbp]
    \centering
    \includegraphics[width=0.8\linewidth, height=7cm]{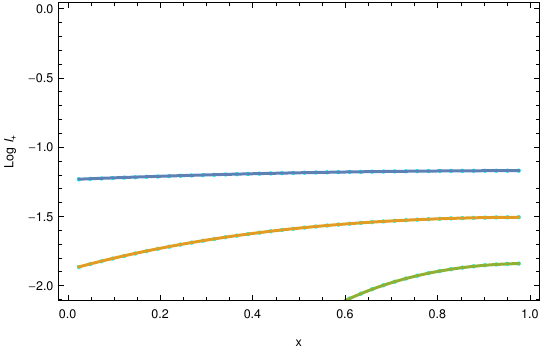}
    \caption{$I_+$ as a function of $x$ for $m=3 \times 10^2$, $3 \times 10^3$ and $3 \times 10^4$
            by exact numerical evaluation
      (blue, orange and green solid lines, respectively)
      and by multipoint (cyan dotted).}
    \label{fig:ExactvsMultidJ}
\end{figure}

Figure \ref{fig:ExactvsMultidPdOmega} shows $dP_m/d\Omega$ calculated exactly (solid
lines) and by multipoint (dotted) for $m=3 \times 10^2$, $3 \times 10^3$ and $3 \times 10^4$.
The abscissa specifies direction ${\hat k}$ along the arc ($x=0$ is
${\dot X}_-$ and $x=1$ is ${\dot X}_+$; ${\hat k}(x)$ is the simple
interpolated quantity $(1-x) {\dot X}_- + x {\dot X}_+$ normalized to
give a unit vector). The two treatments are
in rough agreement bearing in mind that the multipoint is not expected
to be applicable at small $m$ and its relative accuracy asymptotes to $\sim
10$\% at large $m$. Note that the
peak of $dP_m/d\Omega$ shifts from near the tangent lines
to a point midway along
the arc with increasing $m$.

Since the differential
power $dP_m/d\Omega$ depends upon products of bilinears in
$I^\mu_\pm$ we first examine individual norms like
$|\mathbf{I}_\pm|$ (exact versus
multipoint methods) in \autoref{fig:ExactvsMultidI} and \autoref{fig:ExactvsMultidJ}. Both show steep dropoffs away from the expansion directions.

To understand this behavior
simplify the analytic multipoint forms by expanding to lowest
non-vanishing order all angle-dependent terms in small angle
displacements from the tangent directions while leaving the Bessel
functions intact. The details are presented in Section
\ref{sec:expand}.

The absolute value of the mode functions summarizes a lot of the
information of the variation along the arc. We have
\begin{align}
  |\mathbf{I}_-| &= \frac{ 5\left(\theta_--\theta\right) {\rm K}_{-1/3} \left(\frac{5m (\theta_--\theta)^{3}}{6}\right)}{2\sqrt{3} \pi}
\end{align}
for $\theta < \theta_- $ (the angle with respect to the tangent vector ${\dot X}_-$
which has $\theta_-=0.78$ in this case)
and
\begin{align}
  |\mathbf{I}_+| & = \frac{(\theta-\theta_+) {\rm K}_{-1/3}\left( \frac{m (\theta-\theta_+)^3}{3} \right)}{\sqrt{3} \pi}
\end{align}
for $\theta>\theta_+$ (the angle with respect to the tangent vector ${\dot X}_+$ which
has $\theta_+=0.69$ in this case).
The analytic and full numerical expressions for the multipoint results are essentially indistinguishable in this example (see plots in Section \ref{sec:expand}).

Terms like $|\mathbf{I}_-| |\mathbf{I}_+|$ scale with the product of
the Bessel functions, each having its own centers of expansion. When $m (\theta_- - \theta_+) >> 1$ both
terms are exponentially small and the maximum of the product lies
between the two tangent lines. Likewise, a local
maximum of $dP_m/d\Omega$ appears between closely separated tangent
lines. This flux must decrease exponentially with $m$.  Even if the emission
from a pseudocusp exceeds that of a true cusp at given, finite $m$,
the emission (between the tangent lines) becomes subdominant as $m
\to \infty$.

\subsection{Numerical Investigation}

\begin{figure}[H]

\includegraphics[width=1\linewidth, height=6cm]{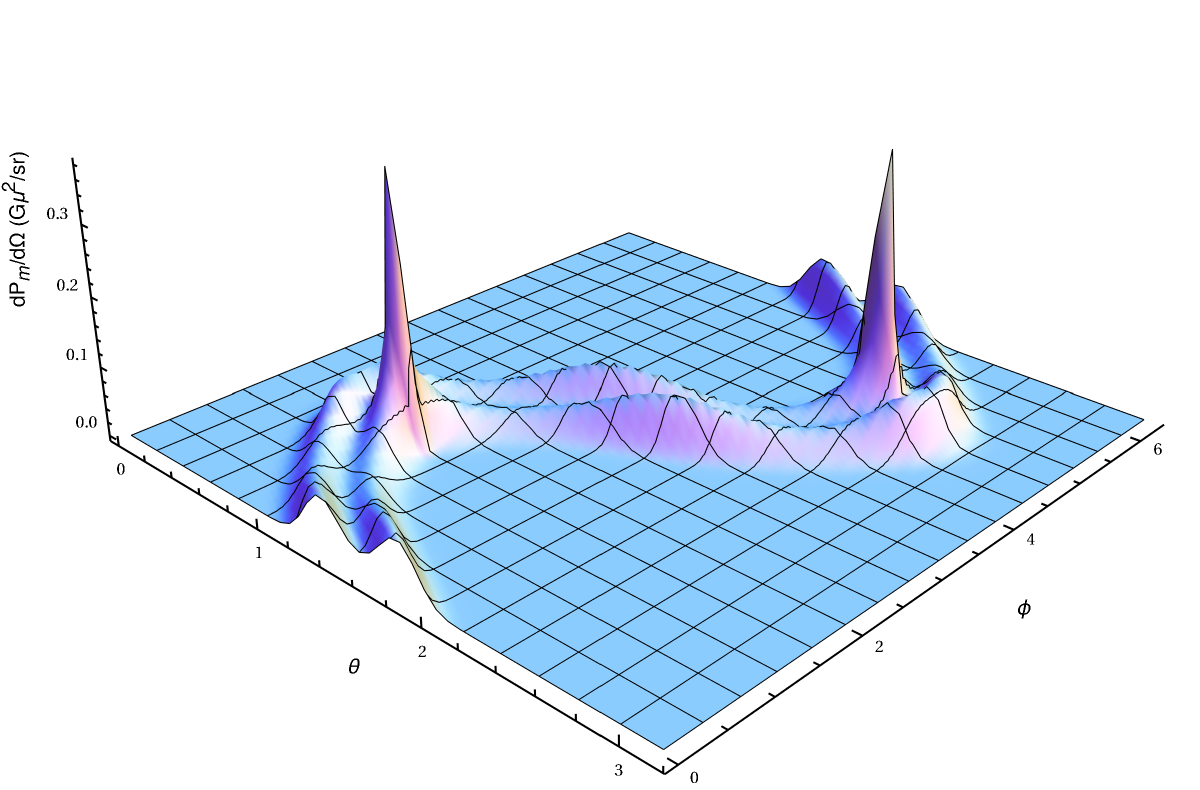} 

\caption{$dP_{m}/d\Omega$ vs. $(\theta,\phi)$ calculated using the multipoint method for $(\alpha,\Phi)=(3/20,-18\pi/25)$ for $m=100$.}
\label{fig:example_dPdOmega_analytic}
\end{figure}

The numerical $I_{\pm}$ from the full multipoint are combined using
Eq. \eqref{dPdOmegaIpm} to give
$dP_{m}/d\Omega$. Figure \ref{fig:example_dPdOmega_analytic} shows the
plot of $dP_{m}/d\Omega$ for the Turok loop with
$(\alpha,\Phi)=(3/20,-18\pi/25)$ at $m=100$ calculated using the
multipoint method. The method qualitatively \footnote{The size of the
discrepancy at the peak is consistent with the relative errors seen in the
transition from low to intermediate $m$ where we would switch from
direct to multipoint methods; it reflects the marginal nature of the
cubic fit as in Case 3.} reproduces the pseudocusps
which appear in the profile of $dP_{m}/d\Omega$ calculated by an exact
numerical method in
\autoref{fig:example_dPdOmega_numerical}.

\subsection{Implications}

\begin{figure}[htbp!]

\includegraphics[width=1\linewidth, height=6cm]{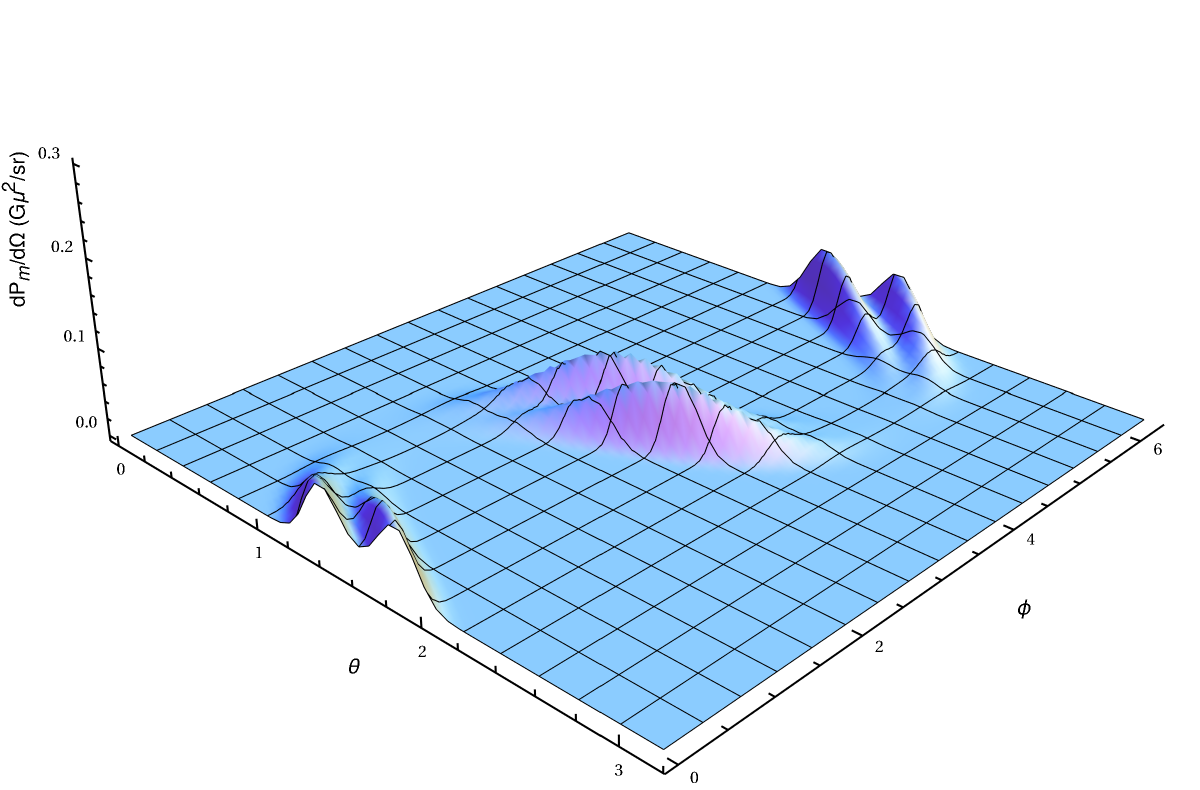} 

\caption{$dP_{m}/d\Omega$ vs. $(\theta,\phi)$ calculated using the single-point method for $(\alpha,\Phi)=(3/20,-18\pi/25)$ for $m=100$.}
\label{fig:example_dPdOmega_BPO}
\end{figure}

Any method that expands {\it only} around a single point like the cusp
will be insufficient for this example in which cusp and pseudocusp
emit in very different directions. In \cite{Blanco_Pillado_2017}, the authors calculate an analytic expression to approximate $dP_{m}/d\Omega$ for points on the celestial sphere close to the cusps making use of the small-angle approximation. To make comparisons between the various treatments, we extend this single-point method to cover all of the celestial sphere. Discrepancies arising from extending the small-angle approximation to regions away from the cusps are expected but, since the contribution from a cusp falls rapidly as the distance from the cusp increases, these are very small as far as the single-point method is concerned. Figure \ref{fig:example_dPdOmega_BPO}
shows the profile of $dP_{m}/d\Omega$ calculated using the
single-point method (extending to angles that cover the whole
celestial sphere) for the Turok loop with
$(\alpha,\Phi)=(3/20,-18\pi/25)$ at $m=100$. Notice the absence of
pseudocusp compared with \autoref{fig:example_dPdOmega_numerical} and
\autoref{fig:example_dPdOmega_analytic}. Local expansion are
insufficient to describe distant parts of the
string. \cite{Blanco_Pillado_2017} overcomes the limitation of the
single-point method by utilizing it only within a small angle of the cusp
and relying on numerical methods outside it. This is feasible when
there is an apriori known center. The multipoint method finds
expansion points as needed for any direction of emission.

We have shown analytically how pseudocusps vary with $m$ and the
results are in qualitative agreement with previous works
\cite{Damour_2001,Blanco_Pillado_2017,Stott_2017}. In Appendix \ref{sec:formalsmallangleexpansion}, we show that the argument for a
typical mode varies like $m |{\mathbf{\delta} . \vecXdot|^{3/2}}/{
  |\vecXddot |}$.  A combination of large $m$, large angle of
deviation, and small mode acceleration leads to large Bessel function
arguments that give exponentially small results \footnote{We have not given results {\it outside} the pair of
lines because the expansion points aren't necessarily fixed and the
type of Bessel function can vary when crossing the tangent curve. Nonetheless,
we expect similar behavior.}. Conversely, when $m$
is small, when the two tangent curves come close to each
other, move slowly and have large acceleration then the results
need not be small.  This is qualitatively in accordance with
\cite{Stott_2017}.

\section{\label{sec:Results}A Survey of \texorpdfstring{$\mathbf{dP_{m}/d\Omega}$}{dP/dO} Using Numerical, Multipoint and Single-Point Methods}

We now seek to quantify
the accuracy of the multipoint method in a systematic, albeit empirical
manner. This is necessary because of the difficulty setting forth apriori
requirements for accuracy (e.g. discussion of linear and cubic expansions in Section \ref{sec:MultipointMethod}).
We will compute $dP_{m}/d\Omega$ over the entire celestial sphere for a few example sets of parameters for the Turok loop described by Eq. \eqref{eq:Turokloop},  using an numerically exact method, the multipoint method and the single-point method. The following procedure defines the evaluation metric:

\begin{itemize}
    \item For each set of parameters $(\alpha,\Phi)$ and each mode number $m$, the maximum value of power emitted per solid angle, $dP_{m,max}/d\Omega$, over the entire celestial sphere is calculated by a numerically exact procedure.
    \item For the same loop and mode number, we find the maximum difference over the entire celestial sphere between the values of $dP_{m}/d\Omega$ computed by numerically exact and by the approximate methods: $\Delta\left(dP_{m}/d\Omega\right)_{max,i}=max\left[|\left(dP_{m}/d\Omega\right)_{exact}-\left(dP_{m}/d\Omega\right)_{i}\right|]$ where $i$ stands for the multipoint method or the single-point method and exact is the numerical evaluation.
    \item We summarize method fidelity in terms of the maximum absolute difference (MAD) $\Delta\left(dP_{m}/d\Omega\right)_{max,i}$ (expressed in terms of $G\mu^2$ per steradian) and the maximum relative difference (MRD) $\Delta\left(dP_{m}/d\Omega\right)_{max,i}/\left(dP_{m,max}/d\Omega\right)$ where $i$ is the method.
\end{itemize}
We plot MAD and MRD vs. $m$ in log-log scale for the two methods and compare the trends for four different illustrative cases involving two or six cusps. Cases with two and six cusps occupy finite areas in \autoref{fig:paramspace} while those with four cusps only occur on boundaries \footnote{To be precise, for most $\alpha$ and $\Phi$ the two tangent curves intersect at two or six points. Consider the separation along the $x=0$ plane. For each value of $\alpha$, there is a specific value of $\Phi$ for which the curves intersect in the plane and form a cusp. These solutions yield loops with four cusps.}. We shall not discuss these boundary cases separately -- the conclusions drawn from them are similar. 

\subsection{\label{ssec:CaseA}Two Cusps, Well-Separated Tangent Curves}

\begin{figure}[!htbp]
    \centering
    \includegraphics[width=0.8\linewidth, height=7cm]{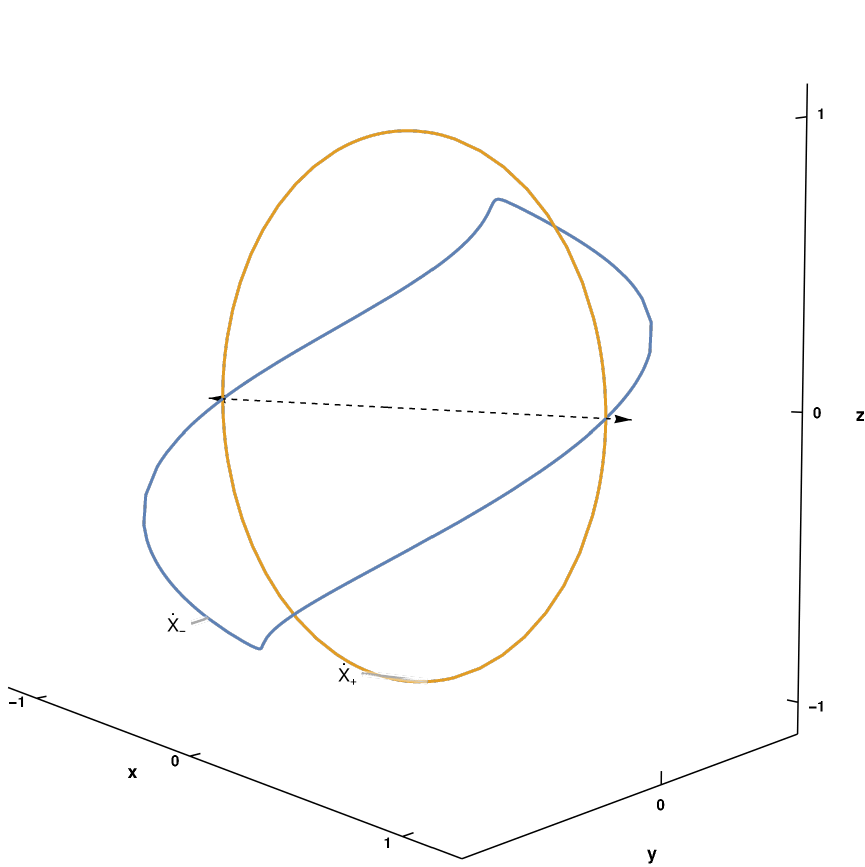}
    \caption{The tangent curves described by $\vecXdotpm$ for $(\alpha,\Phi)=(1/5,-\pi/2).$ The curves intersect at two points to form two cusps. The direction of the velocities of the cusps are shown by the black arrows.}
    \label{fig:tangent_curves_caseA}
\end{figure}

Consider the Turok loop with $(\alpha,\Phi) = (1/5,-\pi/2)$. For this loop, the tangent curves $\mathbf{\Xdot_{-}}$ and $\mathbf{\Xdot_{+}}$ intersect at two points on the unit sphere, forming two cusps, as shown in \autoref{fig:tangent_curves_caseA}. The velocities point along the $+x$ and $-x$ axes respectively i.e. along $(\theta,\phi)=(\pi/2,0),\;(\pi/2,\pi)$. The tangent curves are elsewhere well-separated. We don't expect pseudocusps so the single-point and multipoint method should be equally effective.

\begin{figure*}[!htbp]
\includegraphics[width=0.3\linewidth, height=4cm]{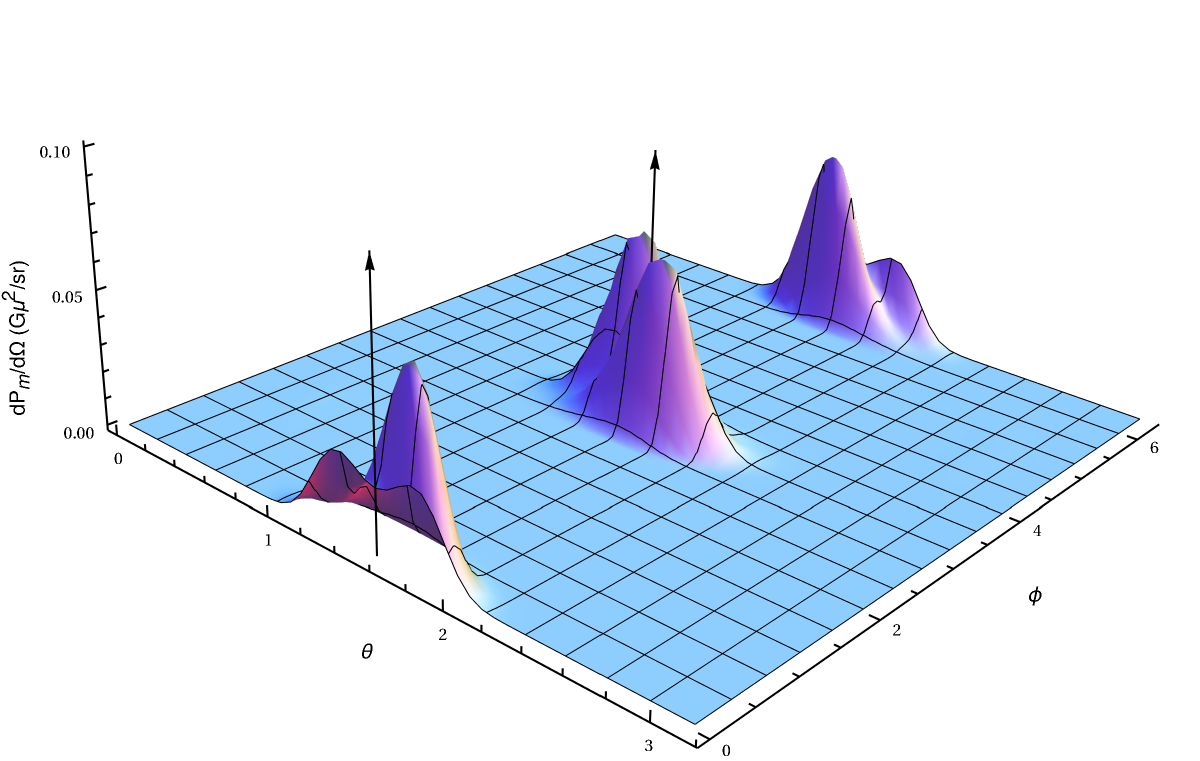} 
\quad
\includegraphics[width=0.3\linewidth, height=4cm]{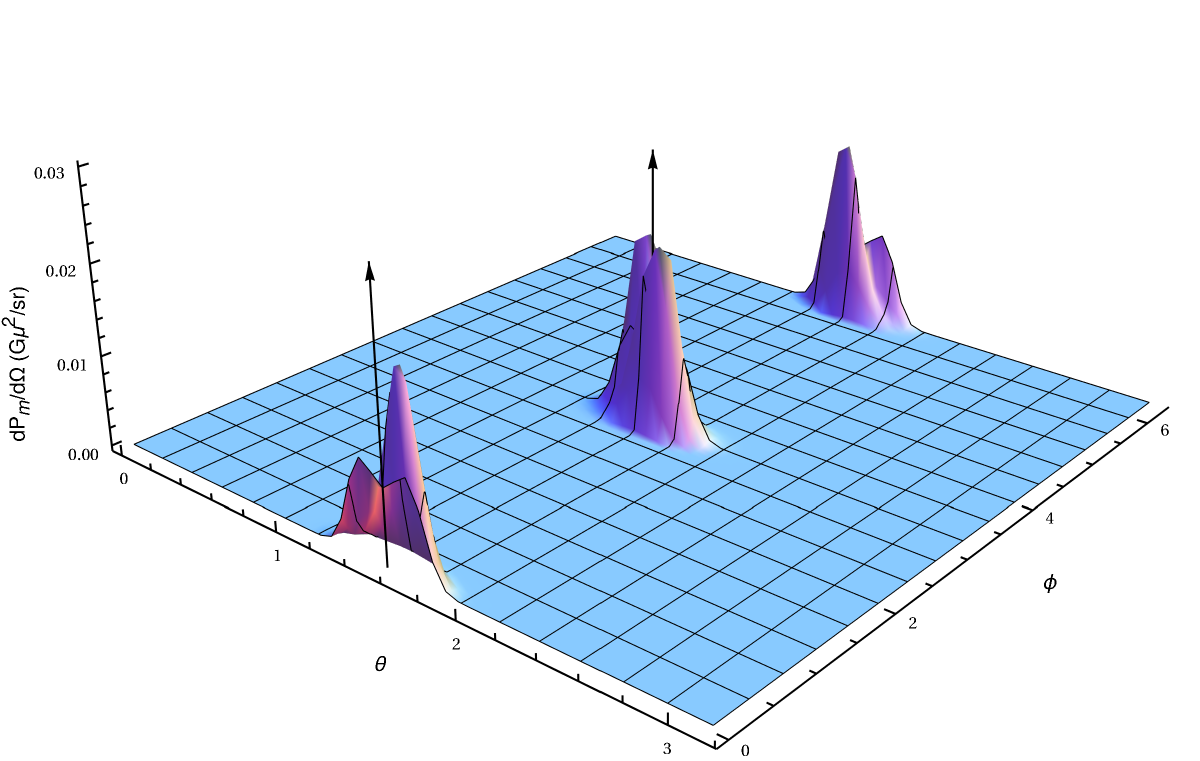} 
\quad
\includegraphics[width=0.3\linewidth, height=4cm]{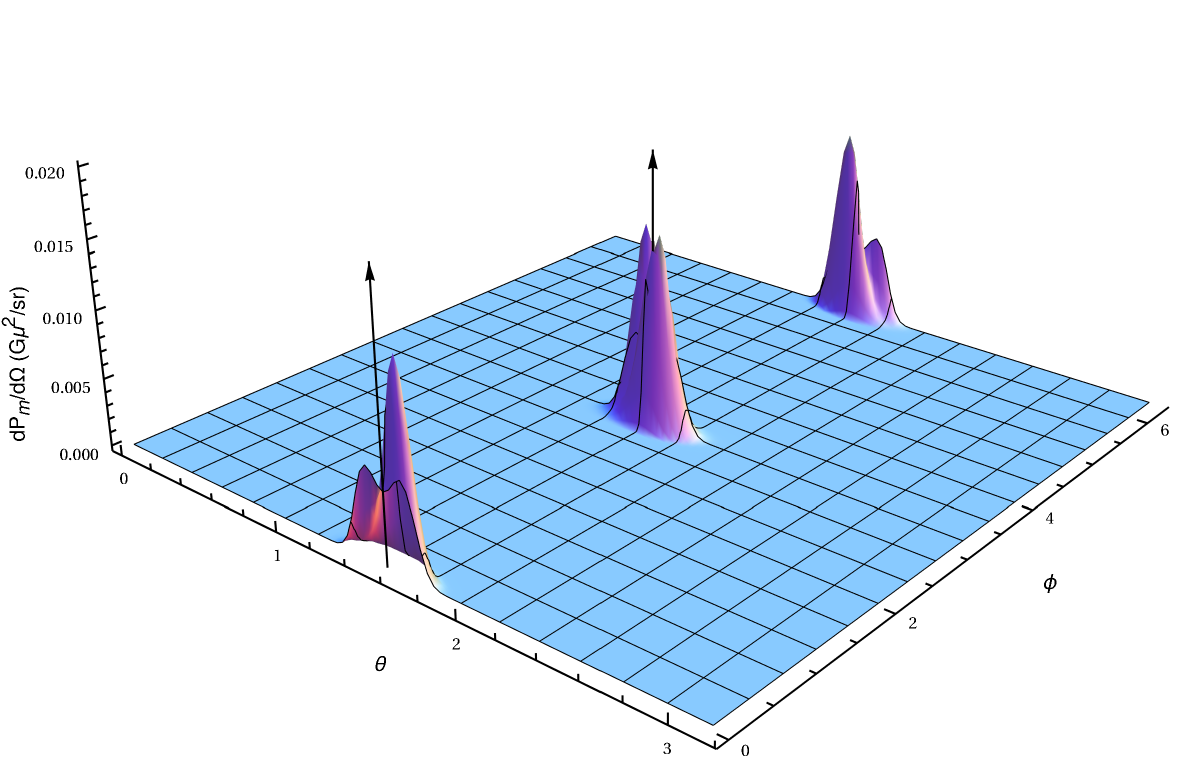} 
\caption{$dP_{m}/d\Omega$ vs. $(\theta,\phi)$ calculated numerically for $(\alpha,\Phi)=(1/5,-\pi/2)$ for $m=100, 500, 1000$.}
\label{fig:numerical_caseA}
\end{figure*}

\begin{figure}[!htbp]
    \centering
    \includegraphics[width=\linewidth, height=10cm]{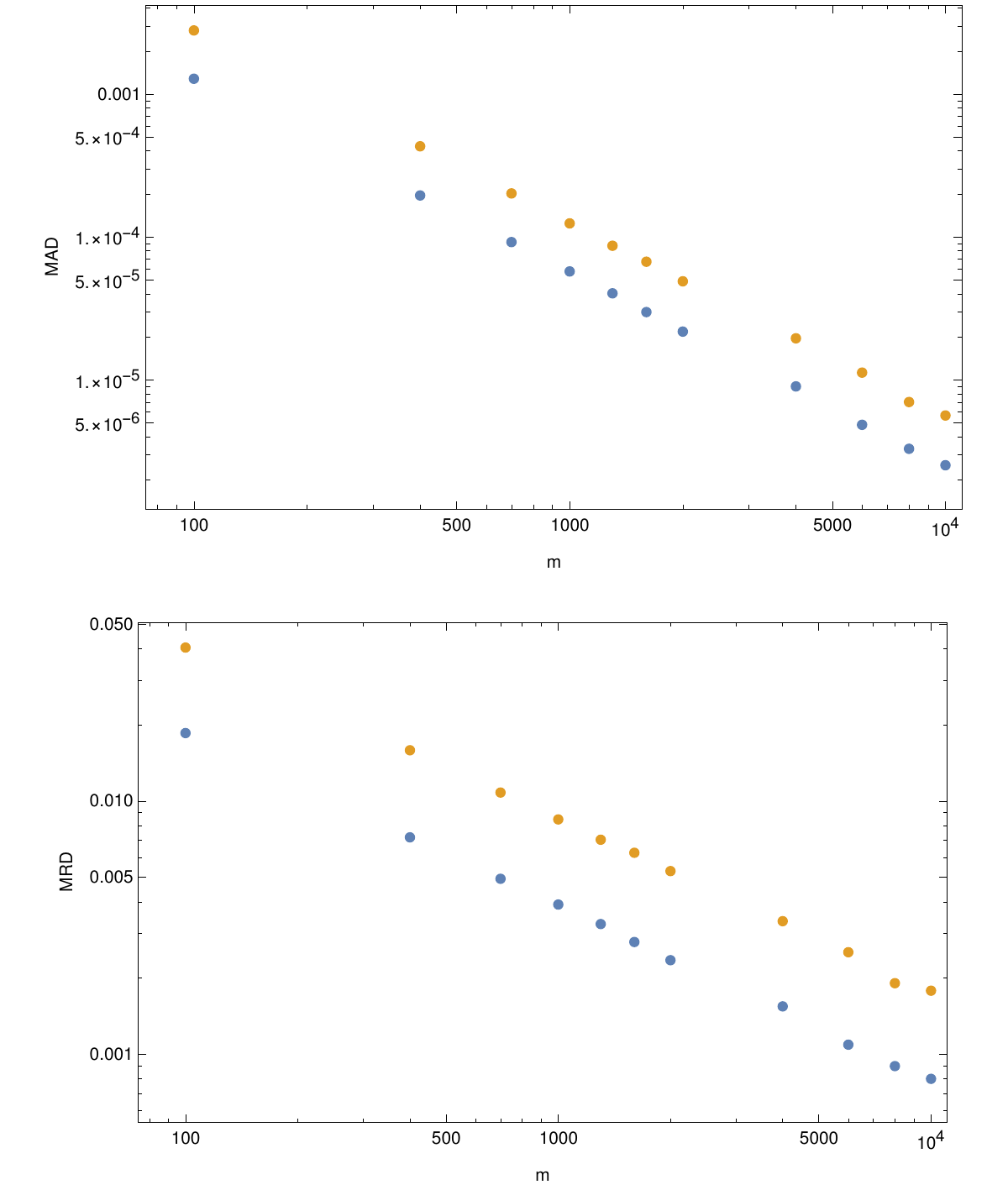}
    \caption{Plot showing log MAD vs. log $m$ and log MRD vs. log $m$ for multipoint method and the single-point method for $(\alpha,\Phi)=(1/5,-\pi/2)$. The blue markers correspond to the multipoint method and the orange markers correspond to the single-point method.}
    \label{fig:MAD_MRD_CaseA}
\end{figure}

Figure \ref{fig:numerical_caseA} shows plots of $dP_{m}(\theta,\phi)/d\Omega$ calculated numerically for three different mode numbers $m=100, 500, 1000$. Most emission comes from a small region near the direction $k.X_{-}=k.X_+=0$. The beam shape is more complicated than a filled cone with solid opening angle $\Delta\Omega \sim m^{-2/3}$ \cite{Damour_2001}. In fact, there are two prominent subpeaks. These scale down as mode number $m$ increases in the sense that the separation of the subpeaks and the width of the subpeaks all shrink together. The beam is asymptotically self-similar. All qualitative features of the plots seen in the numerical calculation are reproduced by both the multipoint method and the single-point method but there are quantitative differences. Figure \ref{fig:MAD_MRD_CaseA} shows MAD and MRD vs. mode number. While both the multipoint method and the single-point method are in good agreement with each other and with the numerical calculation, the former performs marginally better than the latter. The multipoint method reaches a MRD of 0.5\% by mode number $m \approx 700$ while the single-point method reaches the same by mode number $m \approx 2000$. Both methods yield more accurate results for higher modes.

\subsection{\label{ssec:CaseB}Two Cusps with Pseudocusp Effect}
As parameters $(\alpha,\Phi)$ vary many different ``close encounters''
between the two tangent curves may occur.
In some cases the tangent curves approach at a few specific points
(like the ``nub'' in our previous example that corresponds to a local
minimum in the distance of separation and gives rise to a localized
pseudocusp) while in others the curves may be roughly parallel over
some range of $\sigma_\pm$.  If the angle between the two curves that cross
to give a true cusp is small then the possibility arises for an
extended emission region in the vicinity of the cusp itself.

\begin{figure}[!htbp]
    \centering
    \includegraphics[width=0.8\linewidth, height=7cm]{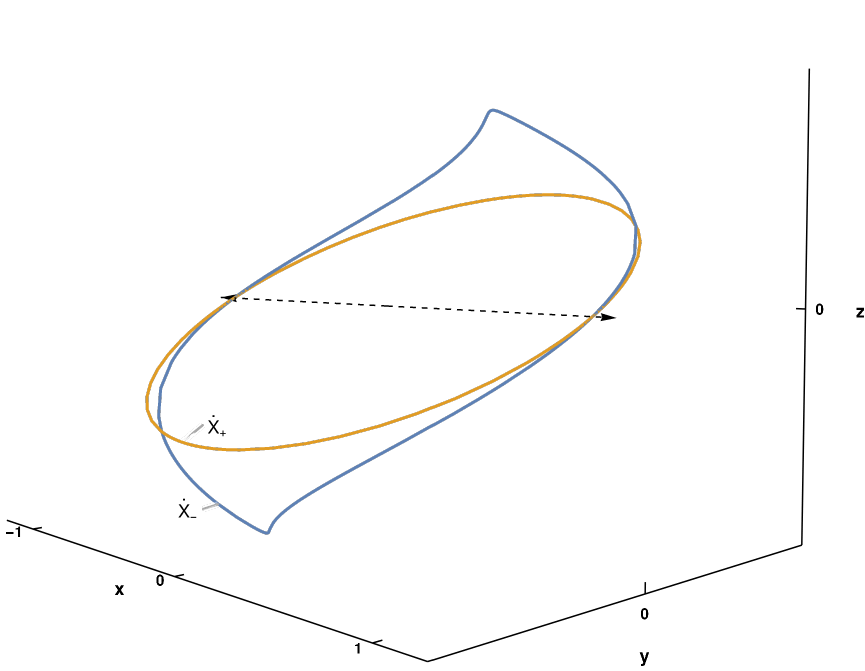}
    \caption{The tangent curves described by $\vecXdotpm$ for $(\alpha,\Phi)=(1/5,3\pi/20).$ There are two cusps with velocities shown by the black arrows. The two tangent curves are furthest apart at the top and bottom in the $x=0$ plane. They come close to each other near the cusp. This case stands in contrast
\autoref{fig:tangent_curves_caseB}.}
    \label{fig:tangent_curves_caseB}
\end{figure}

As an example consider the Turok loop with $(\alpha,\Phi) = (1/5,
3\pi/20)$, shown in \autoref{fig:tangent_curves_caseB}. This loop has
two cusps with velocities pointing along
$(\theta,\phi)=(\pi/2,0),\;(\pi/2,\pi)$. For asymptotically large $m$,
the maximum emission is expected to be along these directions on the
celestial sphere. Unlike the previous example, in the $x=0$ plane the
two tangent vectors are at maximum distance from each other (the
``nub'' on $I_-$ is far from $I_+$) and they approach closely for an
extended range near the cusp only because they form an acute angle at the
point of intersection.

\begin{figure*}[!htbp]

\includegraphics[width=0.3\linewidth, height=4cm]{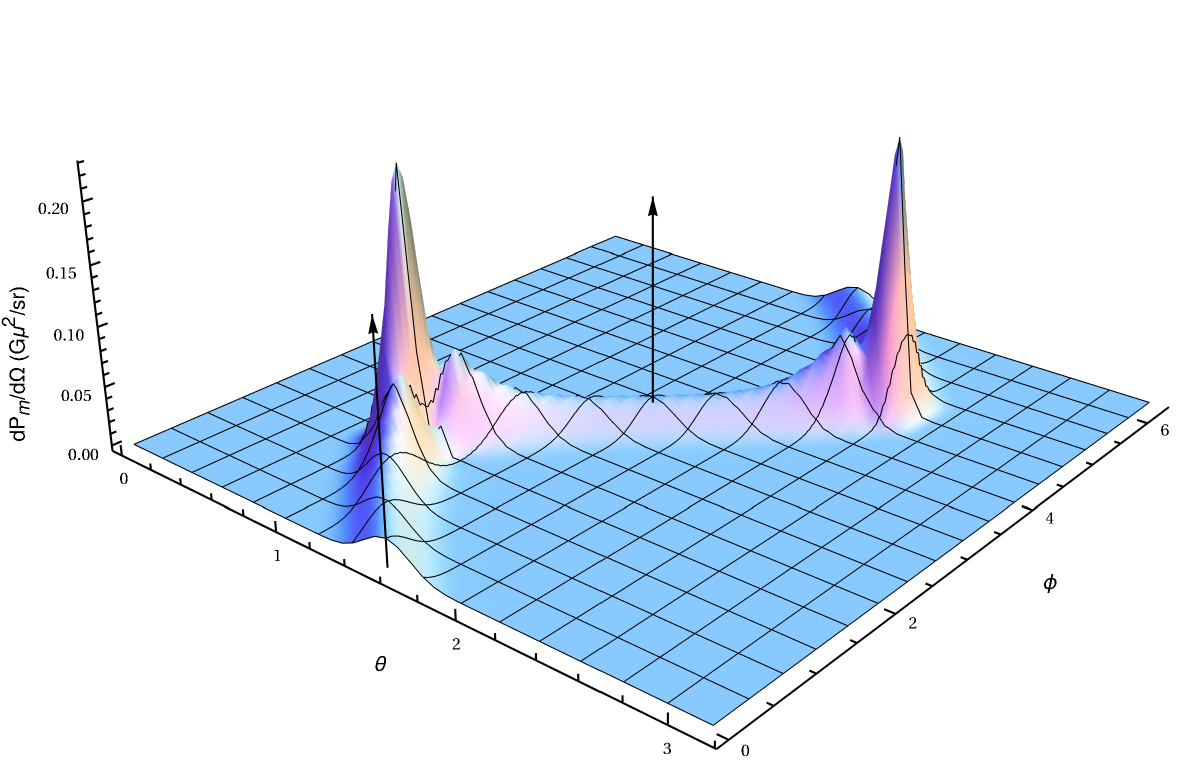} 
\quad
\includegraphics[width=0.3\linewidth, height=4cm]{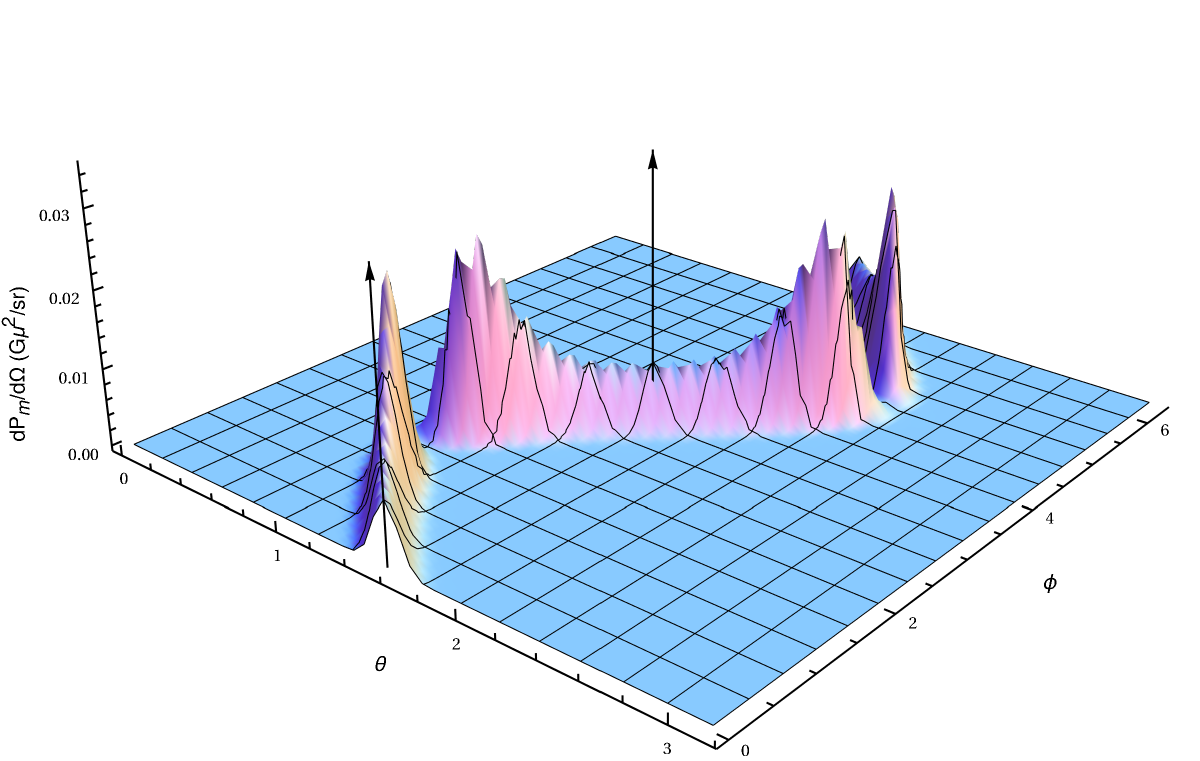}
\quad
\includegraphics[width=0.3\linewidth, height=4cm]{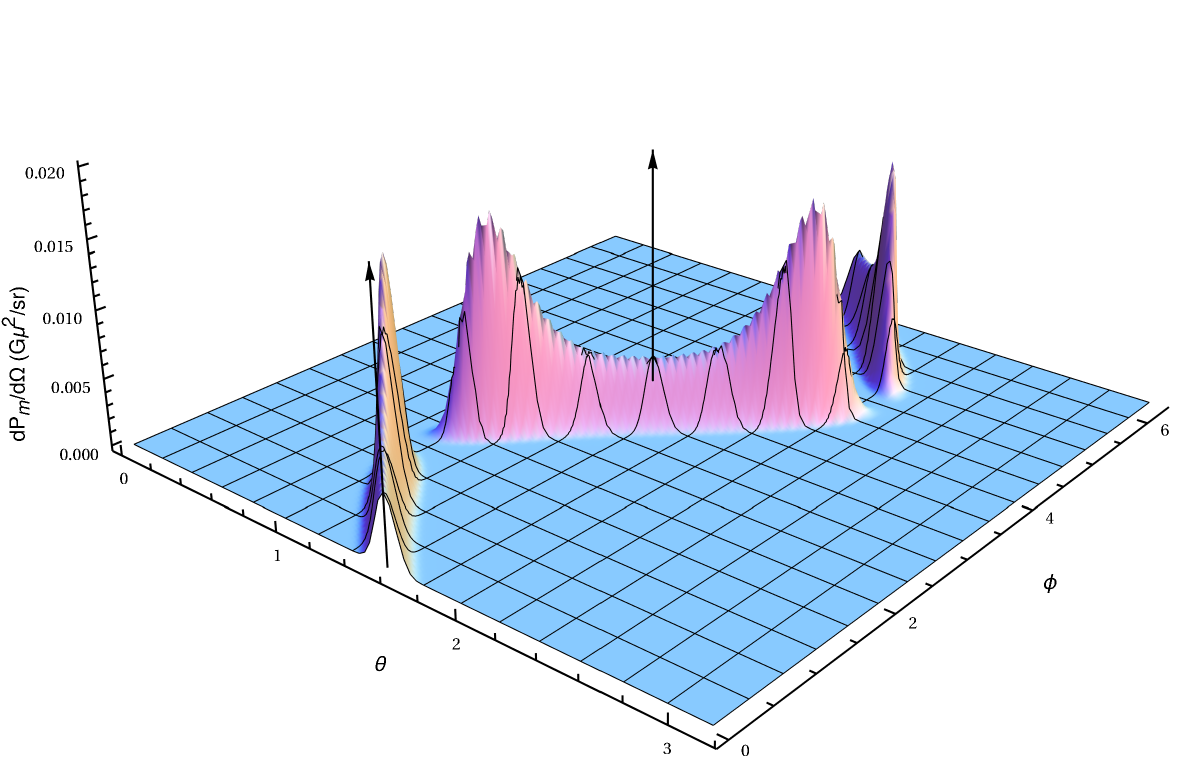}

\caption{$dP_{m}/d\Omega$ vs. $(\theta,\phi)$ calculated numerically for $(\alpha,\Phi)=(1/5,3\pi/20)$ for $m=100, 500, 1000$.}
\label{fig:numerical_caseB}
\end{figure*}

\begin{figure*}[htbp!]

\includegraphics[width=0.3\linewidth, height=4cm]{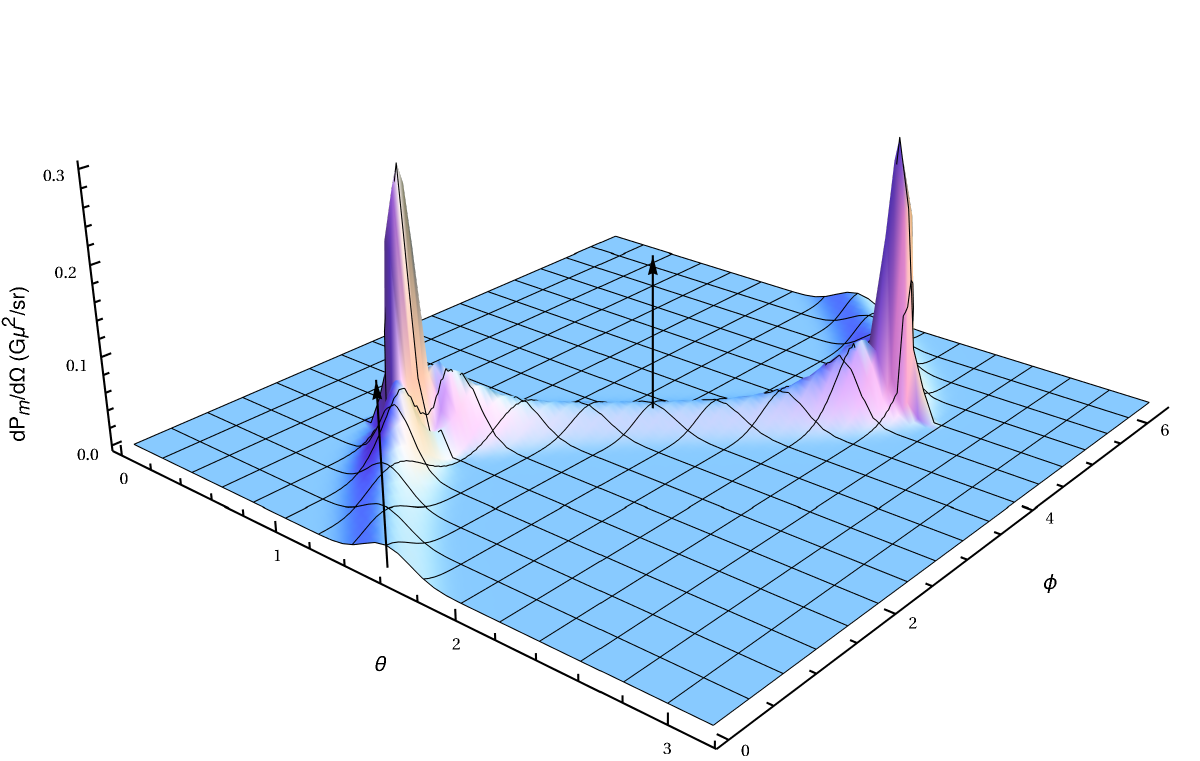} 
\quad
\includegraphics[width=0.3\linewidth, height=4cm]{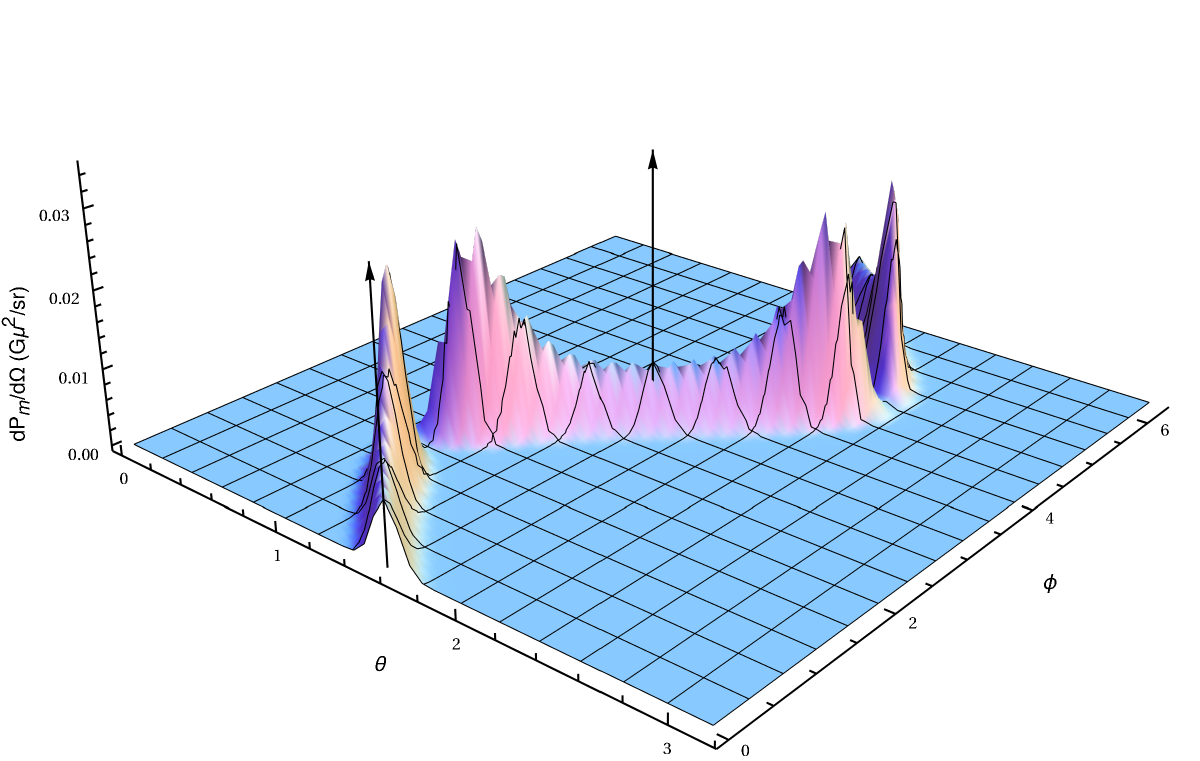}
\quad
\includegraphics[width=0.3\linewidth, height=4cm]{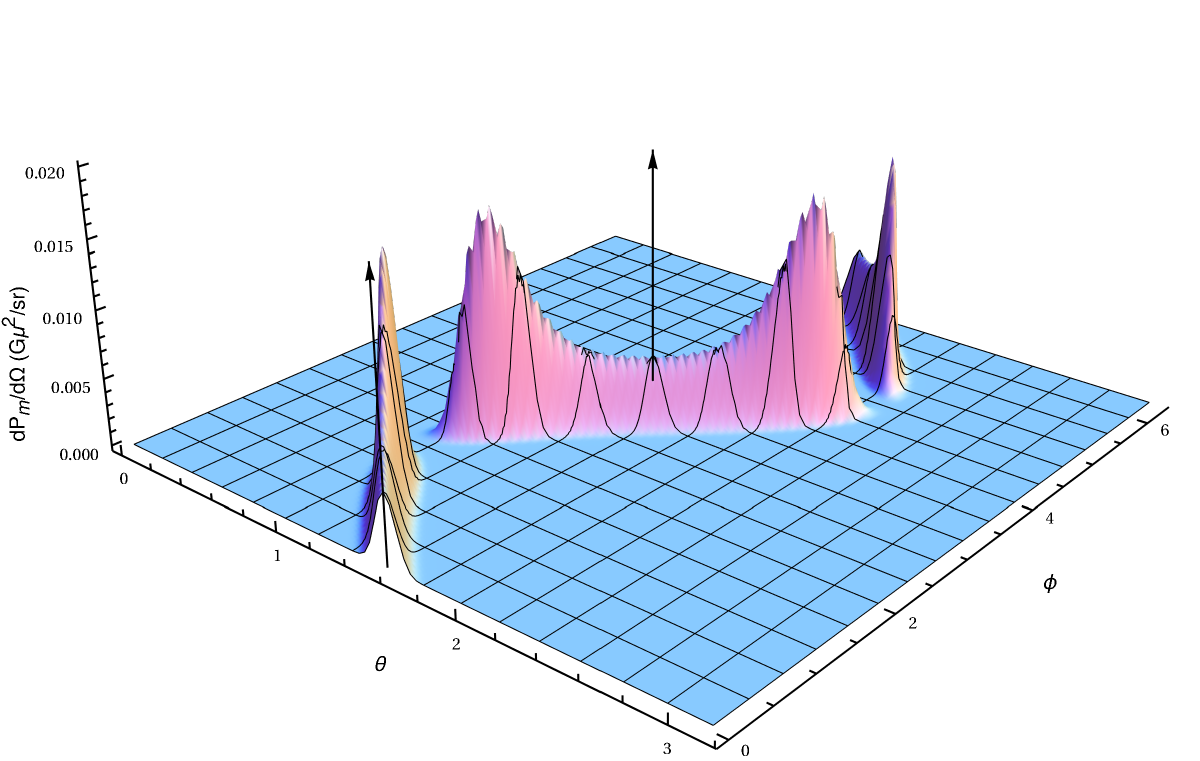}

\caption{$dP_{m}/d\Omega$ vs. $(\theta,\phi)$ calculated using the multipoint method for $(\alpha,\Phi)=(1/5,3\pi/20)$ for $m=100, 500, 1000$.}
\label{fig:analytic_caseB}
\end{figure*}

\begin{figure*}[htbp!]

\includegraphics[width=0.3\linewidth, height=4cm]{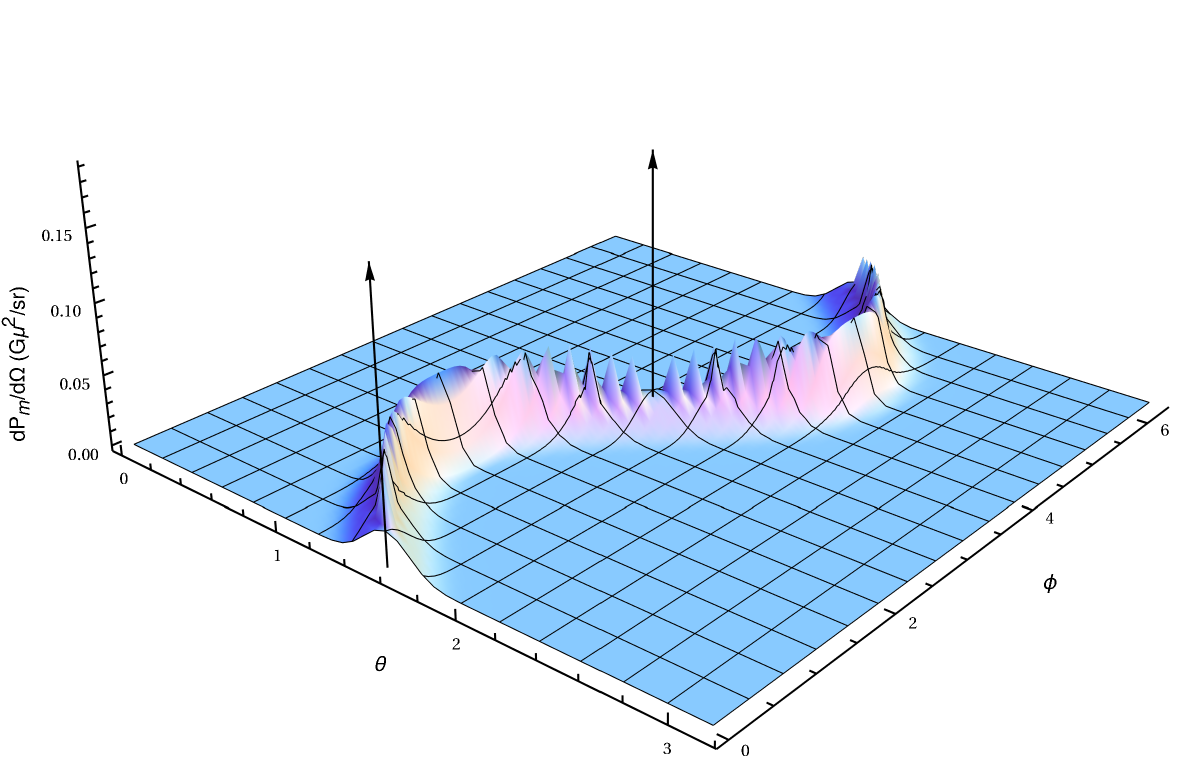} 
\quad
\includegraphics[width=0.3\linewidth, height=4cm]{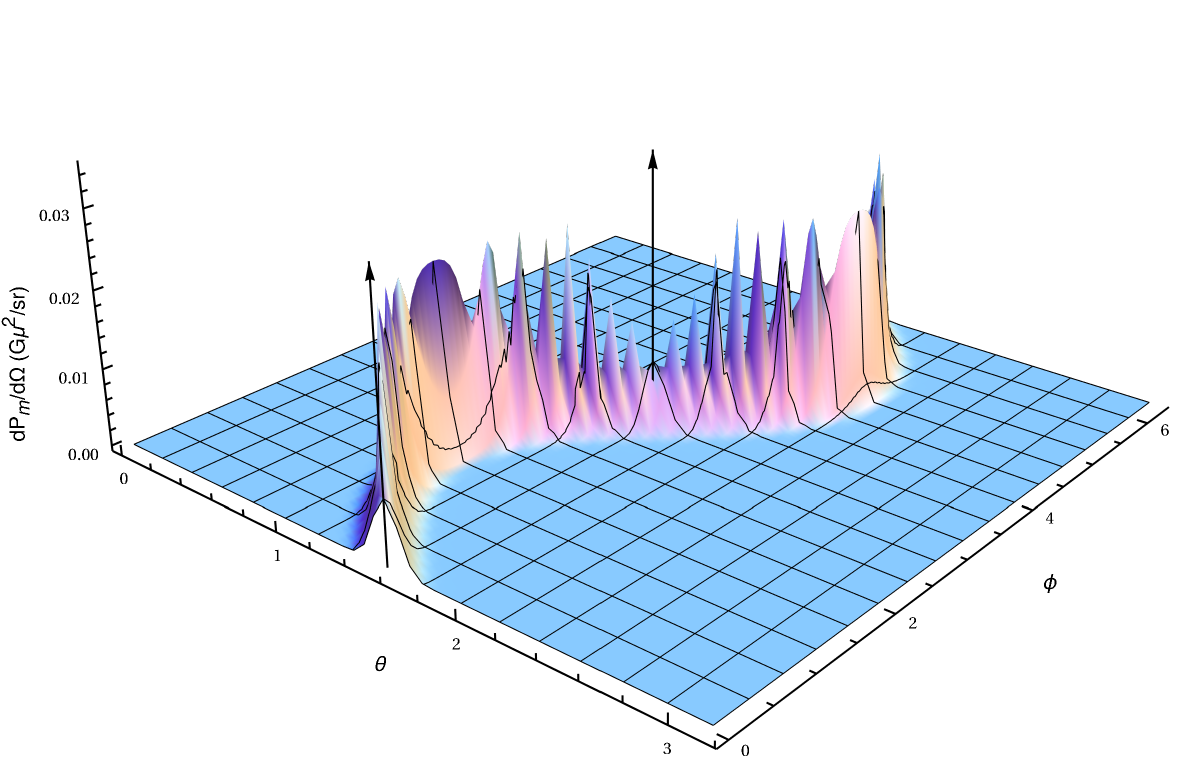}
\quad
\includegraphics[width=0.3\linewidth, height=4cm]{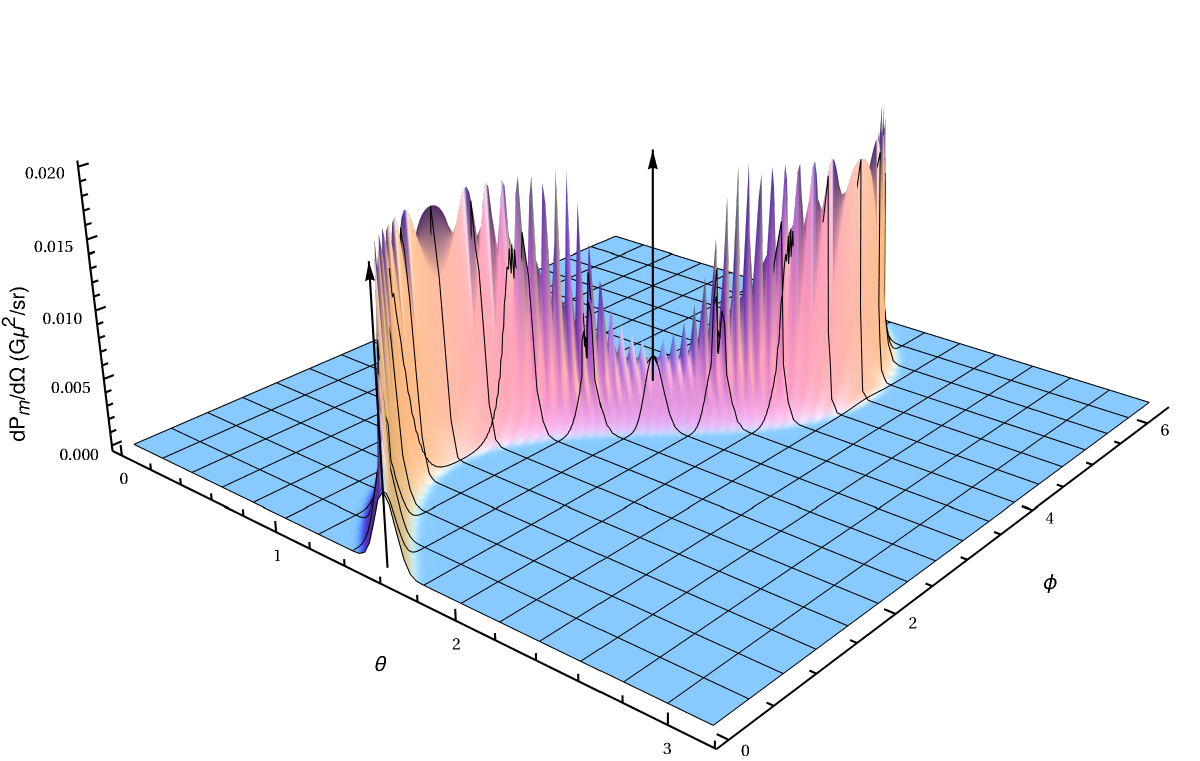}

\caption{$dP_{m}/d\Omega$ vs. $(\theta,\phi)$ calculated using the single-point method for $(\alpha,\Phi)=(1/5,3\pi/20)$ for $m=100, 500, 1000$.}
\label{fig:BPO_caseB}
\end{figure*}

Figures \ref{fig:numerical_caseB}, \ref{fig:analytic_caseB} and \ref{fig:BPO_caseB} show the plots of $dP_{m}/d\Omega$ computed using the numerical, the multipoint and the single-point method for this loop at modes $m=100, 500, 1000$. The emission is maximum at $m=100$, appearing at $\phi \approx \pi/2, 3\pi/2$ near the nub. That region migrates towards the cusps on either side as $m$ increases. Higher $m$ requires smaller angles of separation between the tangent curves.

The multipoint method, depicted in \autoref{fig:analytic_caseB}, reproduces the trends. For $m=100$ the maximum value of $dP_{m}/d\Omega$ near the nub computed using this method is off from the value computed numerically by $\sim 38 \%$
(the size of the discrepancy is consistent with the relative errors seen
in the transition from low to intermediate $m$ where we would
switch from direct to multipoint methods).
As $m$ increases the differences shrink. The single-point method, with center of expansion at the cusp itself, misses the most prominent and distant pseudocusp near the nub but the size of the error decreases as $m$ increases once the cusp becomes prominent.

\begin{figure}[htbp!]
    \centering
    \includegraphics[width=1\linewidth, height=10cm]{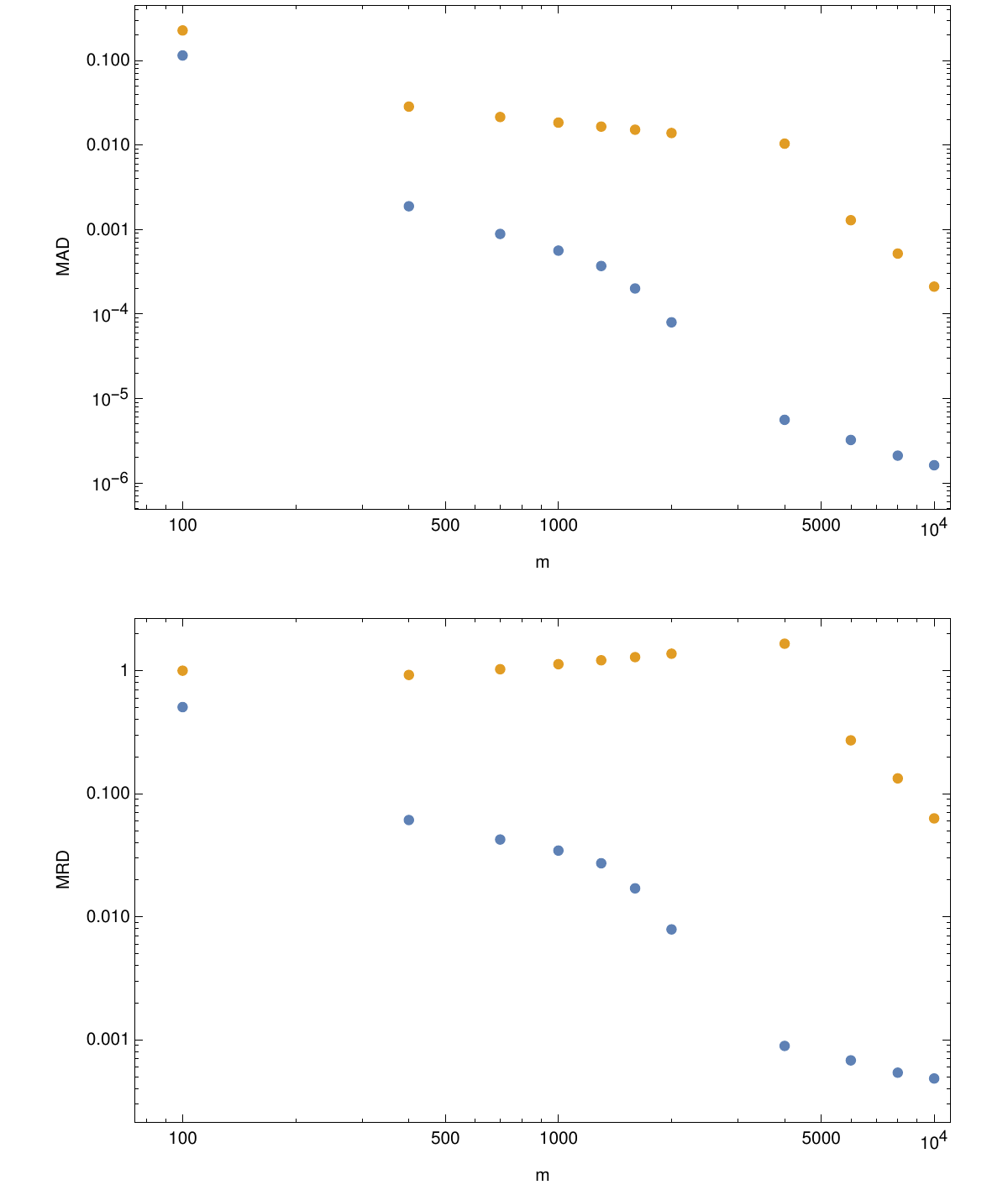}
    \caption{Plot showing log MAD vs. log $m$ and log MRD vs. log $m$ for the multipoint method and the single-point method for $(\alpha,\Phi)=(1/5,3\pi/20)$. The blue markers correspond to the multipoint method and the orange markers correspond to the single-point method.}
    \label{fig:MAD_MRD_CaseB}
\end{figure}

Figure \ref{fig:MAD_MRD_CaseB} shows the MAD and MRD for both methods as a function of the mode number. The multipoint method reaches a MRD of 6\% by mode number $m=400$ while the single-point method reaches the same by mode number, $m=10000$. The dip in the MAD and MRD for both methods at $m=6000$ has a simple explanation. For $m<6000$, the maximum differences between each of the approximate methods and the numerical method occur near the nub. Because the multipoint method better reproduces the pseudocusp there, the errors are smaller compared with those of the single-point method. For $m>6000$ the maximum differences occur closer to the cusps which are picked up by both the methods. Both methods get more accurate near the cusps and the differences decrease. But the multipoint method still fares better overall for this case.

\subsection{\label{ssec:CaseC}Six Cusps with Pseudocusp Effect}

\begin{figure}[htbp!]
    \centering
    \includegraphics[width=0.8\linewidth, height=7cm]{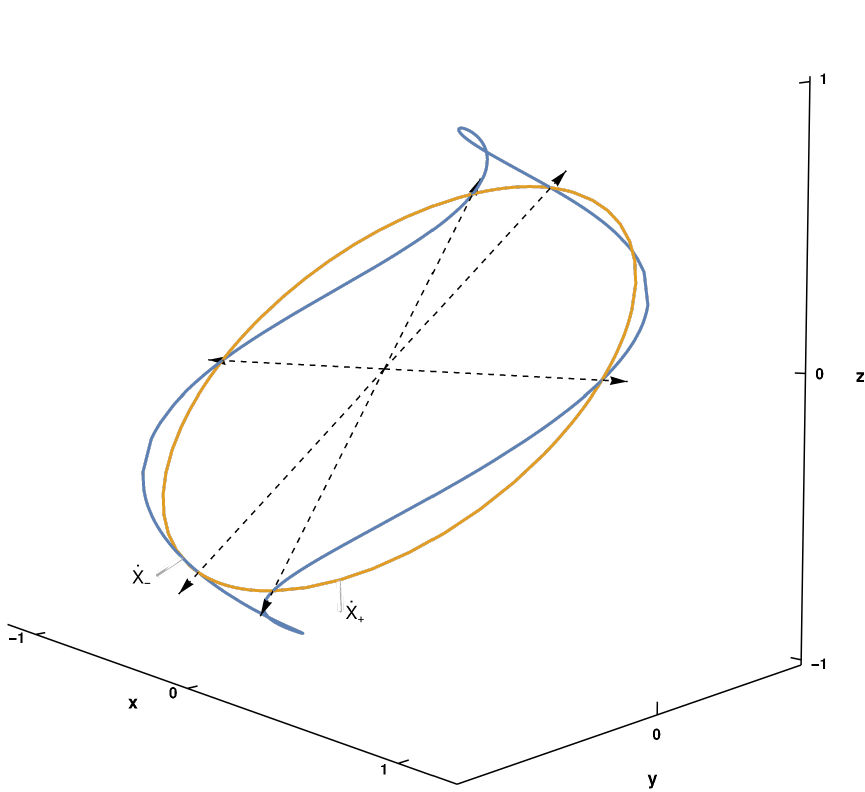}
    \caption{The tangent curves described by $\vecXdotpm$ for $(\alpha,\Phi)=(3/10,\pi/4).$ There are six cusps with velocities shown by the six arrows. The curves also come close to each other in between the cusps.}
    \label{fig:tangent_curves_caseC}
\end{figure}

For the Turok loops with six cusps, there exists a range of parameters where the two tangent vectors are close to each other (see \autoref{fig:paramspace}).
Consider the Turok loop described by $(\alpha,\Phi)=(3/10,\pi/4)$. The tangent curves intersect each other at six points on the unit sphere giving rise to six cusps, as shown in \autoref{fig:tangent_curves_caseC}. In addition to the cusps, the two tangent curves also approach each other along the plane $x=0$ ($\phi=\pi/2, 3\pi/2$). These generate extended emission. We omit detailed renditions of the emission and simply use MAD and MRD to quantify the accuracy.

\begin{figure}[htpb!]
    \centering
    \includegraphics[width=1\linewidth, height=10cm]{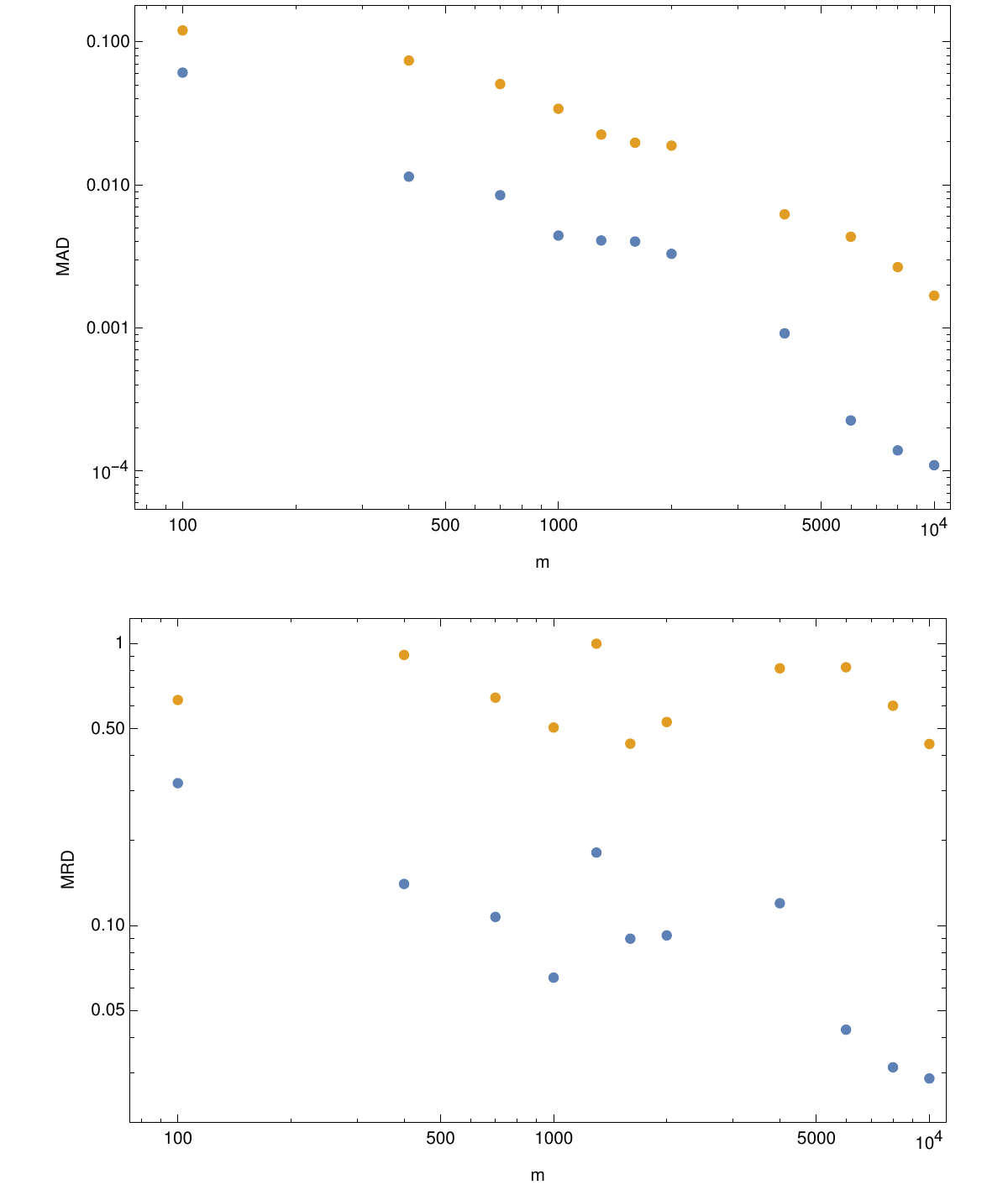}
    \caption{Plot showing log MAD vs. log $m$ and log MRD vs. log $m$ for the multipoint (analytic approximation) method and single-point method for $(\alpha,\Phi)=(3/10,\pi/4)$. The blue markers correspond to the multipoint method and the orange markers correspond to the single-point method.}
    \label{fig:MAD_MRD_CaseC}
\end{figure}

Figure \ref{fig:MAD_MRD_CaseC} shows the MAD and MRD for the two analytic
methods. The MAD decreases more or less consistently with $m$. The MRD
also decreases over a large range of $m$, but since it involves
scaling the MAD by the maximum value of $dP_{m}/d\Omega$, there are
minor variations when the decrease in MAD is not proportional to the
decrease in $dP_{m,max}/d\Omega$. The MRD reaches $\sim 4$\% by
$m=6000$ for the multipoint method but it is an order of magnitude
bigger for the single-point method. The maximum differences in both
cases are found near the pseudocusps for $m<6000$. In summary, the
multipoint method yields more accurate results as it takes into
account the pseudocusps which the single-point method does not
describe.

\subsection{\label{ssec:CaseD}Six Well-Separated Cusps}

\begin{figure}[htbp!]
    \centering
    \includegraphics[width=0.8\linewidth, height=7cm]{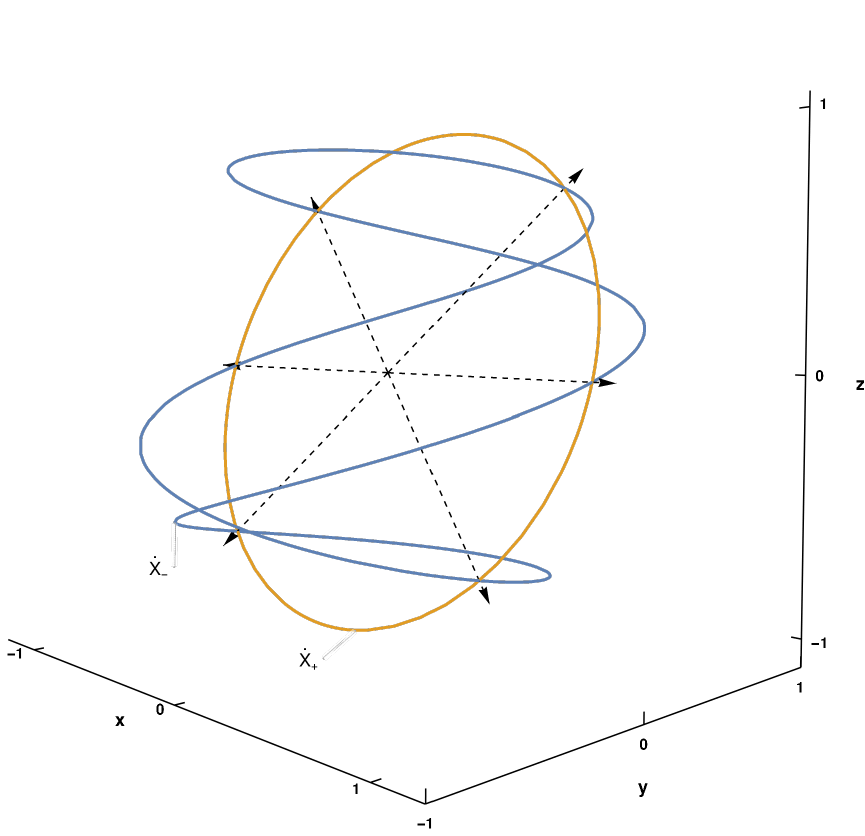}
    \caption{The tangent curves described by $\vecXdotpm$ for $(\alpha,\Phi)=(4/5,2\pi/5).$ There are six cusps which are all well-separated from each other.}
    \label{fig:tangent_curves_caseD}
\end{figure}

For larger values of $\alpha$, the tangent curve $\mathbf{\Xdot_{-}}$ becomes more wiggly and the separation between the cusps increases. Compared to the previous case, larger $\alpha$ create larger separation between the tangent curves in between the cusps. As an example, consider the Turok loop described by $(\alpha,\Phi)=(4/5,2\pi/5)$. Figure \ref{fig:tangent_curves_caseD} shows the two tangent curves for this loop. The six cusps are well-separated on the unit sphere and the distance between the two curves at $x=0 (\phi=\pi/2, 3\pi/2)$ is also larger than in the previous case. 

\begin{figure}[!htbp]
    \centering
    \includegraphics[width=1\linewidth, height=10cm]{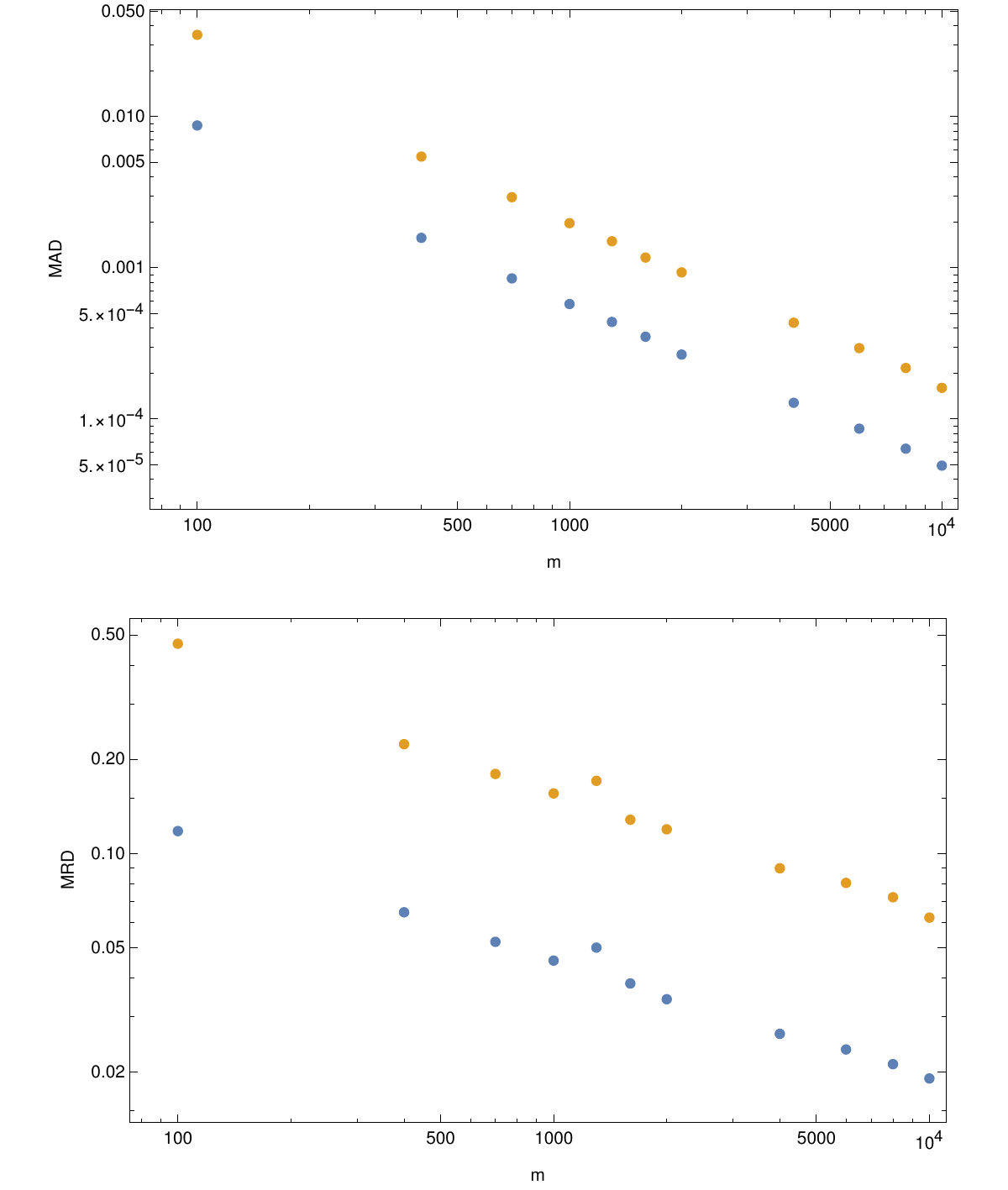}
    \caption{Plot showing log MAD vs. log $m$ and log MRD vs. log $m$ for the multipoint (analytic approximation) method and the single-point method for $(\alpha,\Phi)=(4/5,2\pi/5)$. The blue markers correspond to the multipoint method and the orange markers correspond to the single-point method.}
    \label{fig:MAD_MRD_CaseD}
\end{figure}

We expect that pseudocusps will not play as big a role as in the previous cases (Section \ref{ssec:CaseB} and Section \ref{ssec:CaseC}). The MRDs are still smaller for the multipoint method, as shown in \autoref{fig:MAD_MRD_CaseD}. The MRD for the multipoint method reaches $\sim 6$\% by $m=400$ whereas for the single-point method, it reaches the same range only by $m=10000$.

\subsection{Implications}

For all four cases described, the multipoint method fares better
than the single-point methods in the calculation of
$dP_{m}/d\Omega$. These four cases cover the main situations observed
for Turok loops.  The multipoint method yields lower MRDs
than the single-point method overall, but the difference is marked for
cases involving pseudocusps. The advantage of the multipoint method is
that it takes into account regions of the loop away from the cusps,
which play an important role in the low to intermediate mode number
range. In addition, it improves the accuracy of the beam
shape even when only cusps are present.

\FloatBarrier

\section{\label{sec:Discussion}Discussion}

We have considered the gravitational wave emission from a cosmic
string loop. All calculations reduce to evaluating one-dimensional
integrals over the left- and right-moving modes of the string. Our
goal was to develop a network of techniques to handle the emission
over the entire range of harmonics $m$ and over the whole celestial
sphere. The techniques are directly relevant to the calculation of
waveforms of the outgoing gravitational waves and the evaluation
of fluxes of radiated conserved quantities.

Here, we have concentrated on applying the techniques to calculating the
frequency-dependent emission of energy, momentum and angular
momentum. The SGWB generated by cosmic strings at a range of redshifts
is one of the principal applications. If that signal is experimentally
detected it will probe various cosmological features such as the
radiation-to-matter transition, the number of relativistic degrees of
freedom at different redshifts and other interesting questions related
to the properties of the strings themselves \cite{Auclair_2020}. There
are a wide range of experiments that can potentially probe the SGWB
for frequencies between $10^{-9}$ and $10^6$ Hz. These include pulsar
timing arrays (NANOGRAV \cite{Brazier:2019mmu}, EPTA \cite{EPTA:2016ndq}, IPTA \cite{Verbiest:2016vem}, InPTA \cite{Joshi:2018ogr}, PPTA \cite{Manchester:2012za}, CPTA \cite{Xu:2023wog} and MeerKAT PTA \cite{Miles:2022lkg})
and interferometric ground-based
detector networks \cite{KAGRA:2013rdx} including various individual
interferometers (LIGO and Advanced LIGO
\cite{LIGOScientific:2014pky,LIGOScientific:2016wof}, VIRGO
\cite{VIRGO:2014yos}, KAGRA \cite{KAGRA:2020tym}, IndIGO \cite{Unnikrishnan:2023uou}) and co-located interferometers (Holometer \cite{Holometer:2016ipr}). Proposed
space-based experiments (LISA \cite{Babak:2021mhe}, ASTROD-GW \cite{Ni:2012eh}, BBO \cite{Crowder:2005nr},
DECIGO \cite{Kawamura:2020pcg}, TianGO \cite{Kuns:2019upi} and TianQind \cite{Torres-Orjuela:2023hfd}) and proposed
ground-based detectors (ET
\cite{Punturo:2010zza,Punturo:2010zz,Hild:2010id}, Cosmic Explorer
\cite{Evans:2021gyd,Reitze:2019iox}) should extend the sensitivity and
frequency coverage.

To give an example when the beam modeling of a source becomes directly
relevant to observations we will consider the possibility that a
relatively nearby string loop radiates and is detected in the face of
the confusion of a cosmologically produced SGWB
\cite{depies2009harmonic,khakhalevali2020lisa,Martinez:2020cdh}. This
will illustrate how each of the separate analytic regimes describing
the beam may play a role in a full analysis of the loop's putative
signal.

Today, local string loops have a characteristic size $\ell_{evap} =
200 \mu_{-9}$ pc for tension $\mu_{-9}=G \mu/10^{-9}$. (This size
corresponds to a loop that radiates its entire mass by emission of
gravitational radiation in a period of time equal to the age of the
Universe). In the network scaling solution a fixed number of loops
with size equal to a fraction of the horizon form at each epoch.  The
oldest, surviving loops in the Universe have the highest number
density today.  For loop size $\ell = x \ell_{evap}$ the most numerous
and typical loops have $x$ of order 1. Given an instrument sensitive
to frequency $f$ we infer the harmonic of interest $m \ge 1$. The
characteristic harmonic of the emission of a nearby ({\it not}
cosmologically distant) loop is $m \sim f (G\mu/c^2) \Gamma t_0/2$
where $t_0$ is the age of the Universe and $\Gamma \sim 50$, a
dimensionless constant characterizing the efficacy of the loop's
emission of gravitational waves. Quantitatively, $m \sim \max \{1,
10^{10} (f/{\rm Hz}) \mu_{-9}\}$. We have calculated the expected
number of observable, nearby loops in a homogeneous Universe as a
function of $f$, $\mu$ and $m$ according to two different detection
criteria: (1) the energy flux in the loop exceeds that of the SGWB
flux and (2) an experiment of fixed total duration yields a high S/N result
in a template search for a harmonic source (a particular mode of the
string emission) where the background noise is the SGWB.  In these
idealized analyses neither instrumental effects nor other
astrophysical sources impede the detection of the nearby loop and
the results are ``best case'' scenarios. The
inhomogeneity of the distribution of loop sources in the Universe is
ignored. We will present details elsewhere but simply summarize that the
characteristic {\it range} of harmonics of the detections agree
with the estimate above. An experiment like LISA with $f \sim 10^{-3}$
Hz is potentially sensitive to harmonics emitted by a single loop up
to $m \sim 10^7 \mu_{-9}$. It is important to emphasize that the
numbers of detections forecast differ markedly for the two criteria
and are strong functions of the observing frequency. In addition, all
numbers are sensitive to other factors including local clustering of
string loops (see differing estimates in
\cite{Chernoff:2009tp,Jain:2020dct}) and the degree of enhancement of
superstring numbers over that of normal strings
\cite{Chernoff:2014cba,Chernoff:2017fll}. The important point is
simply that one expects loops to emit in a large range of $m$.

To handle the integrals over a range of $m$ we have discussed three types of techniques:
direct calculation (simple numerical quadrature, FFT-mediated
quadratures and deformed complex contour methods), stationary point and
near-stationary point approximations (multipoint method) and asymptotic
approximations (Laplace's method).

These methods are naturally suitable for covering
complementary ranges of mode numbers $m$ as depicted in
\autoref{fig:3methods}. Direct numerical quadrature is appropriate at
low modes where the relevant integrals are mildly oscillatory. The
FFT-mediated approach has the efficiency of handling multiple modes
together but the accuracy is limited by aliasing effects as $m$
increases.\footnote{We have also validated the capability of the method of
steepest descent to produce exact calculations, equivalent to direct
numerical integration, but not included it in the set of comparisons
and will report the details elsewhere.}

All direct techniques suffer from increasing calculational complexity
as $m$ increases. We have introduced two approximate methods well
suited to larger $m$: the multipoint method and the asymptotic
method. The multipoint method builds upon the existing techniques
\cite{Damour_2001, Blanco_Pillado_2017} but is not restricted to cusps
and, in particular, is ideal for analyzing the emission away from
cusps such as that associated with pseudocusps.

Our multipoint exploration of the pseudocusp phenomena demonstrates that
pseudocusps can dominate cusps at intermediate $m$ but are
subdominant at large $m$. This is in agreement with the previous works
\cite{Damour_2001,Blanco_Pillado_2017,Stott_2017}. The strength of the
pseudocusp emission is influenced by multiple factors -- the angle of
separation between the tangent curves on the celestial sphere and the
velocity and acceleration of the tangent vectors. These set the
magnitude and the variation of $k.\dot{X}$ (refer to Appendix \ref{sec:expand}
for details) which is the key element of the multipoint approximation.

We have investigated the behavior of the multipoint method over a wide
range of $m$ and assessed its accuracy.  The absolute scale of the
beam emission decreases as $m$ increases but the relative errors
plateau at values of $\mathcal{O}(1)$ for large $m$. Depending upon
the application, it may be sufficient to utilize the results
without further worry -- they have small absolute errors but large
relative errors.  In any case, the multipoint method compares
favorably with all other existing techniques not only away from cusps
but also in the direction of cusps. For example, it provides more
accurate cusp beam shapes than existing methods.

We also introduced an asymptotic analysis based on locating the
steepest descent contours in the complex plane and treating the
one-dimensional integral by means of Laplace's method.  This was
particularly effective at large $m$ and yielded decreasing relative
errors. The most accurate treatment of the beam should switch from the
multipoint (or direct) methods to the asymptotic method at large
$m$. The transition is not simple.  It depends not only upon $m$ but
also upon the alignment of the viewing angle and the tangent
directions. The closer the direction is to exact alignment the larger
the mode number $m$ must be for the asymptotic analysis to become
accurate. We have provided an empirical method to determine the
transition by examining the results that are generated by successive
orders of the asymptotic method.

Elsewhere we will return to the problem of constructing time-dependent
waveforms.

\section{\label{sec:Conclusion}Conclusion}
We have introduced two new methods to approximate the emission of gravitational radiation from a cosmic string loop. The asymptotic method makes use of the method of steepest descent to approximate the values of the integrals $I_{\pm}$ and matches the exact answer as $m \to \infty$. The multipoint method performs best for an intermediate range of modes $m_{low} < m < m_{cross}$. It generally gives better results than other approximate methods. It improves upon single-point methods (which focus mainly on the cusps formed on the loop) by including new regions of the loop which contribute to the emission at lower modes. It extends the range of mode numbers for which the emission can be reliably calculated.

A combination of direct numerical methods at low modes, the multipoint
method at intermediate modes and the asymptotic method at high modes
has the potential to provide a complete, calculable model for gravitational
wave emission from a single cosmic string loop.

\section{\label{sec:Acknowledgements} Acknowledgements}
DC thanks Soumyajit Bose, Liam McAllister and Barry Wardell for useful
conversations. NS thanks the Cornell Department of Astronomy for
summer graduate fellowship over June 2023.  \nocite{*}

\bibliography{main}% Produces the bibliography via BibTeX.

\appendix
\section{\label{sec:LoopEmission}Details on the emission of energy, momentum and angular momentum of the Vachaspati-Vilenkin loop}

It is of interest to explore how the different ranges of modes for energy, momentum and angular momentum emerge.
Qualitatively, there are three factors that influence these results.
\begin{itemize}
\item The intrinsic magnitudes of the emitted energy, momentum and angular
  momentum radiated differ.
\item The power emitted mode by mode is non-negative in all directions
  but the momentum and angular momentum emitted are signed quantities.
  Integrations of the latter over the sphere involve cancellations.
\item The low order modes do not have any simple power law scaling
  with $m$ while high order modes do.
\end{itemize}
We quantitatively describe these effects by comparing
the emission of momenta to the power radiated (take $G=\mu=1$ for
notational simplicity).  Define the ratio of $f$ with respect to $g$
by $O(f, g) = \int f d\Omega/\int g d\Omega$.

Consider the y-component of momentum radiated at harmonic $m$
compared to the power radiated at the same
harmonic.  Write $f=\dot{p_{y,m}}$ and $g=dP_{m}/d\Omega$ so that $O(f,
g) = O( |f|, g ) O( f, |f|) $. The first term provides an indication
of the magnitude of $f$ with respect to $g$, the second is sensitive
to the effect of sign cancellations for $f$ during the integration
over the sphere at fixed $m$. For large harmonics ($m \gtrsim 200$) we
find $O(|f|,g) \sim 0.57$ and $O(f, |f|) \sim -0.27$. The net effect
is $O(f,g) \sim -0.15$. For $f=\dot{p_{z,m}}$ the numbers are
essentially the same.

Repeating the analysis for the y and z components of angular momentum
we find $O(|f|,g) \sim 0.04$ and $-0.04$, respectively; $O(f, |f|) \sim -1$;
the net effect on radiated angular momentum is $O(f,g) \sim -0.04$
and $0.04$. Sign cancellations for angular momentum
are negligible whereas the magnitude of the angular momentum radiated
is much smaller than that of the momentum and both are small
compared to the energy radiated.

These results show that the mode-by-mode asymptotics for energy,
momentum and angular momentum differ on account of the intrinsic
magnitudes of the radiated quantities and the degree to which sign
cancellations occur. Next we join results for low and high order
modes. Let $m_*$ be an
approximate dividing point between low and high order modes such
that $m>m_*$ is well-described by the asymptotics. Denote
the spherical quadrature of quantity $I$ by $I_i$
for mode $i$. The cumulative sum up to mode $m$ is $I(\le m)$ (implicitly
assuming $m>m_*$) is $A + B \Delta(m_*,m)$
where $A=\sum_{i=1,m_*} I_i$ accounts for low order modes, $B$ is
a constant determined by the asymptotic transition and $\Delta(m_*,m)
= \zeta(4/3,m_*+1)-\zeta(4/3,m+1)$.  The incomplete Riemann zeta
function is $\zeta(s,a)=\sum_{k=0}^\infty=(k+z)^{-s}$. The total
sum ($m=\infty$) is $I (\le \infty) = A + B \zeta(4/3,m_*+1)$.

Note that $A$ incorporates the cancellations that accrue for a signed
quantity integrated over the sphere and variation from one
harmonic to another and must be found numerically; $B$ accounts for
the magnitude of the emission and for the cancellation effects in
signed quantities and is independent of $m$.  $\Delta$
accounts for the variation with $m$ for $m>m_*$.

Write the following approximations for the cumulative quadratures for
the 5 individual non-zero components
$\left\{ P, \dot{p_y}, \dot{p_z}, \dot{L_y},
  \dot{L_z} \right\} = {\vec A} + {\vec B} \Delta$ for 5-vectors
${\vec A} = \left\{66.7, -4.37, -6.81, -3.82, 3.08 \right\}$ and ${\vec
  B} = \left\{24.2, -3.78, -3.76, -1.00, 0.93 \right\}$.  The net
contributions at fixed large $m$ are proportional to ${\vec B}$.
Consistent with our discussion above $|B_{2,3}/B_1| \sim 0.16$ and
$|B_{4,5}/B_1| \sim 0.04$. The new information is $A_j/B_j$ which is
proportional to the low order over high order contributions.

To estimate the mode number $m$ to reach a fraction $x$ of the total
we set $x I( \le \infty) = I(\le m)$ and solve for $m$. For component
$j$ we find $m \simeq (3/((1-x)((A_j/B_j)+\zeta(4/3,m_*+1) ))^3
-1$. Note the sensitivity of $m$ to $x$ as $x \to 1$, to $A_j/B_j$ and
to the selection of $m_*$.  As an illustration, for $m_*=891$ and
$x=0.9$ we estimate $m$ for the 5 components
$\{930,8600,2800,380,570\}$. This agrees with a numerical evaluation
for each component carried out using the explicit, interpolated and
extrapolated contributions described above.  The treatment of the
magnitude of momentum and of angular momentum follows in a similar
fashion.

\section{\label{sec:AnalyticIntegrals}Analytic Integrals}

The integrals Eq. \eqref{eq:analytic_integrals} were done using Mathematica:
\begin{align}
    I_{1} &\equiv \int_{-\infty}^{\infty}dt\;e^{iA_{m}\left(t^{3}+p t+q \right)} \nonumber\\
    &= \frac{1}{9|p|^{9/2}}e^{i A_{m}q}\left[\sqrt{3}\pi \left(-p^{5}+|p|^{5}\right)J_{-1/3}\left(\frac{2|A_{m}||p|^{3/2}}{3\sqrt{3}}\right)\right.\nonumber\\
    &\left.+\sqrt{3}\pi \left(-p^{5}+|p|^{5}\right)J_{1/3}\left(\frac{2|A_{m}||p|^{3/2}}{3\sqrt{3}}\right)\right.\nonumber\\
    &\left.+3\left(p^{5}+|p|^{5}\right)K_{-1/3}\left(\frac{2|A_{m}||p|^{3/2}}{3\sqrt{3}}\right)\right]\\
    I_{2} &\equiv \int_{-\infty}^{\infty}dt\;t\;e^{iA_{m}\left(t^{3}+p t+q\right)} \nonumber\\
    &=\frac{i}{9}\text{sign}(A_{m}p)e^{i A_{m}q}\left[\pi \left(p-|p|\right)J_{-2/3}\left(\frac{2|A_{m}||p|^{3/2}}{3\sqrt{3}}\right)\right.\nonumber\\
    &\left.+\pi\left(p-|p|\right)J_{2/3}\left(\frac{2|A_{m}||p|^{3/2}}{3\sqrt{3}}\right)\right.\nonumber\\
    &\left.\sqrt{3}\left(p+|p|\right)K_{-2/3}\left(\frac{2|A_{m}||p|^{3/2}}{3\sqrt{3}}\right)\right]
\end{align}
where $J$ and $K$ are Bessel function of the first kind and modified Bessel function of the second kind respectively.

\section{\label{sec:Asymptotics}Asymptotics}
The methodology for a systematic expansion is given in
\url{https://www2.ph.ed.ac.uk/~mevans/amm/lecture04.pdf}. 
Writing $Fn$ as the $n$-th derivative of $F(z)$ at the critical point
and $gn$ as the $n$-th derivative of $g(z)$ we find the following
expressions:
\begin{eqnarray}
  a & = & (5 F3^2)/(24 F2^3)-F4/(8 F2^2)-\\
  & & (F3 g1)/(2 F2^2 g0)+g2/(2 F2 g0)\nonumber\\
  b & = & (385 F3^4)/(1152 F2^6)+(35 F4^2)/(384 F2^4)+\\
  & & (7 F3 F5)/(48 F2^4)+(35 F3 F4 g1)/(48 F2^4 g0)-\nonumber\\
  & & (F5 g1)/(8 F2^3 g0)+\nonumber\\
  & & (35 F3^2 (-3 F4 g0-4 F3 g1))/(192 F2^5 g0)+\nonumber\\
  & & (35 F3^2 g2)/(48 F2^4 g0)-(5 F4 g2)/(16 F2^3 g0)-\nonumber\\
  & & (5 F3 g3)/(12 F2^3 g0)+g4/(8 F2^2 g0)\nonumber
\end{eqnarray}

\section{\label{sec:Symmetry}Symmetry of Vachaspati-Vilenkin and Turok Loops}
The modes of the Turok and Vachaspati-Vilenkin loops have definite
parity under $\sigma \to -\sigma$ for the forms given in the paper.

{\it Turok:} Both $X_\pm$ modes have parity
$-$ in the x-component and $+$ in the y- and z-components,
i.e. $X^x_\pm(\sigma) = -X^x_\pm(-\sigma)$,
$X^y_\pm(\sigma) = X^y_\pm(-\sigma)$ and
$X^z_\pm(\sigma) = X^z_\pm(-\sigma)$. We abbreviate this as $-++$.

{\it Vachaspati-Vilenkin:} The $X_+$ mode has the same form and
hence the same parity as its Turok counter part $-++$. The
$X_-$ mode has parity $-+-$.

The fact that both modes of the Turok loop share the same
parity has direct implications for the emission process.
The symmetry of the Turok loops guarantees that no vector momentum is
radiated. This can be verified by considering the radiated components
with respect to antipodal directions of emission ${\hat k}$ and
$-{\hat k}$; this is equivalent to $\theta \to \pi - \theta$ and $\phi
\to \phi + \pi$ for the spherical polar system. Recall we defined
$\hat k$, $\hat u$ and $\hat v$ as the right-handed orthonormal basis
for analyzing emission by the string loop. Write the the spatial part
of $I^\mu_{\pm}$ as ${\mathbf{I}}_{\pm}$ and of $M^{\mu\nu}_{\pm}$ as
${\mathbf{M}}_{\pm}$. We must examine how the one dimensional
integrals transform when $\hat k$ changes direction. We start with
\begin{eqnarray}
  {\hat k} & \to & -{\hat k} \\
  {\hat u} & \to & -{\hat u} \\
  {\hat v} & \to & {\hat v}
\end{eqnarray}
and infer for harmonic mode number $n$ 
\begin{eqnarray}
  {\mathbf{I}}_{\pm} & \to & (-1)^{1+n} {\mathbf{I}}_{\pm} \\
  {\hat u} \cdot {\mathbf{I}}_{\pm} & \to & (-1)^{n} {\hat u} \cdot {\mathbf{I}}_{\pm} \\
  {\hat v} \cdot {\mathbf{I}}_{\pm} & \to & (-1)^{1+n} {\hat v} \cdot {\mathbf{I}}_{\pm} \\
  ({\hat u} \cdot {\mathbf{I}}_{\pm})
  ({\hat v} \cdot {\mathbf{I}}_{\pm})^* & \to &
  -({\hat u} \cdot {\mathbf{I}}_{\pm})
  ({\hat v} \cdot {\mathbf{I}}_{\pm})^*  .
\end{eqnarray}
These results imply that the each of the component parts of
${dP_m}/{d\Omega}$ in Eq. \eqref{dPdOmegaIpm} is invariant,
i.e. $|{\hat u} \cdot {\mathbf{I}}_{\pm}|$, $|{\hat v} \cdot
{\mathbf{I}}_{\pm}|$ and ${\rm Im}\left( ({\hat u} \cdot
{\mathbf{I}}_{+}) ({\hat v} \cdot {\mathbf{I}}_{+})^* \right) {\rm
  Im}\left( ({\hat u} \cdot {\mathbf{I}}_{-}) ({\hat v} \cdot
{\mathbf{I}}_{-})^*\right)$ are fixed for ${\hat k} \to -{\hat
  k}$. Equivalently, we can observe that $\tau_{ij}$ for indices $ij$ equal to $11$, $12$, $21$, $22$ and
$33$ are invariant whereas $\tau_{ij}$ for indices $13$, $23$, $31$ and $32$ change the
sign. Inserting in Eq. \eqref{eq:EMtensor} the sums $\tau_{pq}^* \tau_{pq}$
and $\tau_{qq}^* \tau_{pp}$ are invariant for $p$ and $q$ ranging
over values $2$ and $3$. This implies $dP_m/d\Omega$ unchanged
by the flip in ${\hat k}$. 

Next observe that ${d\mathbf{\dot{p}}_{m}}/{d\Omega}$ switches
signs since it is ${dP_m}/{d\Omega} \times {\hat k}$. No
net momentum is radiated.

By similar reasoning, for ${\hat k} \to -{\hat k}$ we find
${\mathbf{M}}_{\pm} \to (-1)^{n} {\mathbf{M}}_{\pm}$.
Now
${d{\dot{L}}_{m,u}}/{d\Omega}$ reverses sign
(both $\Re[i (3 \tau_{13}^* \tau_{pp} + 6 \tau_{3p}^* \tau_{p1})]$ and
$\Re[2 \tau_{3pq}^*\tau_{pq} - 2 \tau_{3p}^* \tau_{pqq} - \tau_{pq3}^*\tau_{pq} + (1/2) \tau_{qq3}^* \tau_{pp}]$ reverse sign summed over $p$ and $q$).
By similar reasoning, $\frac{d{\dot{L}}_{m,v}}{d\Omega}$
is invariant (both
$Re[i (3 \tau_{12}^* \tau_{pp} + 6 \tau_{2p}^* \tau_{p1})]$ and
$Re[2 \tau_{2pq}^*\tau_{pq} - 2 \tau_{2p}^* \tau_{pqq} - \tau_{pq2}^*\tau_{pq} + (1/2) \tau_{qq2}^* \tau_{pp}]$ are invariant).

For the emission along ${\hat k}$ we have
${\mathbf{d{\dot L}}}/{dt} = A {\hat u} + B {\hat v}$ where
$A={d{\dot{L}}_{m,u}}/{d\Omega}$ and 
$B={d{\dot{L}}_{m,v}}/{d\Omega}$. For the emission along $-{\hat k}$
we have shown $A \to -A$, $B \to B$, ${\hat u} \to -{\hat u}$ and
${\hat v} \to {\hat v}$. The conclusion is that
${\mathbf{d{\dot L}}}/{dt}$ is the same for ${\hat k}$ and $-{\hat k}$
directions.

The generic Turok loop radiates energy and angular momentum.

By similar arguments in which ${\hat x} \to -{\hat x}$ we can show that
the Vachaspati-Vilenkin loop does not radiate the x-component of momentum
or the x-component of angular momentum. The generic loop radiates
energy and y- and z-components of momentum and angular momentum.

\section{\label{sec:IplusforTurok}Calculation of \texorpdfstring{$I_{+}$}{I+}}
For the Turok loop presented in Section \ref{sec:Turokloop}, we know  $I_{+}$ exactly and outline its evaluation. Following \cite{Vachaspati_Vilenkin_1985,DURRER1989238}, we evaluate $I_{+}$ for the Turok loop as follows.
From Eq. \eqref{Imu}, 
\begin{align}
    I^{\mu}_{+}&=\frac{1}{l}\int_{-l/2}^{l/2}d\sigma_{+}\left\{1,\cos\left(\frac{2\pi\sigma_{+}}{l}\right),\sin\left(\frac{2\pi\sigma_{+}}{l}\right)\cos\Phi,\right.\nonumber\\
    &\left.\sin\left(\frac{2\pi\sigma_{+}}{l}\right)\sin\Phi\right\}e^{-im\varphi}
    \label{IplusTurok}
\end{align}
where 
\begin{align}
    \varphi&=\frac{2\pi}{l}k.X_{+}\nonumber\\
           &=-\frac{2\pi\sigma_{+}}{l}+\sin\theta\cos\phi\sin\left(\frac{2\pi\sigma_{+}}{l}\right)\nonumber\\
           &-\left(\sin\theta\sin\phi\cos\Phi+\cos\theta\sin\Phi\right)\cos\left(\frac{2\pi\sigma_{+}}{l}\right).
\end{align}
Making the change of variables $2\pi\sigma_{+}/l \to \eta$ and defining
\begin{align}
    x&=-\sin\theta\cos\phi\\
    y&=\left(\sin\theta\sin\phi\cos\Phi+\cos\theta\sin\Phi\right),
\end{align}
we can rewrite Eq. \eqref{IplusTurok},
\begin{equation}
    I^{\mu}_{+} = \left\{\Sigma_{m}(x,y),Ic_{m}(x,y),Is_{m}(x,y)\cos\Phi,Is_{m}(x,y)\sin\Phi\right\}
    \label{Iplusfinal}
\end{equation}
where

\begin{align}
    \Sigma_{m}(x,y)&=\frac{1}{2\pi}\int_{-\pi}^{\pi}d\eta\; e^{im(\eta+x\sin\eta+y\cos\eta)}\\
    Ic_{m}(x,y)&=\frac{1}{2\pi}\int_{-\pi}^{\pi}d\eta\; e^{im(\eta+x\sin\eta+y\cos\eta)}\cos\eta\\
    Is_{m}(x,y)&=\frac{1}{2\pi}\int_{-\pi}^{\pi}d\eta\; e^{im(\eta+x\sin\eta+y\cos\eta)}\sin\eta.
\end{align}
The integrals $\Sigma_{m},Ic_{m}$ and $Is_{m}$ can be expressed analytically in terms of Bessel functions. Defining $r=\sqrt{x^{2}+y^{2}}$,

\begin{align}
    \Sigma_{m}(x,y)&=\left(\frac{x-iy}{r}\right)^{m}J_{m}(-mr)\\
    Ic_{m}(x,y)&=\frac{1}{2}\left(\frac{x-iy}{r}\right)^{m}\left[\frac{x-iy}{r}J_{m+1}(-mr)\right.\nonumber\\
    &\left.\qquad\qquad+\frac{x+iy}{r}J_{m-1}(-mr)\right]\\
    Is_{m}(x,y)&=-\frac{i}{2}\left(\frac{x-iy}{r}\right)^{m}\left[\frac{x-iy}{r}J_{m+1}(-mr)\right.\nonumber\\
    &\left.\qquad\qquad-\frac{x+iy}{r}J_{m-1}(-mr)\right].
\end{align}

In summary for any $\alpha$, $\theta$ and $\phi$, Eq. \eqref{Iplusfinal} gives the exact value of $I_{+}$.

\section{\label{sec:formalsmallangleexpansion} Small angle expansion at special point}
We analyze the behavior of $\mathbf{I_\pm}$ for directions $\hat k$
that lie in the symmetry plane of tangent vectors $\vecXdotpm$. The
arc of $\hat k$ are in the $x=0$ plane in the example discussed in the
text. In this section we suppress all $\pm$ subscripts since the
results are valid equally to each mode.

The center of expansion $\sigma^*$ satisfies $k.\dot{X}(\sigma^*)=0$
when $\hat k$ points directly to the tangent curve in question; it
satisfies $k.\ddot{X}(\sigma^*)=0$ and $k.X^{(3)}(\sigma^*)<0$
elsewhere along the arc.

We consider situations in which the expansion point on each tangent
curve is fixed as $\hat k$ varies. Depending upon the local geometry
there may be one local expansion point on one side of the tangent
vector and two on the other side. The expansions below apply when
there is a single local point that does not vary as $\hat k$ varies.
We will suppress writing out $\sigma^*$ explicitly or including the
superscript $*$ in terms like $\Xdot^{*\mu}$.

We work to lowest non-vanishing order in $\delta$. This is effectively
a small angle approximation for $\mathbf{I}$ about
the tangent curve direction. The arc lies in the $x=0$ plane, the
varying angle is $\theta$.

For the generic case with $\vecXddot$ non-vanishing, the coefficients of the multipoint method reduce to
\begin{align}
    A_{m} &= -\frac{2\pi m}{l}\left(-\frac{1}{6}|\ddot{X}|^{2}\right),\\
    B_{m} &= 0,\\
    C_{m} &= -\frac{2\pi m}{l}\delta_{\mu}\Xdot^{\mu},\\
    D_{m} &= -\frac{2\pi m}{l}\left(\Xdot_{\mu}X^{\mu}\right),\\
    p &= \frac{C_{m}}{A_m},\\
    &= - 6 \frac{\mathbf{\delta} . \vecXdot}{\vecXddot . \vecXddot}.
\end{align}
There are two terms that
must be calculated $I=I_1 \vecXdot + I_2 \vecXddot $.

For the calculations for the first term with $I_1$, define
\begin{align}
  \beta &= \frac {2^{3/2} e^{i D_m}}{3^{1/2}}
  \sqrt{ \frac{|\mathbf{\delta} . \vecXdot|}{| \vecXddot . \vecXddot |} }\\
  \rho &= \frac{2^{5/2} \pi m}{3} 
  \frac{|\mathbf{\delta} . \vecXdot|^{3/2}}{ |\vecXddot . \vecXddot|^{1/2}}.
\end{align}

It will be useful to introduce the abbreviation
\begin{equation}
  \rm{L}_n(x) \equiv \frac{\pi}{3^{1/2}} \left(
  \rm{J}_{-n}\left(x\right) + \rm{J}_{n}\left(x\right) \right)
\end{equation}
and, since $p$ can have both signs, the $p$-dependent function
\begin{equation}
  \rm{M}_n(x) \equiv \left\{ \rm{K}_n(x), \rm{L}_n(x) \right\} 
\end{equation}
where the first entry is for $p>0$ and the second for $p<0$.

The result is
\begin{equation}
  {I}_1 = \beta \rm{M}_{-1/3} (\rho) .
\end{equation}

For the second term with $I_2$ letting
\begin{equation}
  \gamma = i \frac{|p|^{1/2}}{3^{1/2}} \beta 
\end{equation}
we have
\begin{equation}
  {I}_2 = \gamma \rm{M}_{-2/3} \left( \rho \right) 
\end{equation}
and $\mathbf{I}=I_1 \vecXdot + I_2 \vecXddot$.

\section{\label{sec:3methods_scenarios} Details on the efficiency of FFT, multipoint and the asymptotic methods}

An FFT of length $N$ has a Nyquist frequency $N/2$; harmonics with
$m<N/2$ can be represented for an evenly sampled time series. Using
the FFT as a quadrature technique, however, requires uneven sampling
of the waveform (as explained in \cite{Allen_Shellard_1992,
  Blanco_Pillado_2017} and reviewed in Section
\ref{sec:directreal}). A transform of length $M=cN$ is utilized where
$c$ is typically $8-16$. Aliasing effects degrade the accuracy of the
quadratures as $m$ increases at fixed $M$. Increasing $c$ yields
exponential convergence at fixed $m$.

Consider a loop of given configuration and the task of calculating
$I_{\pm}$ for a given direction of emission.  There are two pieces:
(1) root-finding to generate unevenly spaced points $\sigma_{i}$
corresponding to a set of evenly spaced points $\bar{x}_{i}=2\pi
\left(k.X_{\pm}(\sigma_{i})-k.X_{\pm}(0)\right)$ and (2) performing
the FFT itself. The root-finding for $\sigma_{i}$ must be done $M$
times with cost $\mathcal{O}\left(M \right)$. FFT cost is
$\mathcal{O}\left(M \log M\right)$. In our implementation and at the
$M$ we have studied the time of root-finding dominates that of the
highly optimized FFT and all other mathematical calculations of
$I_{\pm}$. The cost for the $M$ studied scales with $M$ and the
cost per mode is constant, dominated by the root-finding cost (with
implicit multiplier $c$).

Now consider the costs for the same loop configuration and direction
of emission in the context of the approximate methods.  The multipoint
method requires estimating the expansion points. The expansion points
are independent of $m$ so this need be done only once. The rest of the
cost of finding $I_{\pm}$ is evaluation of special functions. Finding
a large range of $m$ implies the cost per mode is constant, dominated
by evaluation of the special functions which are typically very fast.

The situation for the complex asymptotic method is similar.  One finds
the critical points and the corresponding values of the coefficients
$a,b,$ etc. once. The cost per mode is constant, dominated by
evaluation of the special functions.

Assume we intend to find $I_{\pm}$ for a large range of modes.

If we are interested in minimizing total time independent of
error considerations, we should minimize the number of modes
treated by the FFT to $1\le m < m_{low}$. The rest should be done
with the approximate techniques.

If multipoint provides an acceptable level of error then we should
divide the FFT, multipoint and complex asymptotic regimes as discussed
in the text. Typically, FFT $1 \le m < m_{OK}$, multipoint $m_{OK} < m
< m_{cross}$ and complex asymptotic $m_{cross} < m$.

If we are interested in driving errors below that which can be
provided by multipoint then we should extend the FFT until it's
maximum error intersects that of the complex asymptotic method as
discussed in the text. Typically, FFT $1 \le m < m_{OK}$ and complex
asymptotic $m_{OK} < m$ where $m_{OK}>m_{cross}$.

\section{\label{sec:expand}Expansion of $\alpha=3/20$ and $\Phi=-18\pi/25$}

\begin{figure}[htbp]
    \centering
    \includegraphics[width=0.8\linewidth, height=7cm]{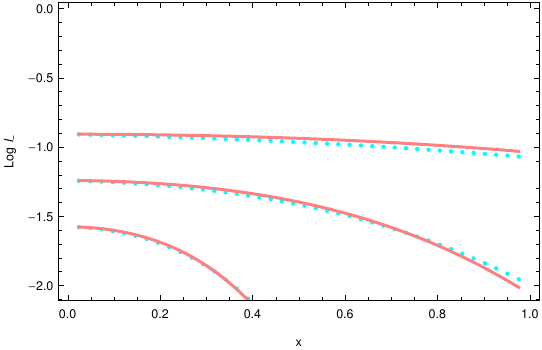}
    \caption{$I_-$ as a function of $x$ for $m=3 \times 10^2$, $3 \times 10^3$ and $3 \times 10^4$
      for multipoint and for analytic simplified multipoint
      (cyan dotted and red solid lines, respectively).}
    \label{fig:MultivsApproxMultiI.pdf}
\end{figure}

\begin{figure}[htbp]
    \centering
    \includegraphics[width=0.8\linewidth, height=7cm]{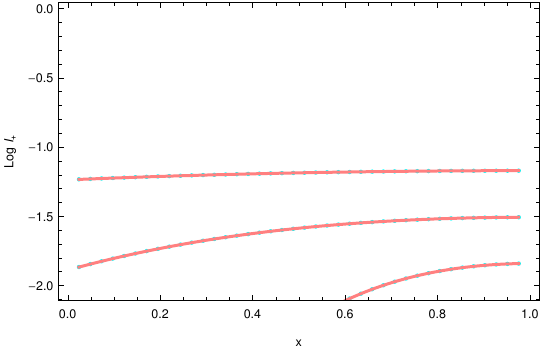}
    \caption{$I_+$ as a function of $x$ for $m=3 \times 10^2$, $3 \times 10^3$ and $3 \times 10^4$
      for multipoint and for analytic simplified multipoint
      (cyan dotted and red solid lines, respectively).}
    \label{fig:MultivsApproxMultiJ.pdf}
\end{figure}

The expansion point $\sigma_-=1/4$ for $I_-$ implies ${\hat
  k}_-=\{0,0.7,0.7141\}$ and $\theta=0.7754$.
To lowest order we have for the important
parameters for the multipoint expansion of $I_-$
\begin{align}
  A_m & = \frac{16\pi^{3}m}{75}\\
  B_m & = 0 \\
  p & = \frac{75d\theta^{2}}{16 \pi^2} \\
  q & = \frac{75}{32 \pi^2}.
\end{align}
At $\sigma_-=1/4$, we have
\begin{align}
    \vecXdot &= \left\{0, 7/10,\sqrt{51}/10\right\},\\
    \vecXddot &= \left\{4\pi/5,0,0\right\}.
\end{align}
Thus the integral $I_-$ becomes,
\begin{align}
  \mathbf{I}_- &= I_{1}\vecXdot+I_{2}\vecXddot\\
  I_{1}        &= \beta {\rm K}_{-1/3}(\rho)\\
  I_{2}        &= \gamma {\rm K}_{-2/3}(\rho)
\end{align}
where 
\begin{align}
  \beta &= \frac{5 |d\theta| e^{i m \pi/2}}{2\sqrt{3} \pi} \\
  \rho & = \frac{5 m |d\theta|^3}{6} .
\end{align}
for $d\theta <0$. The expansion point $\sigma_+=-1/4$ for $I_+$ implies ${\hat k}_+ = \{0,0.6374, 0.7705\}$ and $\theta=0.6912$. To lowest order we have 
\begin{align}
  A_m & = \frac{4 m \pi^3}{3}\\
  B_m & = 0 \\
  p & = \frac{3 d\theta^{2}}{4\pi^{2}} \\
  q & = -\frac{3}{8 \pi^2}
\end{align}
and
\begin{align}
    \vecXdot &= \left\{0, \sin \epsilon, \cos \epsilon\right\},\\
    \vecXddot &= \left\{2\pi,0,0\right\}.
\end{align}
for $\epsilon=11\pi/50$. Thus we get
\begin{align}
  \mathbf{I}_+ &= I_{1}\vecXdot+I_{2}\vecXddot\\
  I_{1}        &= \beta {\rm K}_{-1/3}(\rho)\\
  I_{2}        &= \gamma {\rm K}_{-2/3}(\rho)
\end{align}
where 
\begin{align}
  \beta &=\frac{ d\theta e^{-i m \pi/2}}{\sqrt{3} \pi} \\
  \rho & = \frac{m d\theta^3}{3}
\end{align}
for $d\theta>0$.
Figure \ref{fig:MultivsApproxMultiI.pdf} and Fig. \ref{fig:MultivsApproxMultiJ.pdf} show the approximations are essentially indistinguishable from the full
multipoint calculations.


\end{document}